\documentclass[twocolumn]{aastex63}

\usepackage{gensymb}
\usepackage{amsmath}
\usepackage{array}
\usepackage{graphicx}

\newcolumntype{L}{>{\raggedright\arraybackslash}p{8.8cm}}

\graphicspath{{./}{figures/}}

\shorttitle{CatWISE2020}
\shortauthors{Marocco et al.}
\begin{document}
\title{The CatWISE2020 Catalog}
\correspondingauthor{Federico Marocco}
\email{federico@ipac.caltech.edu}

\author[0000-0001-7519-1700]{Federico Marocco}
\affiliation{Jet Propulsion Laboratory, California Institute of Technology, 4800 Oak Grove Drive, M/S 169-327, Pasadena, CA 91109, USA}
\affiliation{IPAC, Mail Code 100-22, California Institute of Technology, 1200 E. California Blvd., Pasadena, CA 91125, USA}

\author{Peter R. M. Eisenhardt}
\affiliation{Jet Propulsion Laboratory, California Institute of Technology, 4800 Oak Grove Drive, M/S 169-327, Pasadena, CA 91109, USA}

\author[0000-0002-7564-9643]{John W. Fowler}
\affiliation{230 Pacific St., Apt. 205, Santa Monica, CA 90405, USA}

\author[0000-0003-4269-260X]{J. Davy Kirkpatrick}
\affiliation{IPAC, Mail Code 100-22, California Institute of Technology, 1200 E. California Blvd., Pasadena, CA 91125, USA}

\author[0000-0002-1125-7384]{Aaron M. Meisner}
\affiliation{NSF’s National Optical-Infrared Astronomy Research Laboratory, 950 N Cherry Ave, Tucson, AZ 85719, USA}

\author[0000-0002-3569-7421]{Edward F. Schlafly}
\affiliation{Lawrence Berkeley National Laboratory, One Cyclotron Road, Berkeley, CA 94720, USA}

\author{S. Adam Stanford}
\affiliation{Department of Physics, University of California Davis, One Shields Avenue, Davis, CA 95616, USA}

\author{Nelson Garcia}
\affiliation{IPAC, Mail Code 100-22, California Institute of Technology, 1200 E. California Blvd., Pasadena, CA 91125, USA}

\author{Dan Caselden}
\affiliation{Gigamon Applied Threat Research, 619 Western Avenue, Suite 200, Seattle, WA 98104, USA}

\author[0000-0001-7780-3352]{Michael C. Cushing}
\affiliation{Department of Physics and Astronomy, University of Toledo, 2801 West Bancroft St., Toledo, OH 43606, USA}

\author{Roc M. Cutri}
\affiliation{IPAC, Mail Code 100-22, California Institute of Technology, 1200 E. California Blvd., Pasadena, CA 91125, USA}

\author[0000-0001-6251-0573]{Jacqueline K. Faherty}
\affiliation{Department of Astrophysics, American Museum of Natural History, Central Park West at 79th Street, NY 10024, USA}

\author{Christopher R. Gelino}
\affiliation{IPAC, Mail Code 100-22, California Institute of Technology, 1200 E. California Blvd., Pasadena, CA 91125, USA}

\author{Anthony H. Gonzalez}
\affiliation{Department of Astronomy, University of Florida, 211 Bryant Space Center, Gainesville, FL 32611, USA}

\author[0000-0002-4939-734X]{Thomas H. Jarrett}
\affiliation{Department of Astronomy, University of Cape Town, Private Bag X3, Rondebosch, 7701, South Africa} 

\author{Renata Koontz}
\affiliation{University of California, Riverside, 900 University Ave, Riverside, CA 92521, USA}

\author{Amanda Mainzer}
\affiliation{Lunar and Planetary Laboratory, University of Arizona, Tucson, AZ 85721, USA}

\author{Elijah J. Marchese}
\affiliation{University of California, Riverside, 900 University Ave, Riverside, CA 92521, USA}

\author{Bahram Mobasher}
\affiliation{University of California, Riverside, 900 University Ave, Riverside, CA 92521}

\author[0000-0002-5042-5088]{David J. Schlegel}
\affiliation{Lawrence Berkeley National Laboratory, Berkeley, CA, 94720, USA}

\author{Daniel Stern}
\affiliation{Jet Propulsion Laboratory, California Institute of Technology, 4800 Oak Grove Drive, M/S 169-327, Pasadena, CA 91109, USA}

\author{Harry I. Teplitz}
\affiliation{IPAC, Mail Code 100-22, California Institute of Technology, 1200 E. California Blvd., Pasadena, CA 91125, USA}

\author[0000-0001-5058-1593]{Edward L. Wright}
\affiliation{Department of Physics and Astronomy, UCLA, 430 Portola Plaza, Box 951547, Los Angeles, CA 90095-1547, USA}

\begin{abstract}

The CatWISE2020 Catalog consists of 1,890,715,640 sources over the entire sky selected from WISE and NEOWISE survey data at 3.4 and 4.6 \micron\ (W1 and W2) collected from 2010 Jan. 7 to 2018 Dec. 13. This dataset adds two years to that used for the CatWISE Preliminary Catalog \citep{Eisenhardt2020}, bringing the total to six times as many exposures spanning over sixteen times as large a time baseline as the AllWISE catalog. The other major change from the CatWISE Preliminary Catalog is that the detection list for the CatWISE2020 Catalog was generated using \textit{crowdsource} \citep{Schlafly2019}, while the CatWISE Preliminary Catalog used the detection software used for AllWISE. These two factors result in roughly twice as many sources in the CatWISE2020 Catalog. The scatter with respect to \textit{Spitzer} photometry at faint magnitudes in the COSMOS field, which is out of the Galactic plane and at low ecliptic latitude (corresponding to lower WISE coverage depth) is similar to that for the CatWISE Preliminary Catalog. 
The 90\% completeness depth for the CatWISE2020 Catalog is at W1=17.7\,mag and W2=17.5\,mag, 1.7\,mag deeper than in the CatWISE Preliminary Catalog. From comparison to \textit{Gaia}, CatWISE2020 motions are accurate at the 20\,mas\,yr$^{-1}$ level for W1$\sim$15\,mag sources, and at the $\sim100$\,mas\,yr$^{-1}$ level for W1$\sim$17\,mag sources. This level of precision represents a 12$\times$ improvement over AllWISE.  
The CatWISE catalogs are available in the WISE/NEOWISE Enhanced and Contributed Products area of the NASA/IPAC Infrared Science Archive. 
 
\end{abstract}


\vspace{0.5in}

\section{Introduction \label{sec:intro}}

The CatWISE Preliminary Catalog \citep{Eisenhardt2020}, which was released via the NASA/IPAC Infrared Science Archive in August 2019, consists of 900,849,014 sources over the entire sky selected from \textit{WISE} \citep{Wright2010} and NEOWISE \citep{Mainzer2014} survey data at 3.4 and 4.6 \micron\ (W1 and W2) collected from 2010 to 2016. This dataset includes four times as many exposures and spans over ten times as large a time baseline as the AllWISE catalog \citep{Cutri2013}. CatWISE adapts AllWISE software to measure the sources in co-added images created by the unWISE team from six month subsets of these data, each representing one coverage of the inertial sky, or epoch \citep{Meisner2018}. The CatWISE Preliminary Catalog includes the measured motion of sources in 8 epochs over the 6 year span of the data, which are ten times more accurate than those from AllWISE. The CatWISE Preliminary Catalog has been used to identify some of the coldest brown dwarfs known to date \citep{Marocco2019,Marocco2020,Meisner2020}.

Nevertheless, further significant improvements are possible. The most important caveat for the CatWISE Preliminary Catalog is that the number of sources per square degree has relatively small variation over the sky. This is likely a consequence of the source detection methodology used for the Preliminary Catalog \citep{Eisenhardt2020}, which, while optimal for isolated point sources, results in significant incompleteness in high source density regions such as the Galactic plane. 

\begin{figure}
    \centering
    \includegraphics[width=0.49\textwidth]{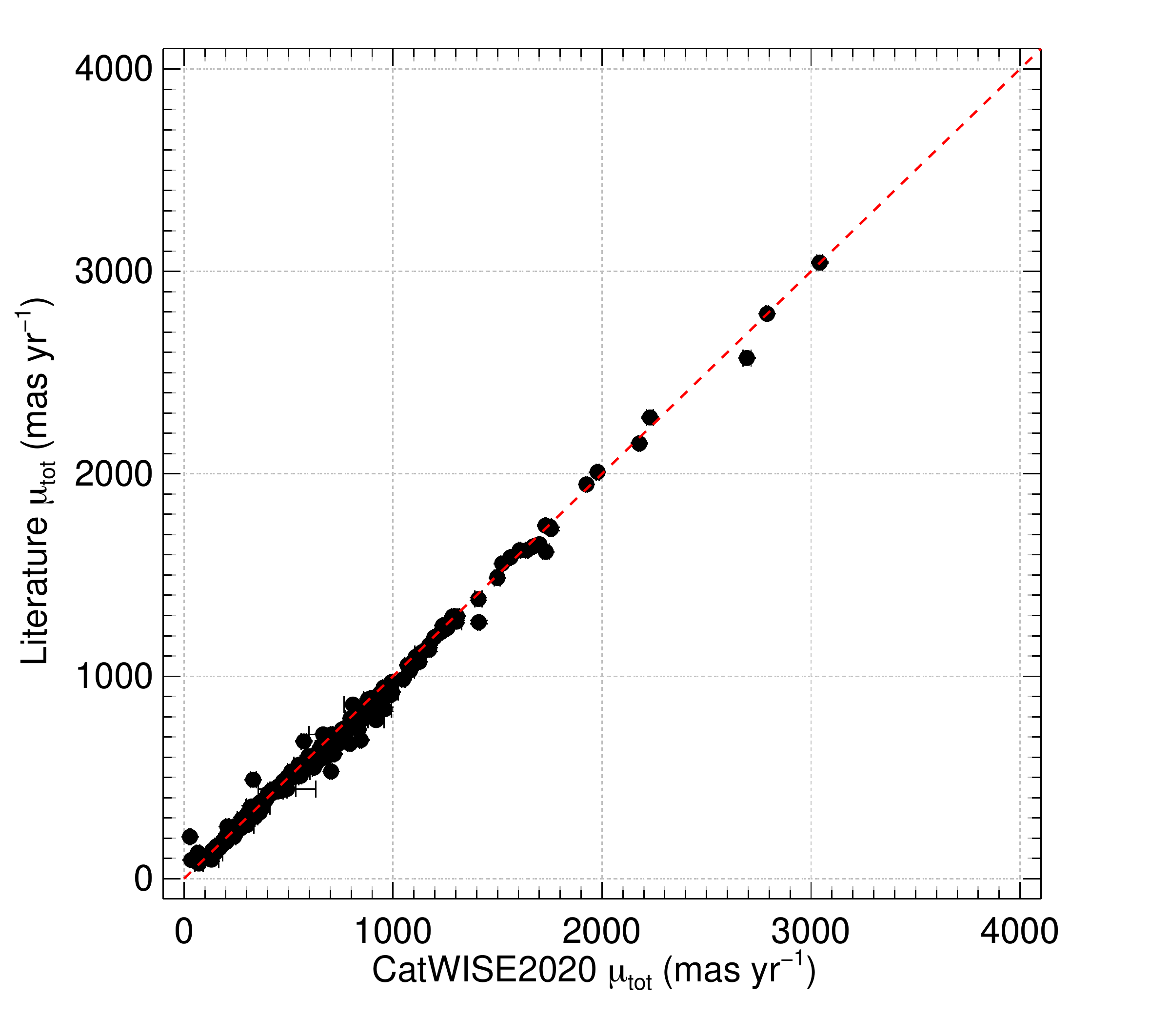}
    \caption{A comparison of CatWISE2020 total measured motion to values reported in the literature, for 224 ultra-cool dwarfs within 20\,pc of the Sun. All have S/N $\geq$3 CatWISE2020 motion measurements. Vertical error bars are typically smaller than the symbols.}
    \label{fig:fast_movers}
\end{figure}

The CatWISE2020 Catalog addresses this issue by using an updated version  of the unWISE catalog \citep{Schlafly2019} as the detection list. In addition, the CatWISE2020 Catalog includes two more years of survey data from NEOWISE than does the Preliminary Catalog, increasing the number of epochs to 12 and the time span to over 8 years. As a result, the CatWISE2020 Catalog has more than twice as many sources as the CatWISE Preliminary Catalog (five times as many in the Galactic plane; see \S\ref{sec:mdet}), and even better astrometric performance for faint sources. Figure~\ref{fig:fast_movers} shows a comparison of CatWISE2020 total proper motions $\left( {\rm i.e.}\ \mu_{\rm tot} = \sqrt{ \mu_\alpha^{*\,2} + \mu_\delta^2}\right)$\footnote{Throughout this manuscript, we adopt the notation $\mu_\alpha^* = \mu_\alpha \cos \delta$ and $\Delta\alpha^* = \Delta\alpha \cos \delta$.} for a sample of 224 ultra-cool dwarfs within 20\,pc from the Sun to values reported in the literature \citep[and references therein]{Kirkpatrick2020}. The agreement is excellent across the whole motion range, and across a broad range of magnitudes, since the sample includes objects as bright as W2$\sim$7.3\,mag and as faint as W2$\sim$16.7\,mag. The CatWISE2020 Catalog is therefore an excellent resource to identify ultra-cool dwarfs in the Solar neighborhood.

\citet{Eisenhardt2020} presents a detailed description of the CatWISE Preliminary Catalog. Here we describe updates to the processing steps used for the CatWISE2020 Catalog relative to the Preliminary Catalog (\S\ref{sec:processing}), assess the astrometric and photometric performance of CatWISE2020 Catalog using comparisons to {\it Gaia} and {\it Spitzer} data (\S\ref{sec:performance}), and provide information on accessing the CatWISE2020 data products (\S\ref{sec:access}).  The Appendix summarizes known issues in the CatWISE2020 Catalog.

\section{CatWISE2020 Processing Updates \label{sec:processing}}

A full description of the processing steps for the CatWISE Preliminary Catalog is given in \citet{Eisenhardt2020}, and we only describe changes for CatWISE2020 processing here.  Coordinates in the CatWISE2020 Catalog are in the ICRS system at epoch MJD=57170 (2015 May 28), while in the CatWISE Preliminary Catalog they are in J2000 at epoch MJD=56700.

\subsection{unWISE Coadds \label{sec:unwise}}
The unWISE coadds follow the atlas tile footprint established by the WISE All-Sky Release, dividing the sky into 18,240 overlapping square tiles, each $\sim$1.56\,deg on a side, aligned with right ascension and declination coordinate grid. Each tile is identified by a \textit{coadd\_id}, in the form RRRRsDDD, where RRRR is the tile center R.A. in deci-degrees, s is the tile center Dec sign, with ``p'' for ``+'' and ``m'' for ``--'', and DDD is the tile center Dec in deci-degrees. For example, the tile containing the South Ecliptic Pole and centered near R.A.\,=\,89.0\,deg and Dec\,=\,--66.7\,deg is 0890m667. For further detail, we refer the reader to section II.1.a in the Explanatory Supplement to the AllWISE Data Release Products \citep{Cutri2013}.

The CatWISE2020 data products are based on the combination of W1 and W2 exposures in the two sky coverages used for the AllWISE data release \citep{Cutri2013} and in the ten additional sky coverages from the NEOWISE 2019 data release\footnote{\url{ http://wise2.ipac.caltech.edu/docs/release/neowise/neowise\_2019\_release\_intro.html}}, while the CatWISE Preliminary data products use AllWISE and the six additional sky coverages from the NEOWISE 2017 Data Release\footnote{\url{http://wise2.ipac.caltech.edu/docs/release/neowise/neowise\_2017\_release\_intro.html}}. The full-depth unWISE coadds from \citet{Meisner2018a} are used for both source detection and aperture photometry in the CatWISE Preliminary Catalog. The CatWISE2020 pipeline uses the full-depth unWISE coaddition of the AllWISE and NEOWISE 2019 Data Release for aperture photometry, while source detection is described in \S\ref{sec:mdet}. 

The CatWISE Preliminary pipeline used the 8 individual unWISE epoch coadds from \citet{Meisner2018c} for point source photometry and astrometry, with an adjustment to the world coordinate system (WCS) for the AllWISE epochs, as described in \citet{Eisenhardt2020}. The CatWISE2020 pipeline uses 12 unWISE epoch coadds constructed using the methodology given in \citet{Meisner2019}. We assumed that the adjustment to the AllWISE epochs had been applied to this set of 12 unWISE coadds, and did not discover the adjustment was not applied until after the CatWISE2020 Catalog and Reject Table had been generated and imported into the IRSA database. This results in small astrometric offsets for sources in the CatWISE2020 Catalog, for which we provide a table of corrections (see \S\ref{sec:astrom_perf}).

\subsection{Detection\label{sec:mdet}}

\begin{figure}
    \centering
    \includegraphics[width=0.49\textwidth]{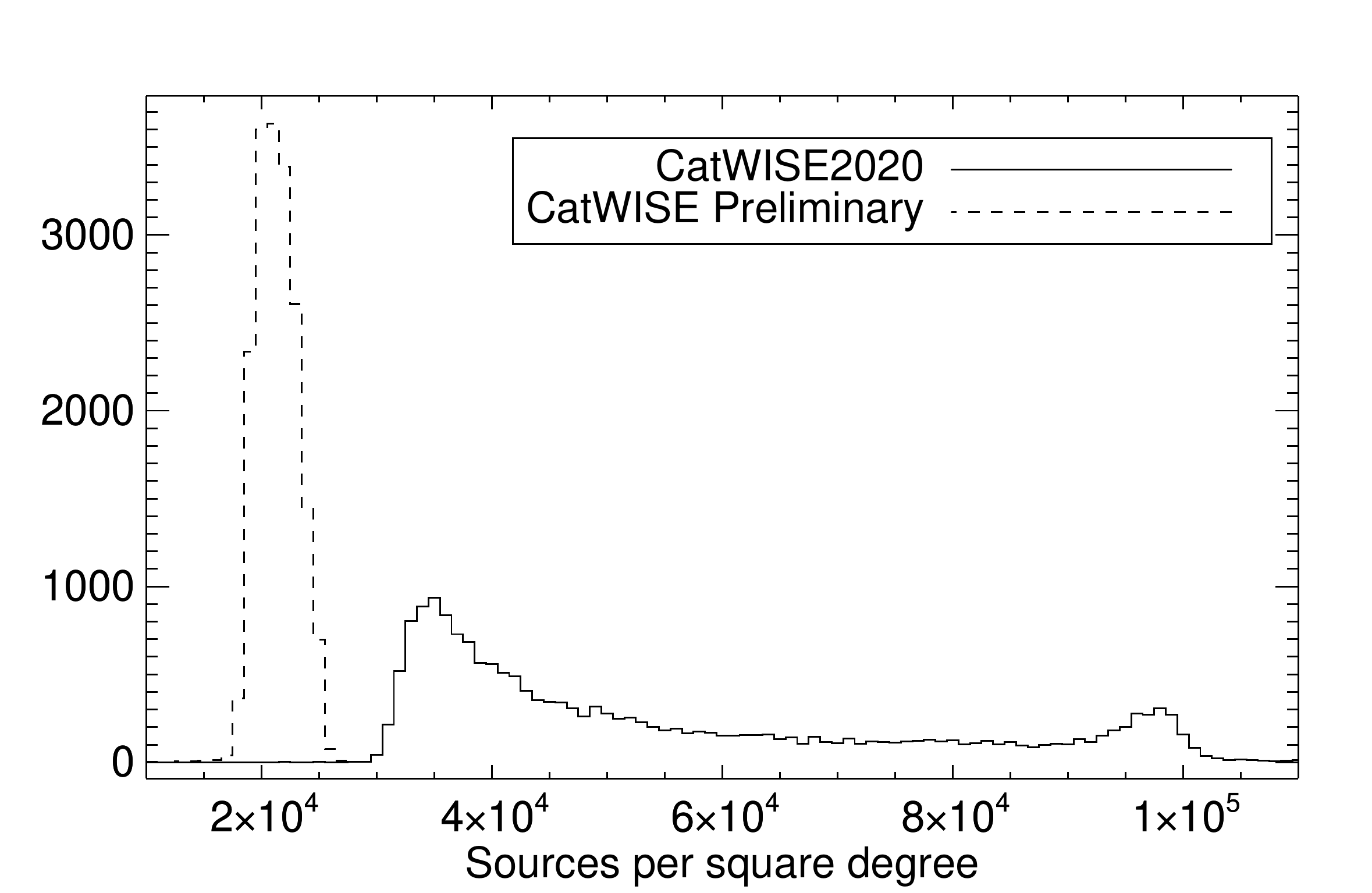}
    \caption{Number of tiles with a given source density, in bins of 1000 sources per square degree, for the CatWISE2020 Catalog compared to the CatWISE Preliminary Catalog. Deeper data, and a more effective detection of sources in crowded fields, result in a broader distribution and a greater number of sources per square degree in the CatWISE2020 Catalog.}
    \label{fig:nsrc_histo}
\end{figure}

\begin{figure*}
    \includegraphics[width=\textwidth]{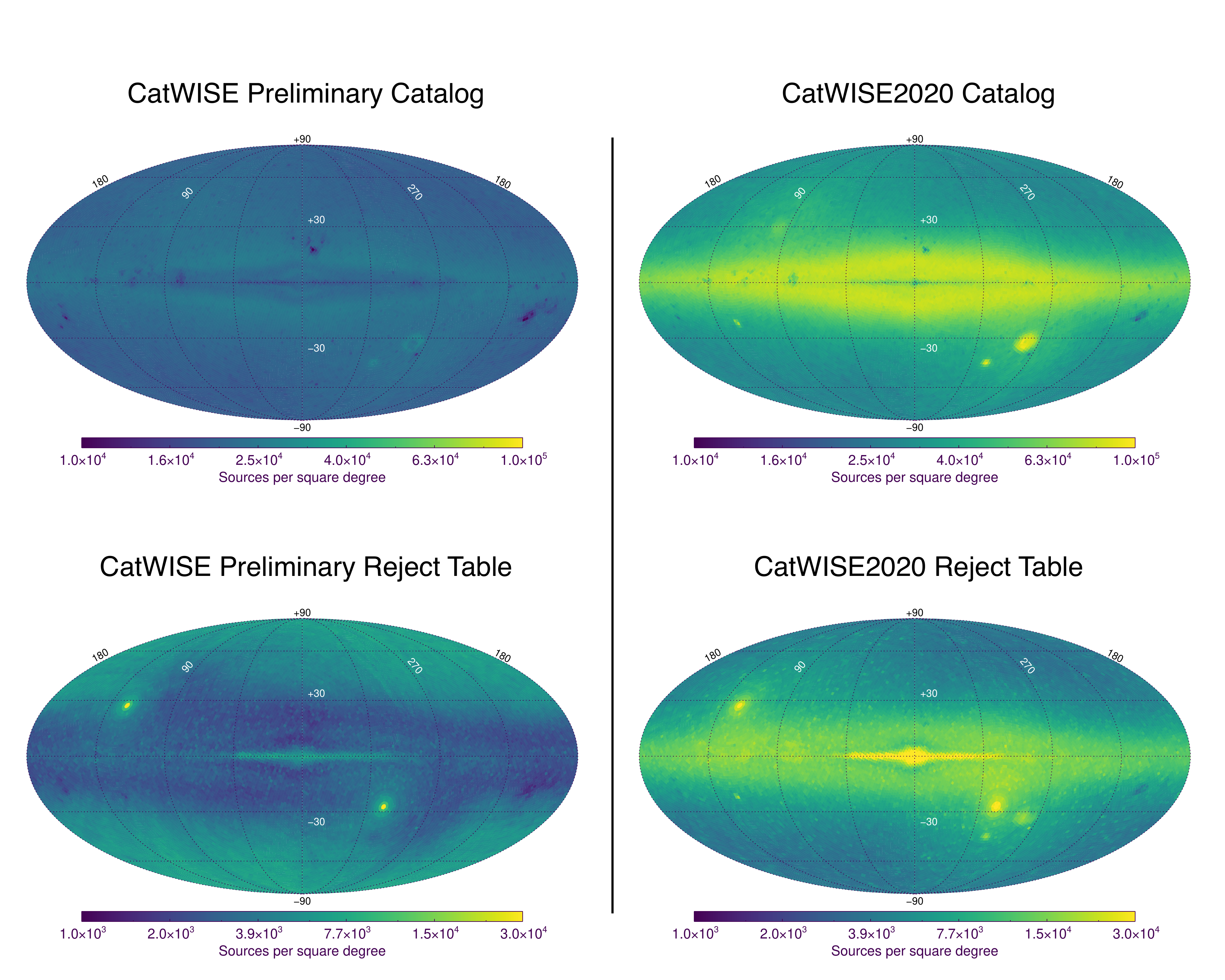}
    \caption{CatWISE Preliminary and CatWISE2020 Catalog (top) and Reject Table (bottom) source density, plotted in Galactic Coordinates. See \S\ref{sec:2020} for details.}
    \label{fig:source_density}
\end{figure*}

The histogram of the number of detected sources per square degree in the CatWISE Preliminary Catalog is narrow (Figure \ref{fig:nsrc_histo}). This means that the source density is no higher in the Galactic plane than at the Galactic poles; in fact, it is slightly lower in the Galactic plane, as the upper left panel of Figure \ref{fig:source_density} shows. The unWISE Catalog \citep{Schlafly2019} uses a crowded-field point-source photometry code called \textit{crowdsource} \citep{Schlafly2018} which detects far more sources in high density regions. 

For the CatWISE2020 Catalog we therefore decided to replace the  CatWISE Preliminary Catalog detection step, which uses MDET, the Multiband Detection software of \citet{MarshJarrett2012}, with an updated version of the unWISE Catalog (hereafter UUC)\footnote{The updated unWISE Catalog is available at \url{https://faun.rc.fas.harvard.edu/unwise/neo5/band-merged}}. The unWISE Catalog of \citet{Schlafly2019} is based on the NEOWISE 2018 Data Release, while the UUC used for the CatWISE2020 detection list is based on the NEOWISE 2019 Data Release. The unWISE Catalog measures source fluxes and static positions in the full-depth coadded image, with the measurements carried out independently in W1 and W2, while the CatWISE2020 pipeline characterizes sources jointly in both bands, measuring their fluxes, positions, and motions in epoch coadds. 

CatWISE2020 processing begins with the band-merged UUC, where photometry for W1 and W2 sources within 2\farcs4 is matched. Sources are measured in order of decreasing signal-to-noise ratio (S/N) within each $\sim1.56$\,deg$\times1.56$\,deg tile on the sky. To generate a single S/N-ordered detection list from the unWISE photometry in both bands, the CatWISE2020 pipeline determines an average flux uncertainty $\sigma_{avg}$ for each band from the mean of the flux uncertainty of sources in the tile whose flux $F$ is within $\pm 0.5\%$ of the median flux in the band. The S/N in each band is calculated from  $F/\sigma_{avg}$, and the W1 and W2 S/N values are root-sum-squared to determine a combined S/N for each source in the tile. 

Figure \ref{fig:nsrc_histo} shows the resultant histogram of source densities for the CatWISE2020 Catalog and Reject Table compared to the CatWISE Preliminary Catalog and Reject Table, and Figure  \ref{fig:source_density} compares the distribution of source density over the sky. With more than twice as many sources overall, the CatWISE2020 Catalog has a higher source density across the whole sky. The catalog source density still drops in the central part of the Galactic plane and the Galactic center. Source are successfully detected in these regions, but a significant fraction of them fail one or more of the criteria for inclusion in the catalog, and therefore go in the reject table. Further details on the selection criteria and a more in-depth discussion of the resulting source density are presented in \S\ref{sec:2020}.

One feature arising from the use of the UUC is the fact that some nearby, resolved galaxies are split in multiple pieces, leading to spurious localized overdensity of sources in the CatWISE2020 Catalog. A detailed discussion of the \textit{crowdsource} treatment of extended galaxies is given in \citet[\S4.4 and \S6.8]{Schlafly2019}, and here we briefly summarize the most important points. Splitting of extended galaxies only happens when the angular size of the galaxy exceeds several arcseconds, due to the 6'' FWHM of WISE. Large galaxies listed in the HyperLEDA catalog \citep{Makarov2014} received special treatment in the unWISE catalog. Elliptical regions around these galaxies are flagged in the unWISE bit masks \citep{Meisner2019}, and \textit{crowdsource} rejects candidate new sources in these regions if they significantly overlap with a neighboring source. Galaxies in the HyperLEDA catalog are therefore less likely to be split into multiple detections, but heterogeneity in the catalog results in a somewhat sky position-dependent effectiveness of this mitigation. Ultimately, this is a trade off between completeness (specifically, the ability to successfully deblend real, blended sources) and reliability (specifically, the need to minimize the fragmenting of resolved sources). Completeness and reliability of the CatWISE2020 Catalog are discussed in further detail in \S\ref{sec:CandR}.

\subsection{Source Measurement and Uncertainties  \label{sec:wphot}}

As for the CatWISE Preliminary Catalog, the CatWISE2020 pipeline used an adapted version of the WPHOT software developed for the AllWISE pipeline to carry out source photometry, astrometry, and motion estimation, processing separately epochs taken with ascending vs. descending survey scan directions, and then merging the results \citep{Eisenhardt2020}. For the CatWISE2020 pipeline however, we did not allow WPHOT to add new sources (``active deblending") to improve the $\chi^2$ of the fit to a source, as the \textit{crowdsource} algorithm (\S\ref{sec:mdet}) already provides a much more complete set of detected sources than does the MDET algorithm used for the CatWISE Preliminary Catalog.
 
In the CatWISE Preliminary Catalog, a minimum motion uncertainty of 10 mas yr$^{-1}$ was enforced, while for the CatWISE2020 Catalog, this floor was reduced to 1 mas yr$^{-1}$. This affects the $\chi^2$ values when comparing \textit{Gaia} and CatWISE2020 motions (see \S\ref{sec:astrom_4tiles}).

\subsection{The CatWISE2020 Catalog \label{sec:2020}} 

The final CatWISE2020 data products consist of a Catalog and a Reject Table. Like the CatWISE Preliminary Catalog, CatWISE2020 Catalog sources are required to:

1) be from the tile where that source is furthest from the tile edge (i.e. flagged as ``primary'') 

\noindent and

2a) have W1 S/N $\geq 5$ with no identified artifacts (a value of 0 in the left character of {\it ab\_flags}) 

\noindent or

2b) have W2 S/N $\geq 5$ with no identified artifacts (a value of 0 in the right character of {\it ab\_flags}).

\begin{figure}
    \centering
    \includegraphics[width=0.49\textwidth=]{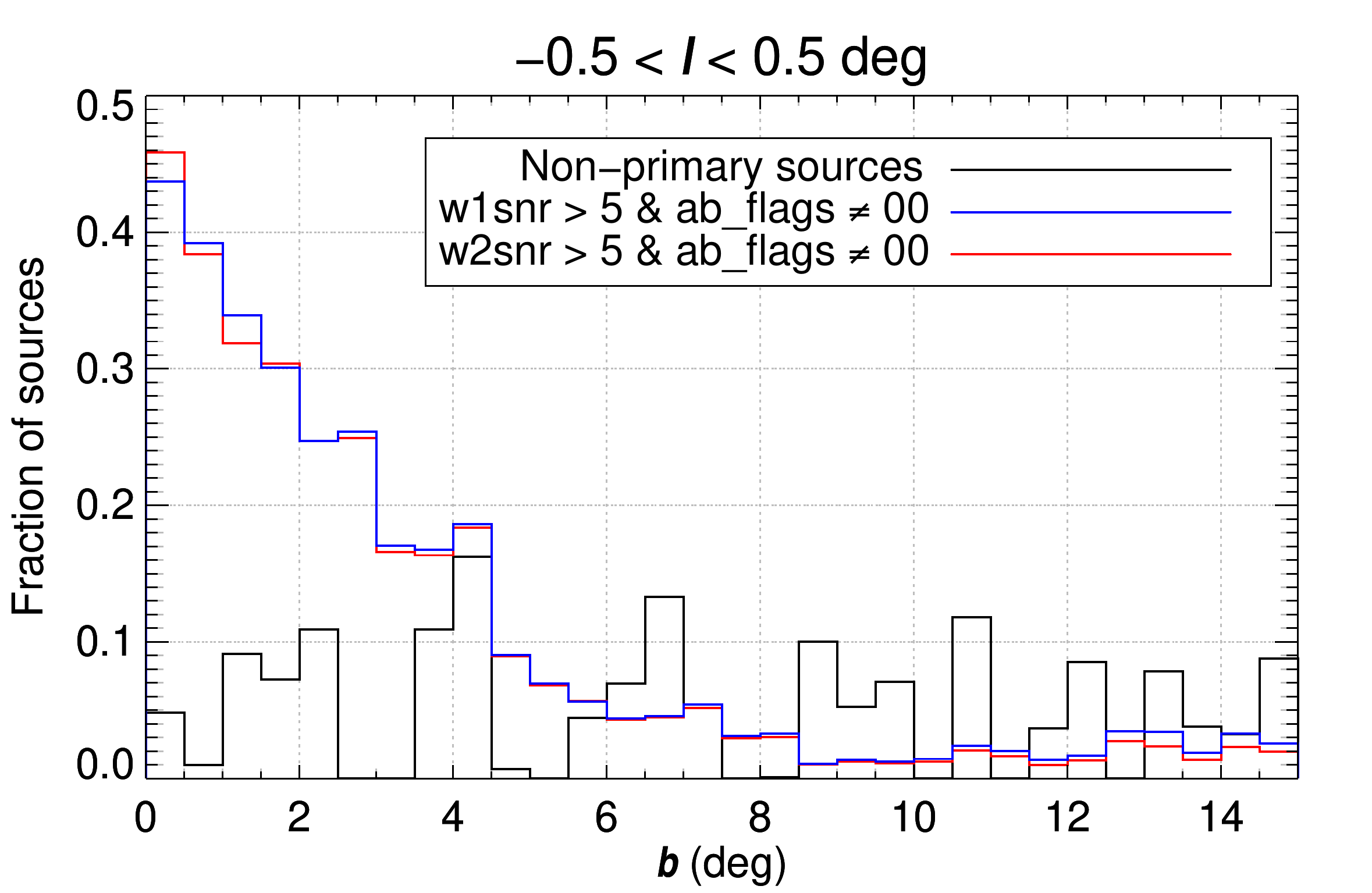}
    \includegraphics[width=0.49\textwidth]{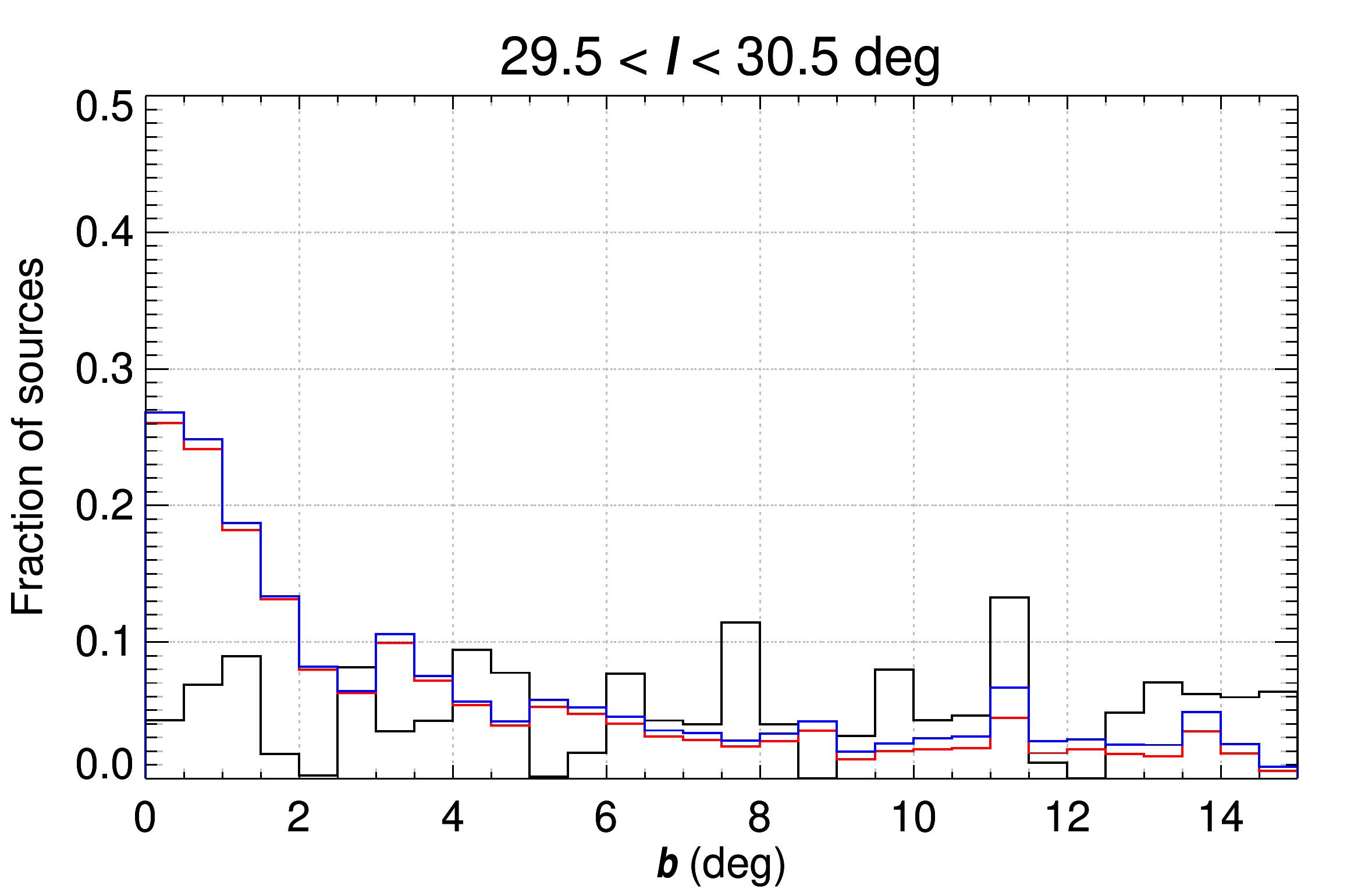}
    \includegraphics[width=0.49\textwidth]{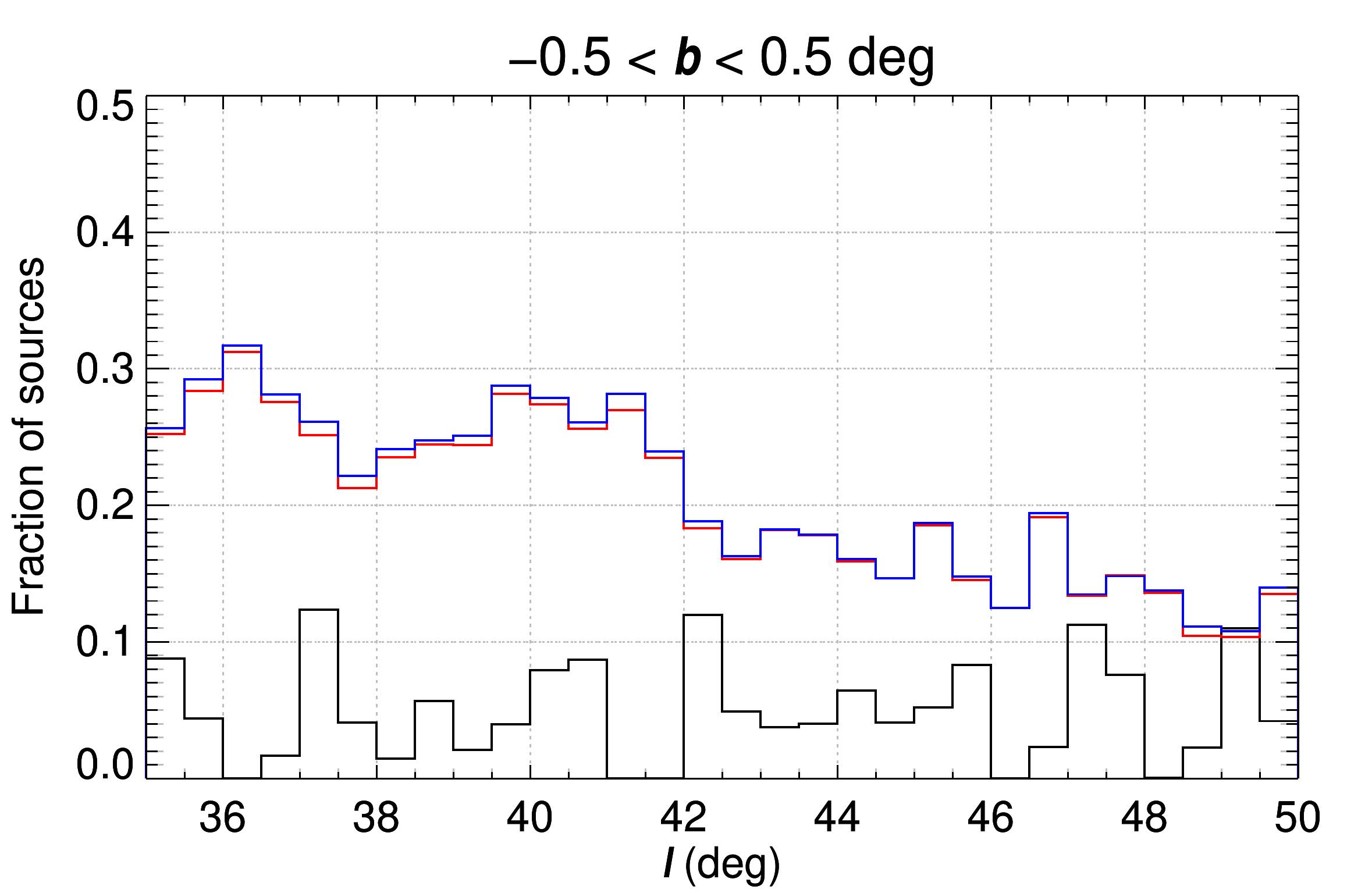}
    \caption{Fraction of non-primary sources (black), and sources with S/N$>$5 in at least one band but rejected because are flagged as artifacts in the same band (blue and red), in three $1\times15$\,deg strips.}
    \label{fig:rejects_glat}
\end{figure}

There are 1,890,715,644 sources that meet these criteria. The 341,799,385 sources that fail to meet these criteria go into the Reject Table. Individual tile reject tables typically contain 11,000 sources, although near the Galactic center they can contain over 120,000 sources due to the large number of artifacts. The larger number of artifacts in dense regions is illustrated in Figure~\ref{fig:source_density} (bottom-right panel), as well as in Figure~\ref{fig:rejects_glat}, where we show the fraction of non-primary sources (i.e. those that fail criteria 1), and the fraction of primary sources with S/N$\geq$5 in at least one band but flagged as artifacts in the same band (i.e. those that pass criteria 1 but fail criteria 2a or 2b) in three 1\,deg$\times15$\,deg strips. The three strips were chosen to trace the border of the overdensity of reject sources around the Galactic Center. They are:

1) $-0.5 < l < 0.5$\,deg and $0 < b < 15$\,deg; 

2) $29.5 < l < 30.5$\,deg and $0 < b < 15$\,deg; 

3) $-0.5 < b < 0.5$\,deg and $35 < l < 50$\,deg. 

The fraction of sources rejected on account of being flagged as artifacts grows rapidly as a function of Galactic latitude, spiking at $b < 6$\,deg. The incidence of artifacts is also a function of Galactic longitude, with a clear increase towards the Galactic Center. The increase in this case is not as dramatic as the increase as a function of $b$. The fraction of non-primary sources spikes whenever our chosen strips cross a tile boundary, which are defined to be parallel to the equatorial grid, and, therefore, are slanted when converted to galactic coordinates. There is no net trend for the fraction of non-primary sources with either $l$ or $b$, as expected since the overlap region between tiles does not vary significantly in this part of the sky. 

Other noteworthy features of the source density map for the Reject Table (Figure~\ref{fig:source_density}, bottom row), are the two very dense circles corresponding to the celestial poles. This is purely a geometric effect, resulting from the increasing overlap between adjacent tiles as a function of declination. While tiles on the celestial equator overlap only by $\sim 7\%$, those near the poles overlap by as much as $\sim 56\%$. The number of duplicate sources grows accordingly, and since duplicates are removed from the catalog and placed in the Reject Table, the resulting source density appears artificially high.

The source density in the CatWISE2020 Catalog increases around the ecliptic poles ($l\sim96$\,deg, $b\sim30$\,deg, and $l\sim276$\,deg, $b\sim-30$\,deg). This is a consequence of the greater depth of the WISE data around the ecliptic poles, which is in turn a result of the survey strategy of WISE \citep{Wright2010}.

The individual CatWISE2020 Catalog and Reject files for the 18,240 tiles were transferred to the NASA/IPAC Infrared Science Archive\footnote{\url{https://irsa.ipac.caltech.edu}} (IRSA), where they were merged into the IRSA database. Information regarding access to the catalog is provided in \S\ref{sec:access}.
 
\newpage

\subsection{CatWISE2020 Catalog Description \label{sec:columns}}

There  are 187 formatted columns of information about each source in the CatWISE2020 Catalog. The CatWISE2020 Reject Table adds a column to indicate whether the source is primary in its tile (see \S\ref{sec:2020}). Descriptions of the columns can be obtained from IRSA\footnote{\url{https://irsa.ipac.caltech.edu/data/WISE/CatWISE/gator\_docs/catwise\_colDescriptions.html}}.  Most of the columns have the same names as in the AllWISE Catalog, and are described in \S II.1.a of the AllWISE Explanatory Supplement \citep{Cutri2013}. Table A1 of \citet{Eisenhardt2020} provides information about selected columns in the CatWISE catalogs that augments the information provided by IRSA. 

 The Galactic coordinates (\textit{glon} and \textit{glat}) for sources in the CatWISE Preliminary Catalog were calculated incorrectly and were not included in the IRSA release. In the CatWISE2020 Catalog, these columns are more accurate and now are included in the IRSA release, but should not be used for astrometry. Two columns ({\it w1fitr} and {\it w2fitr}) remain excluded from the IRSA release of CatWISE2020 data products, for reasons explained in Table A1 of \citet{Eisenhardt2020}. Finally, the CatWISE2020 Catalog and Reject Table include a new column ({\it unwise\_objid}) which provides the updated unWISE Catalog (\S\ref{sec:mdet}) identification corresponding to the source. Note that these identifications end in ``r02" to avoid confusion with unrelated sources in the unWISE Catalog of \citet{Schlafly2019}.
 
CatWISE2020 source designations should have the prefix CWISE for objects in the CatWISE2020 Catalog, and CWISER for objects in the CatWISE2020 Reject Table. The designation for each source, based on its coordinates for the J2000 equinox following the IAU truncation convention and without the leading CWISE or CWISER prefix, is given by the field {\it source\_name}. For example, the quasar 3C~273 is CWISE J122906.70+020308.6.

\section{Performance Characterization \label{sec:performance}}
The additional data, longer baseline, and different detection software lead to an improvement on some key performance parameters for the CatWISE2020 Catalog with respect to the CatWISE Preliminary Catalog. Following the performance characterization strategy adopted in \citet{Eisenhardt2020}, we focus on the completeness and reliability of the CatWISE2020 Catalog at both the bright (W1,W2 $<$ 8\,mag, \S\ref{sec:bright}) and faint end (W1,W2 $>$ 12\,mag, \S\ref{sec:faint}), the photometric performance (\S\ref{sec:photom_perf}), and the astrometric performance (\S\ref{sec:astrom_perf}). \textit{Spitzer} data were used as external truth for photometric comparison, while \textit{Gaia} DR2 was used for astrometric comparison. 

\newpage

\subsection{Completeness and Reliability  \label{sec:CandR}}

\subsubsection{Bright Sources \label{sec:bright}}

The CatWISE2020 Catalog completeness and reliability for sources with W1 or W2 $<$8\,mag were assessed using an updated version of the WISE Bright Star List (BSL) as a truth set. The list was generated by the WISE team for artifact flagging (see \S 4.4.g.vi in the WISE All-Sky Release Explanatory Supplement; \citealp{Cutri2012}), and updated for our performance assessment of the CatWISE Preliminary Catalog (see \citealt{Eisenhardt2020}).

The CatWISE2020 Catalog completeness was determined as the percentage of BSL sources that have astrometric matches in the CatWISE2020 Catalog as a function of BSL magnitude. Differential CatWISE2020 Catalog reliability was determined as the percentage of sources  that have astrometric matches in the BSL as a function of CatWISE2020 magnitude. A relatively large matching radius of 5\farcs5 (corresponding to two WISE pixels) was used, to account for the poorer centroiding accuracy expected for highly saturated sources.

\begin{figure}
    \centering
    \includegraphics[width=0.49\textwidth]{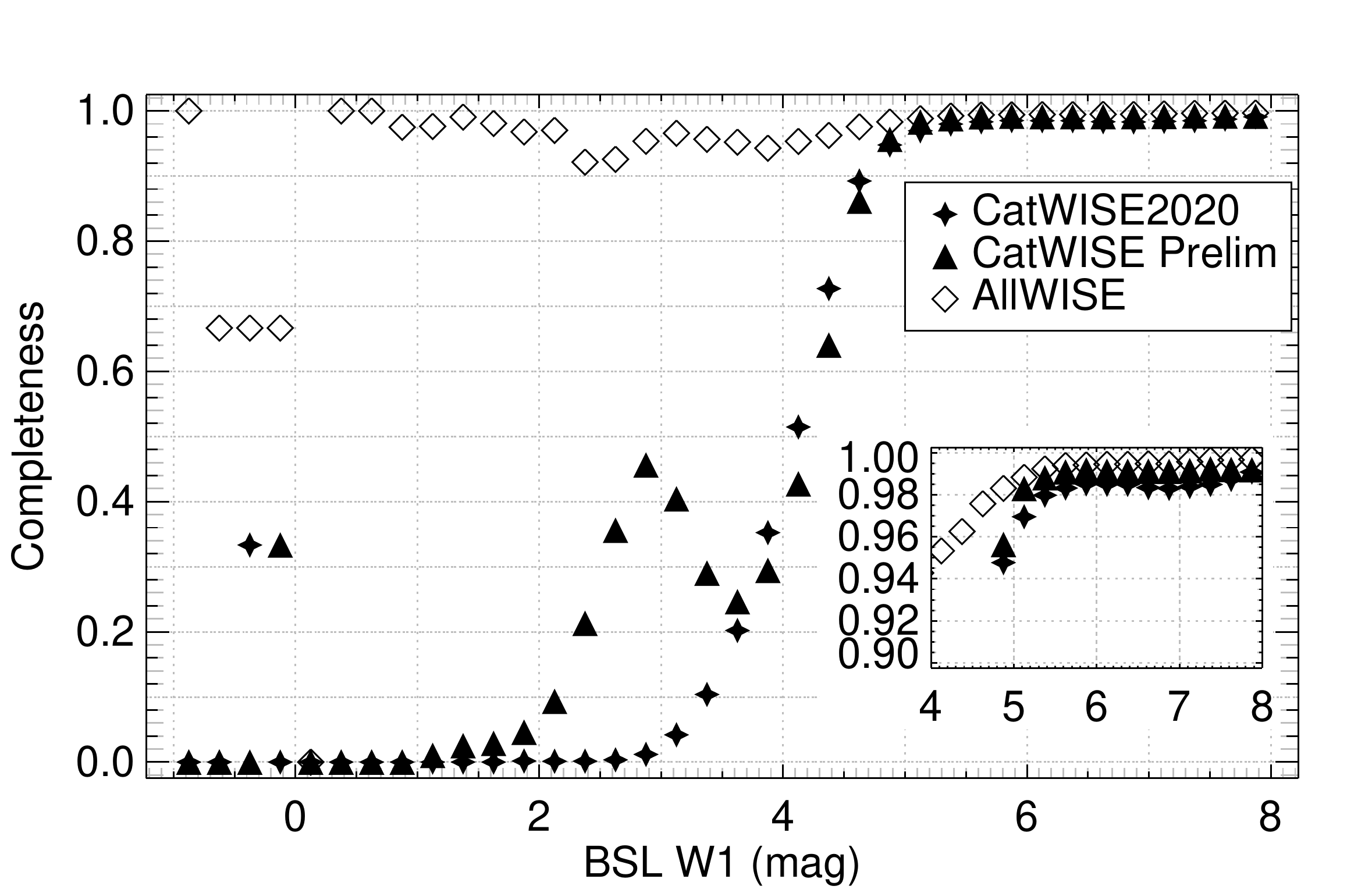}
    \includegraphics[width=0.49\textwidth]{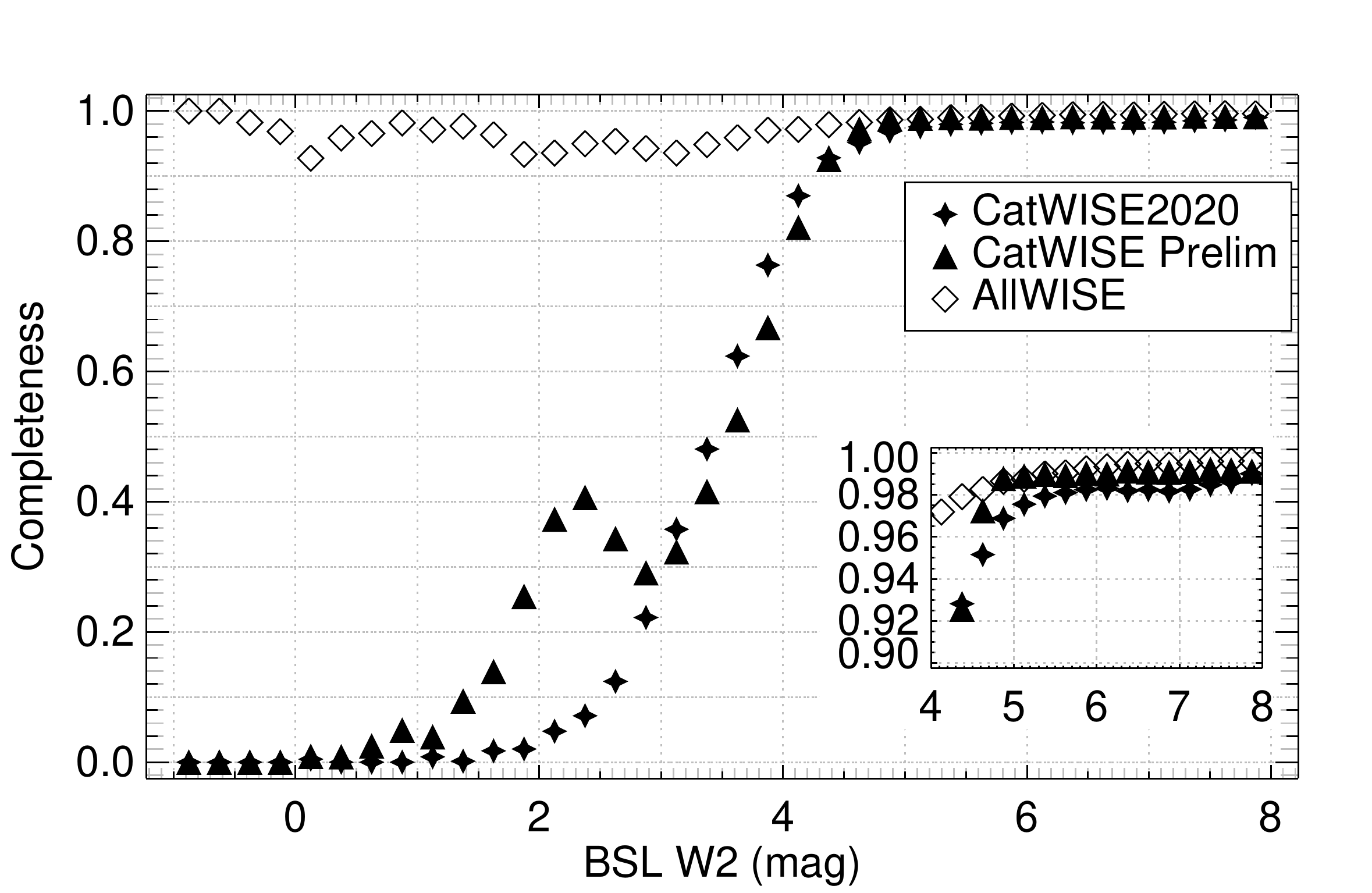}
    \caption{Differential completeness of the CatWISE2020 Catalog as a function of the Bright Star List's W1 (top) and W2 (bottom), compared to AllWISE and the CatWISE Preliminary Catalog.}
    \label{fig:brightCompleteness}
\end{figure}

Figure~\ref{fig:brightCompleteness} shows the results for completeness. The CatWISE2020 Catalog appears slightly less complete for bright sources than both the CatWISE Preliminary and AllWISE Catalogs. While the CatWISE Preliminary Catalog achieves $\sim99\%$ completeness in the BSL W1$\sim5.5-8$\,mag and BSL W2$\sim5-8$\,mag ranges, the CatWISE2020 Catalog has slightly lower completeness, peaking at 98\% in the 5.5$<$W1$<$6.25\,mag range. The CatWISE2020 Catalog completeness drops sharply for stars brighter than $\sim4.5$ mag, in a similar fashion to the CatWISE Preliminary Catalog, falling to $\sim 50\%$ by W1$\sim4.3$\,mag and W2$\sim3.6$\,mag. Missing bright stars belong predominantly in two categories: (1) bright variable stars (e.g. AGBs and Mira-type variables); (2) blended stars in crowded, nebulous fields. Detailed assessment in two test fields (2828p000 and 2657p288, which includes the Galactic Center) reveals that the missing bright stars are either missed at the detection stage, or are detected but then dropped by the PSF-fitting software because of the poor quality of the PSF fit. There are 3742 bright stars within the footprint of tile 2828p000 ($l=33.28; b=-0.54$), 65 of which are missing from the CatWISE2020 Catalog and Reject Table, and 10504 bright stars in the Galactic Center tile, 637 of which are missing. Of the 65 stars missing in the 2828p000 tile, 37 are missed at the detection stage, and the remaining 28 at the PSF-fitting stage. In the Galactic Center tile, 616 stars are missed at the detection stage, and 21 stars are missed at the PSF-fitting stage. Visual inspection reveals that stars missed at the detection stage are almost exclusively partly blended stars in fields with significant nebulosity due to interstellar dust. We refer the reader to \S4.4 in \citet{Schlafly2019} for a more detailed discussion on the impact of nebulosity on \textit{crowdsource}, and the possible incompleteness in such regions. On the other hand, stars missed at the PSF-fitting stage are predominantly very bright stars (W1$\lesssim5$\,mag or W2$\lesssim4$\,mag), and we speculate that their large saturated cores result in poor PSF fitting, leading to the sources being discarded by WPHOT. 

As discussed in \citet{Eisenhardt2020}, AllWISE completeness remains above 90\% even for stars as bright as 0.25 mag, and should therefore be the catalog of choice for bright star science that requires a complete sample. 

\begin{figure}
    \centering
    \includegraphics[width=0.49\textwidth]{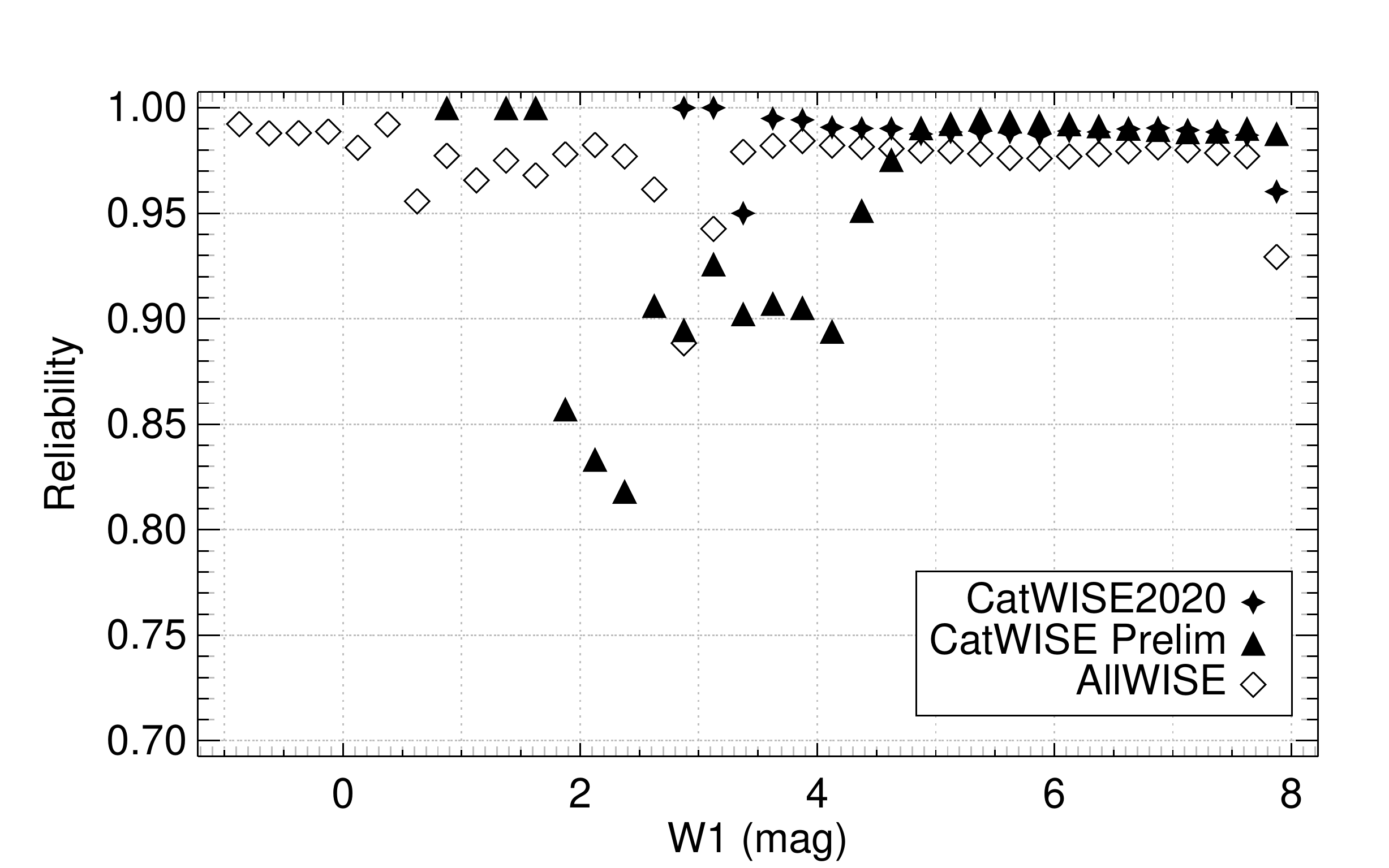}
    \includegraphics[width=0.49\textwidth]{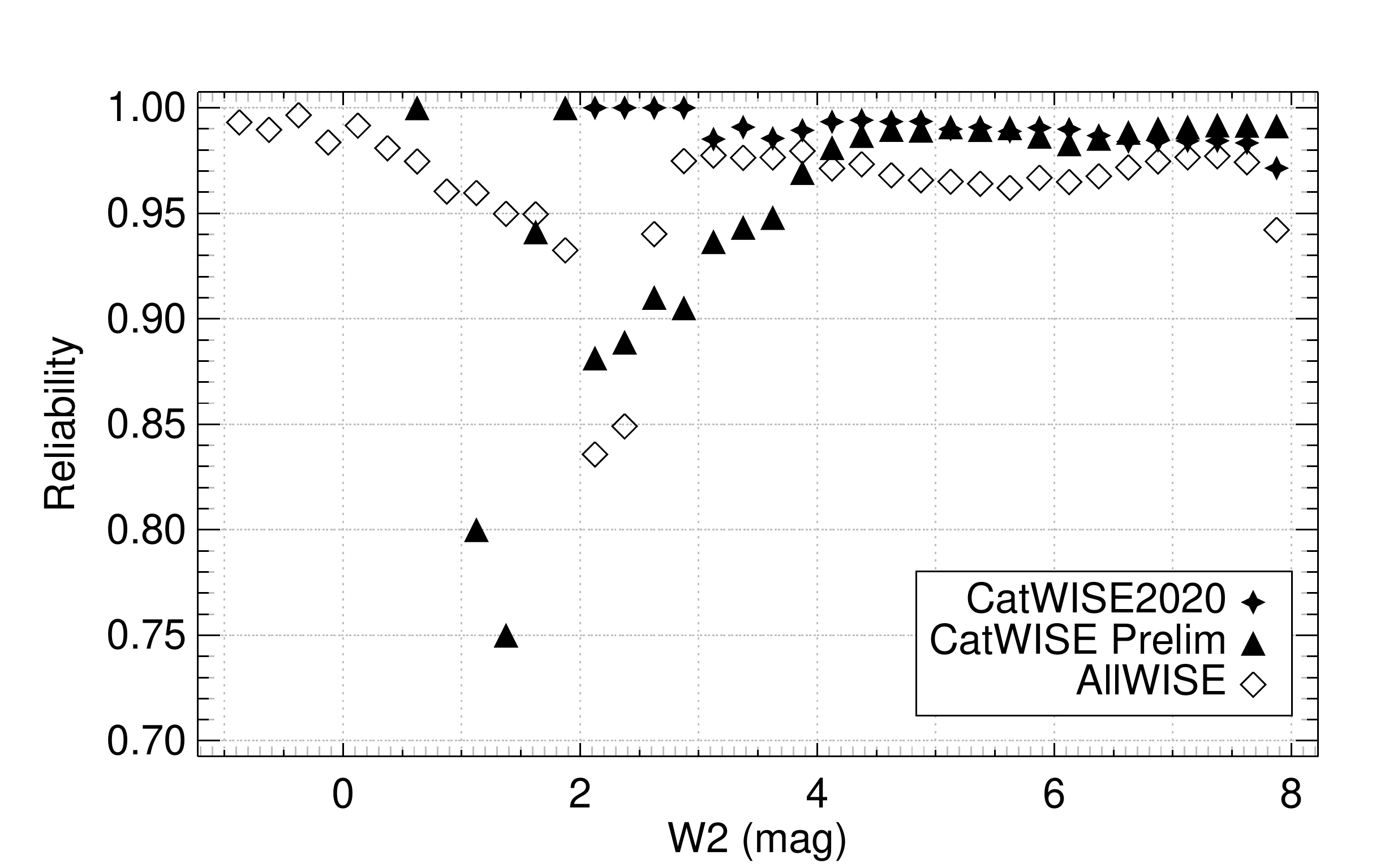}
    \caption{Differential reliability of the CatWISE2020 Catalog as a function of CatWISE2020 W1 (top) and W2 (bottom), compared to AllWISE and the CatWISE Preliminary Catalog.}
    \label{fig:brightReliability}
\end{figure}

The CatWISE2020 Catalog achieves comparable or better reliability than the CatWISE Preliminary and AllWISE Catalogs for stars in the $2.8 <$W1$<8.5$\,mag and $2<$W2$<8.5$\,mag range, with reliability consistently above 97\%, as can be seen in Figure~\ref{fig:brightReliability}.

\newpage

\subsubsection{Faint Sources \label{sec:faint}}

Completeness and reliability were assessed for faint sources in an area of $\sim94$ deg$^2$ by comparison with the {\it Spitzer} South Pole Telescope Deep Field survey \citep[SSDF;][]{Ashby2013}.

\begin{figure}
    \centering
    \includegraphics[width=0.49\textwidth]{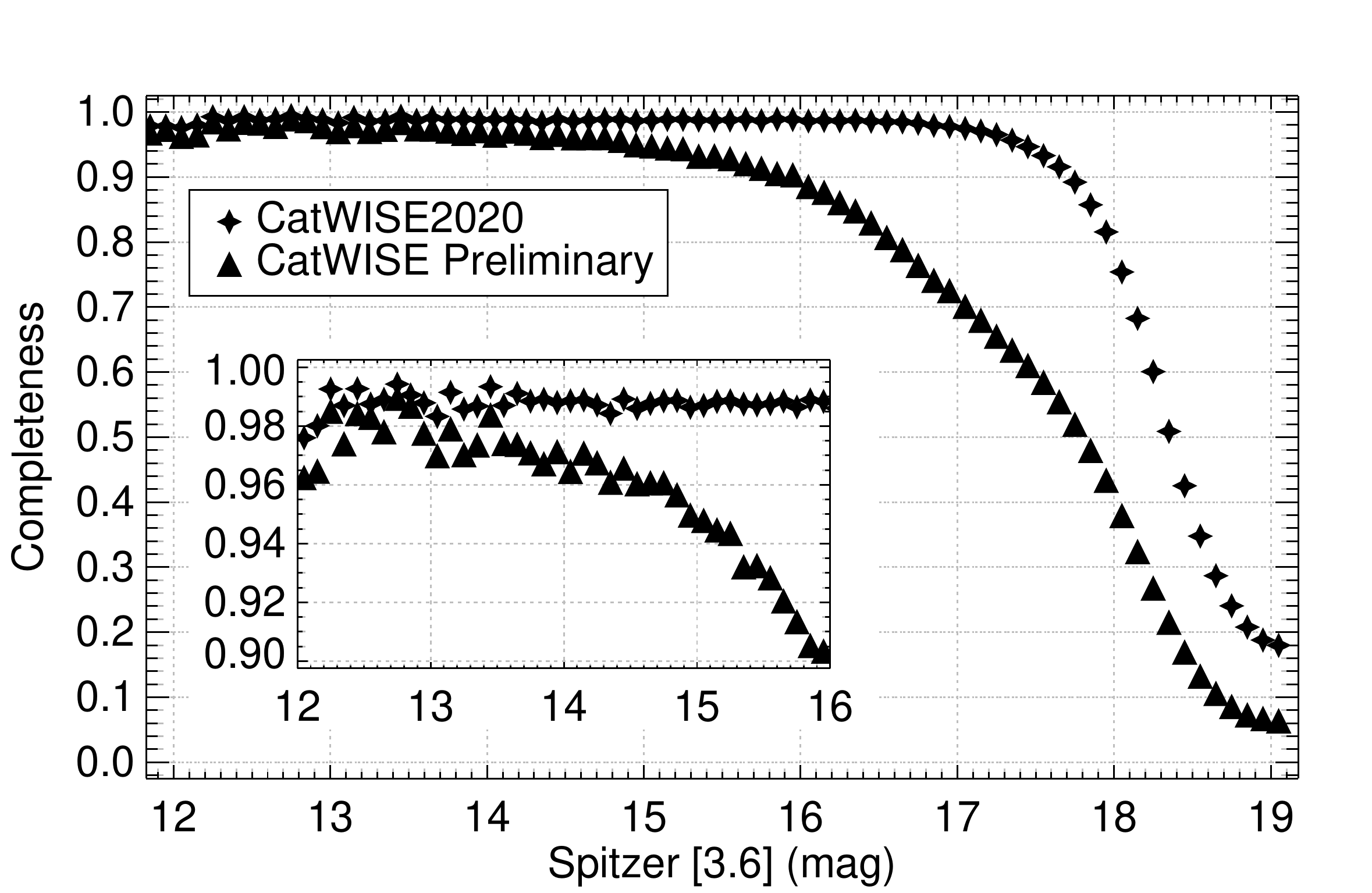}
    \includegraphics[width=0.49\textwidth]{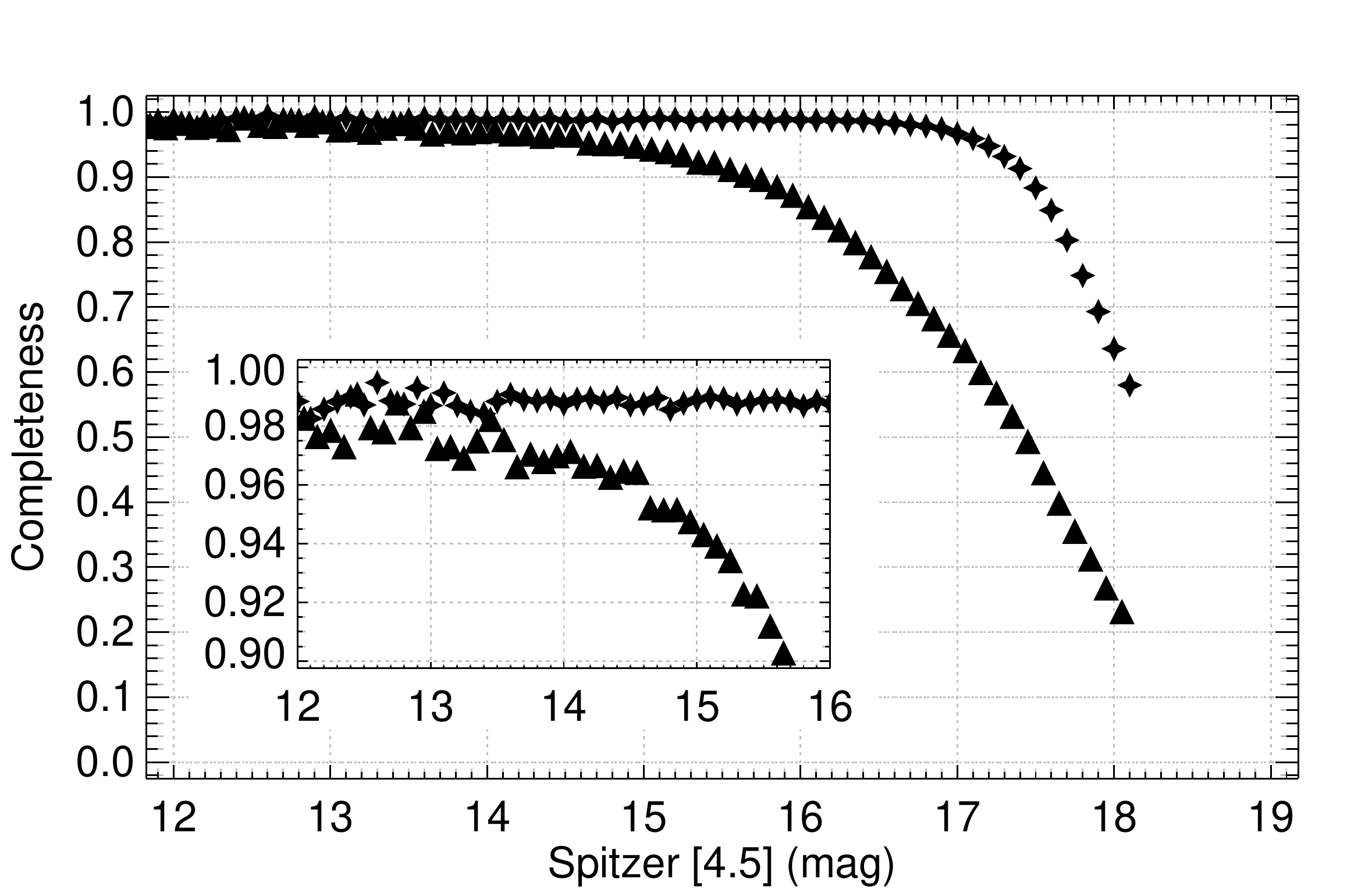}
    \caption{Completeness of the CatWISE2020 and Preliminary Catalog vs. {\it Spitzer} $3.6 \mu$m (top) and $4.5 \mu$m  (bottom) magnitude for sources in the SSDF.}
    \label{fig:SSDFcompleteness}
\end{figure}

\begin{figure}
    \centering
    \includegraphics[width=0.49\textwidth]{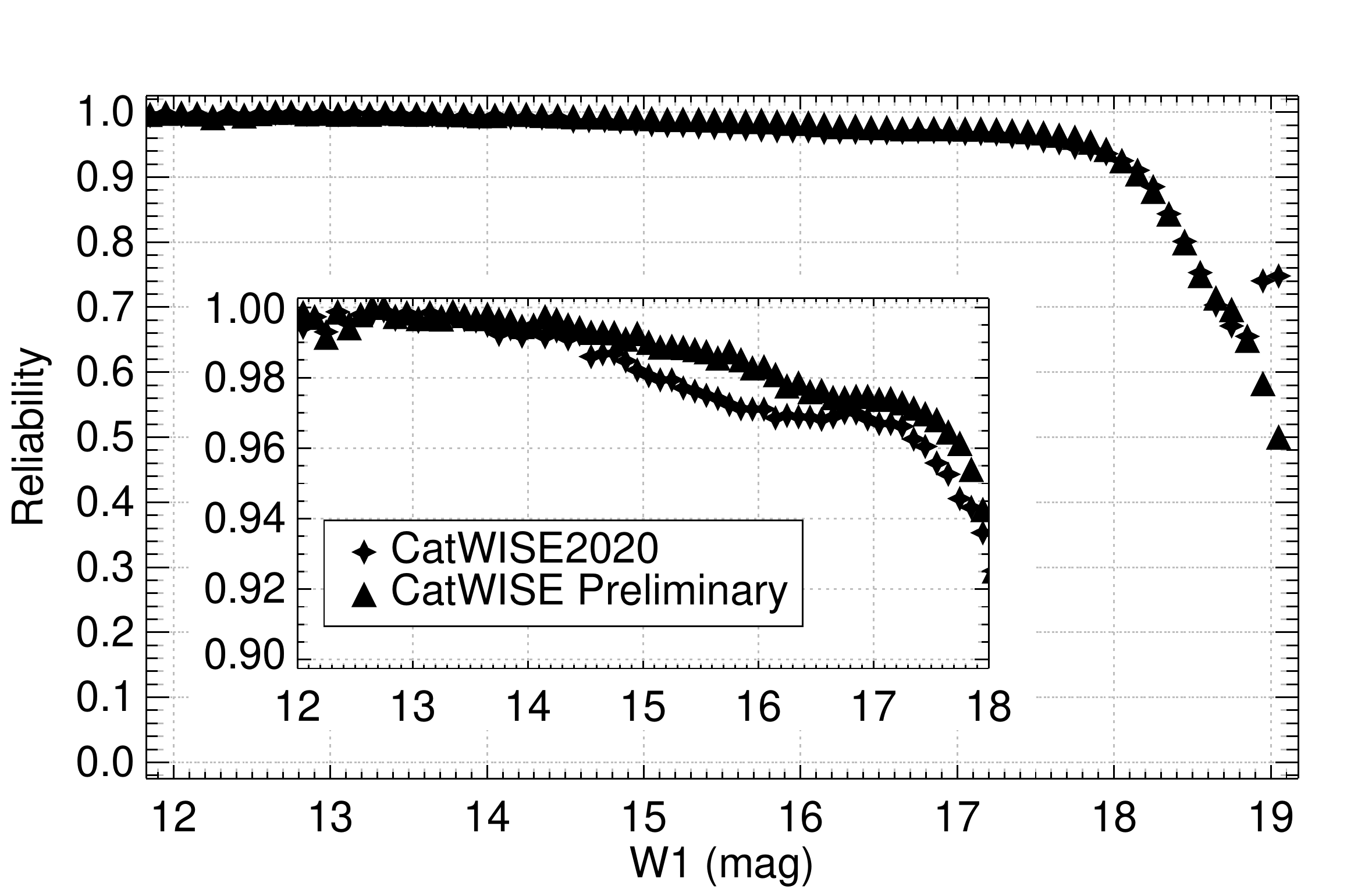}
    \includegraphics[width=0.49\textwidth]{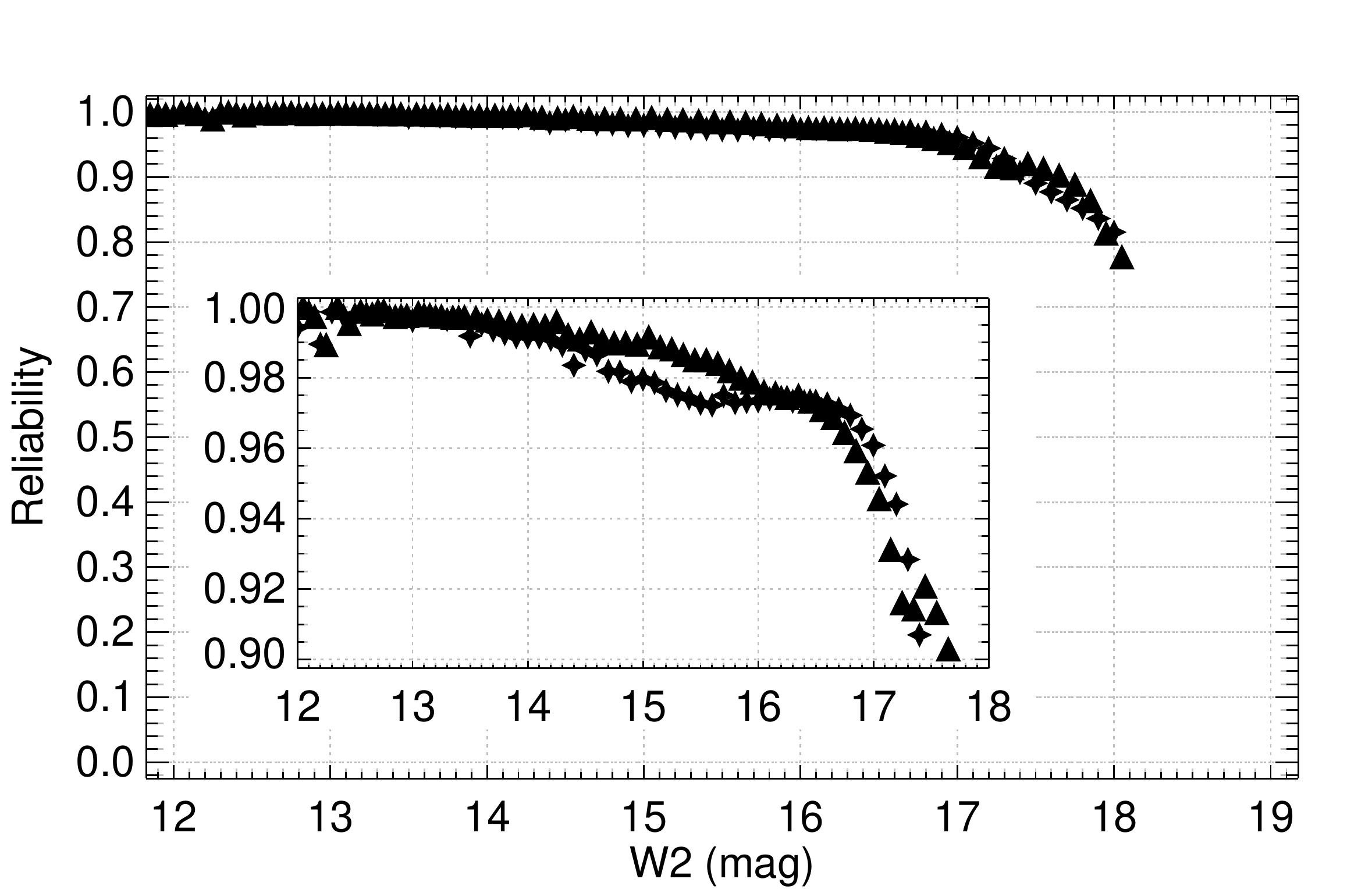}
    \caption{Reliability of the CatWISE2020 and  Preliminary Catalog as a function of W1 (top) and W2 (bottom), for sources in the SSDF.}
    \label{fig:SSDFreliability}
\end{figure}

The analysis followed the same method described in \S\ref{sec:bright}, except that a smaller matching radius of 2\farcs5 was used. The results are shown in Figures~\ref{fig:SSDFcompleteness} and \ref{fig:SSDFreliability}.

The CatWISE2020 Catalog consistently achieves greater completeness than the CatWISE Preliminary Catalog with typical completeness of 99\% across the $12 <$[3.6]$< 17.1$\,mag and $12 <$[4.5]$< 17$\,mag range. Figure~\ref{fig:SSDFcompleteness} shows that the 50\% completeness limit for the CatWISE2020 Catalog is [3.6]\,=\,18.3\,mag, while in [4.5] the completeness remains above 55\% all the way down to [4.5]\,=\,18.1\,mag, the coverage depth of the SSDF (cf. 17.8\,mag and 17.4\,mag for the CatWISE Preliminary Catalog).

The CatWISE2020 Catalog reliability is better than 99\% for sources brighter than 14.5\,mag in both W1 and W2, similar to the CatWISE Preliminary Catalog. For fainter sources, the CatWISE2020 Catalog reliability is slightly worse than the CatWISE Preliminary Catalog, in particular in the 14.5--16\,mag range in both W1 and W2 (and down to 16.8\,mag in W1), with reliability of $\sim$1\% below that achieved by the CatWISE Preliminary Catalog. At fainter magnitudes, the CatWISE2020 Catalog remains a fraction of a percent less reliable than the CatWISE Preliminary Catalog in W1, while it becomes a fraction of a percent more reliable than the CatWISE Preliminary Catalog in W2. 

Given that in the same brightness range the CatWISE Preliminary Catalog completeness significantly deteriorates compared to the CatWISE2020 Catalog, CatWISE2020 performance is overall superior to CatWISE Preliminary for faint stars over the range assessed.

\subsection{Photometric properties  \label{sec:photom_perf}}

The CatWISE2020 Catalog photometric depth was assessed using the SSDF and the COSMOS field. 

\begin{figure*}
        \centering
        \includegraphics[width=\textwidth]{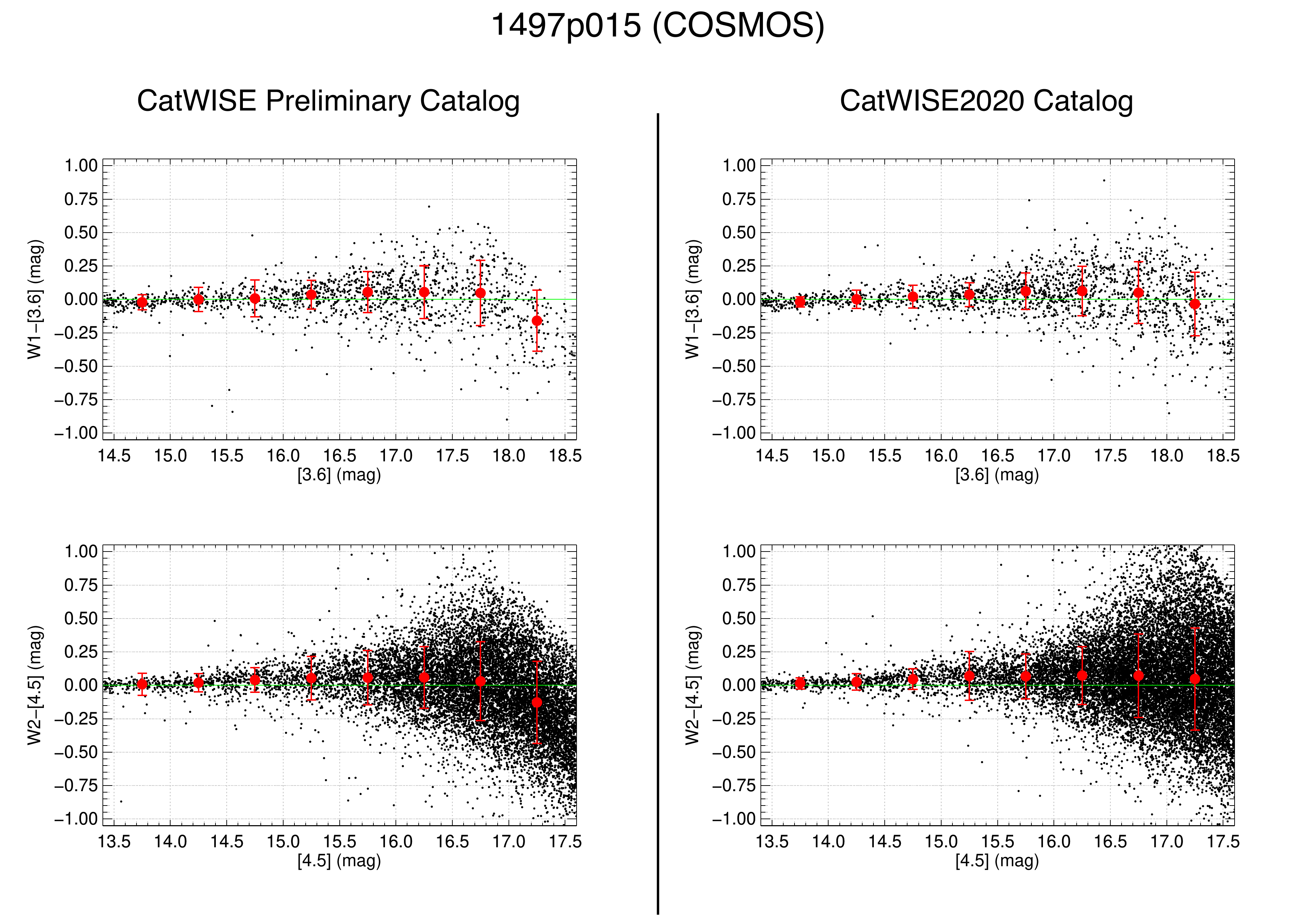}
        \caption{Comparison of CatWISE Preliminary Catalog (left) and CatWISE2020 Catalog (right) photometry to {\it Spitzer} photometry for COSMOS. {\it Top}: Difference between CatWISE W1 PSF and {\it Spitzer} S-COSMOS 2\farcs9 radius aperture photometry at [3.6], for sources with $-0.1 < [3.6] - [4.5] < 0$ and $< 10$\% flux increase from the 1\farcs9 to 2\farcs9 aperture. Median differences and standard deviations in 0.5 mag bins are shown by the red points and error bars. {\it Bottom}: Comparison for CatWISE W2 and {\it Spitzer} [4.5] photometry. No restriction on {\it Spitzer} source color needs to be applied in this case (see \S\ref{sec:photom_perf}).}
        \label{fig:CatWISE_vs_COSMOS}
    \end{figure*}

\begin{figure*}
        \centering
        \includegraphics[width=\textwidth]{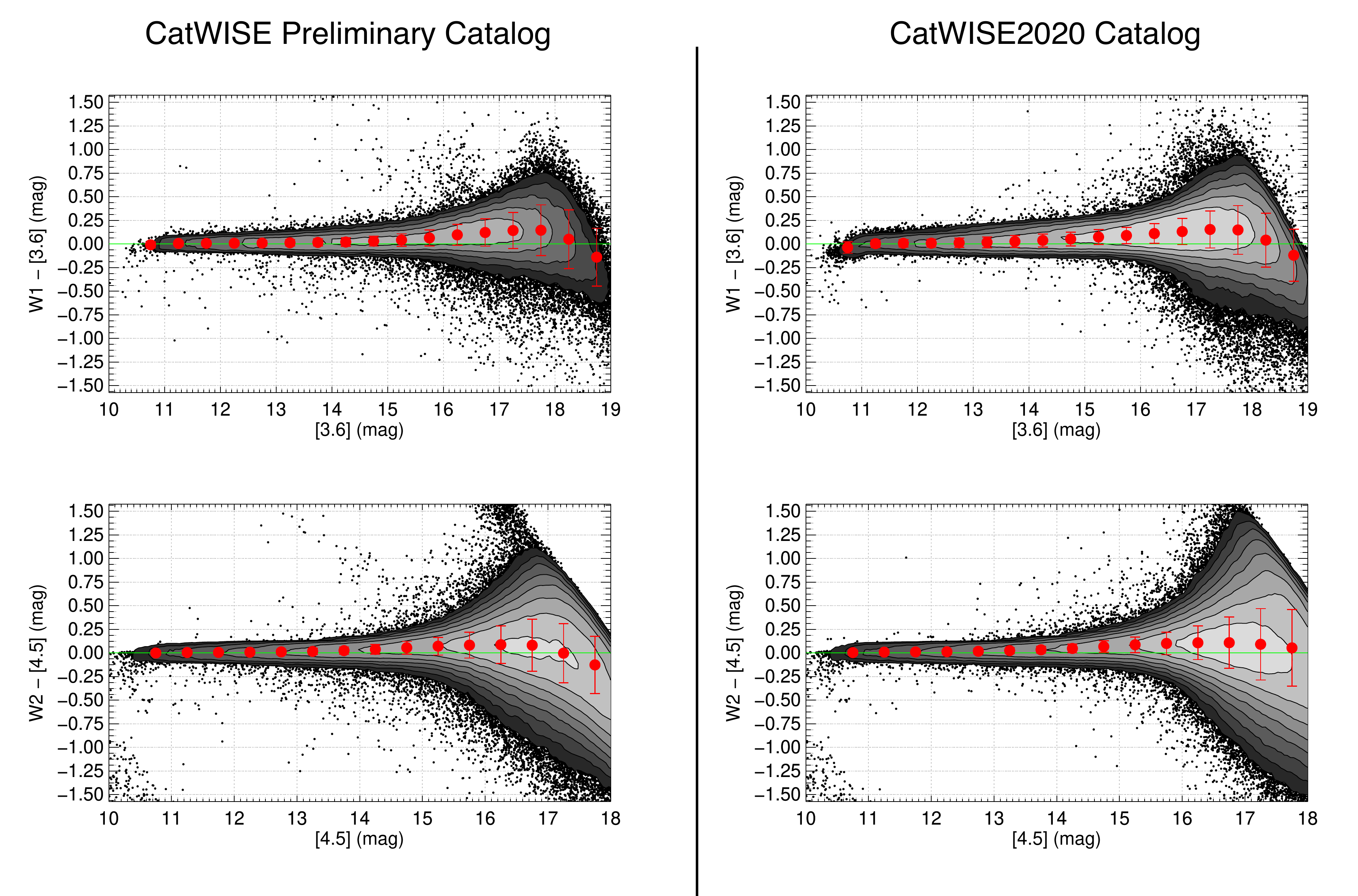}        
        \caption{Comparison of CatWISE photometry to {\it Spitzer} photometry for the SSDF, using the same methodology as in Figure \ref{fig:CatWISE_vs_COSMOS}. The outer contour represents a source density of 10 sources per $0.05 \times 0.05$ mag bin, with each additional contour showing a factor of two increase in source density.} 
        \label{fig:CatWISE_vs_SSDF}
    \end{figure*}

Figure~\ref{fig:CatWISE_vs_COSMOS} compares CatWISE Preliminary Catalog (left) and CatWISE2020 Catalog (right) PSF-fitting photometry to 2\farcs9 radius aperture photometry from the {\it Spitzer} S-COSMOS program \citep{Sanders2007}. These observations were obtained using long integration times (20 minutes), so the S-COSMOS data are much deeper than CatWISE. The closest CatWISE source within 2\farcs75 was taken as the match to the S-COSMOS source. Because the CatWISE photometry is measured via point source fitting, S-COSMOS sources were required to have $<10\%$ flux increase between the 1\farcs9 and 2\farcs9 radius apertures. In addition, because the W1 band is significantly bluer than the [3.6] band, to minimize spurious color-related effects S-COSMOS sources at [3.6] were required to have $ -0.1 \leq [3.6] - [4.5] \leq 0$. These are the same criteria used in \citet{Eisenhardt2020} and \citet{Cutri2012}. Figure~\ref{fig:CatWISE_vs_SSDF} presents the same comparison to photometry from the SSDF survey.

The comparison between the CatWISE Preliminary Catalog and CatWISE2020 Catalog and {\it Spitzer} photometry is consistent in both bands, in both fields. CatWISE2020 photometry becomes increasingly fainter than {\it Spitzer} beyond 16th mag, up to $\sim 0.1$\,mag fainter in W1 at 17.5\,mag, and $\sim0.06$\,mag fainter in W2 at 16.5\,mag. This effect was also observed in CatWISE Preliminary photometry. The measured scatter reaches 0.217 mag, equivalent to a S/N of 5, at [3.6] = 17.41 mag and [4.5] = 16.50 mag. Adjusting for the mean offsets in  W1--[3.6] and W2--[4.5] at these magnitudes, and compensating for the different depth of the three different surveys, the S/N=5 limits for the CatWISE2020 Catalog are W1\,=\,17.43\,mag and W2\,=\,16.47\,mag (cf. W1\,=\,17.41\,mag and W2\,=\,16.33\,mag for the CatWISE Preliminary Catalog).

\subsection{Astrometric properties}
\label{sec:astrom_perf}

\subsubsection{Full-sky Astrometric Assessments \label{sec:astrom_full_sky}}

The astrometric performance of the CatWISE2020 Catalog over the entire sky was assessed by comparing to \textit{Gaia} DR2 \citep{Brown2018, Lindegren2018}. Following the method described in \citet{Eisenhardt2020}, in each tile we selected the 10 brightest sources in each 0.5\,mag bin over the $10<$W1$<17.5$\,mag range. This resulted in a sample of 150 sources per tile, uniformly distributed on the sky. The full sample for astrometric comparison consists of 2,735,892 sources. These sources were cross-matched with \textit{Gaia} DR2 using a 5\farcs5 radius (corresponding to two {\it WISE} pixels), requiring the \textit{Gaia} counterpart to have measured proper motions. This returned 2,179,410 unique matches. The \textit{Gaia} counterparts were propagated to the CatWISE2020 epoch (MJD=57170) using \textit{Gaia} astrometry, and then the median difference and standard deviation between the CatWISE2020 motion-fit and \textit{Gaia} position and motion values were computed. 

\begin{figure}
    \centering
    \includegraphics[width=0.49\textwidth]{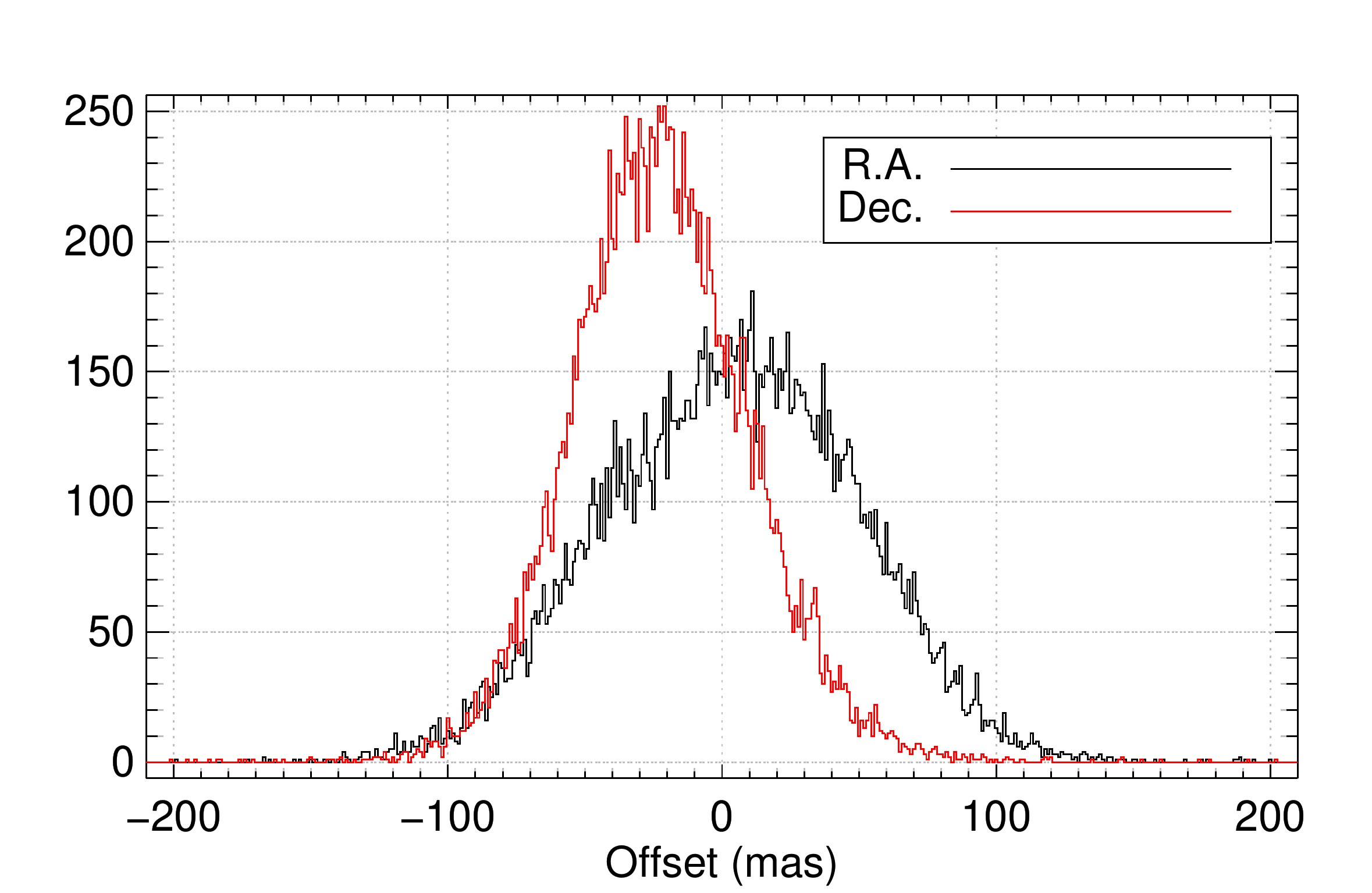}
    \includegraphics[width=0.49\textwidth]{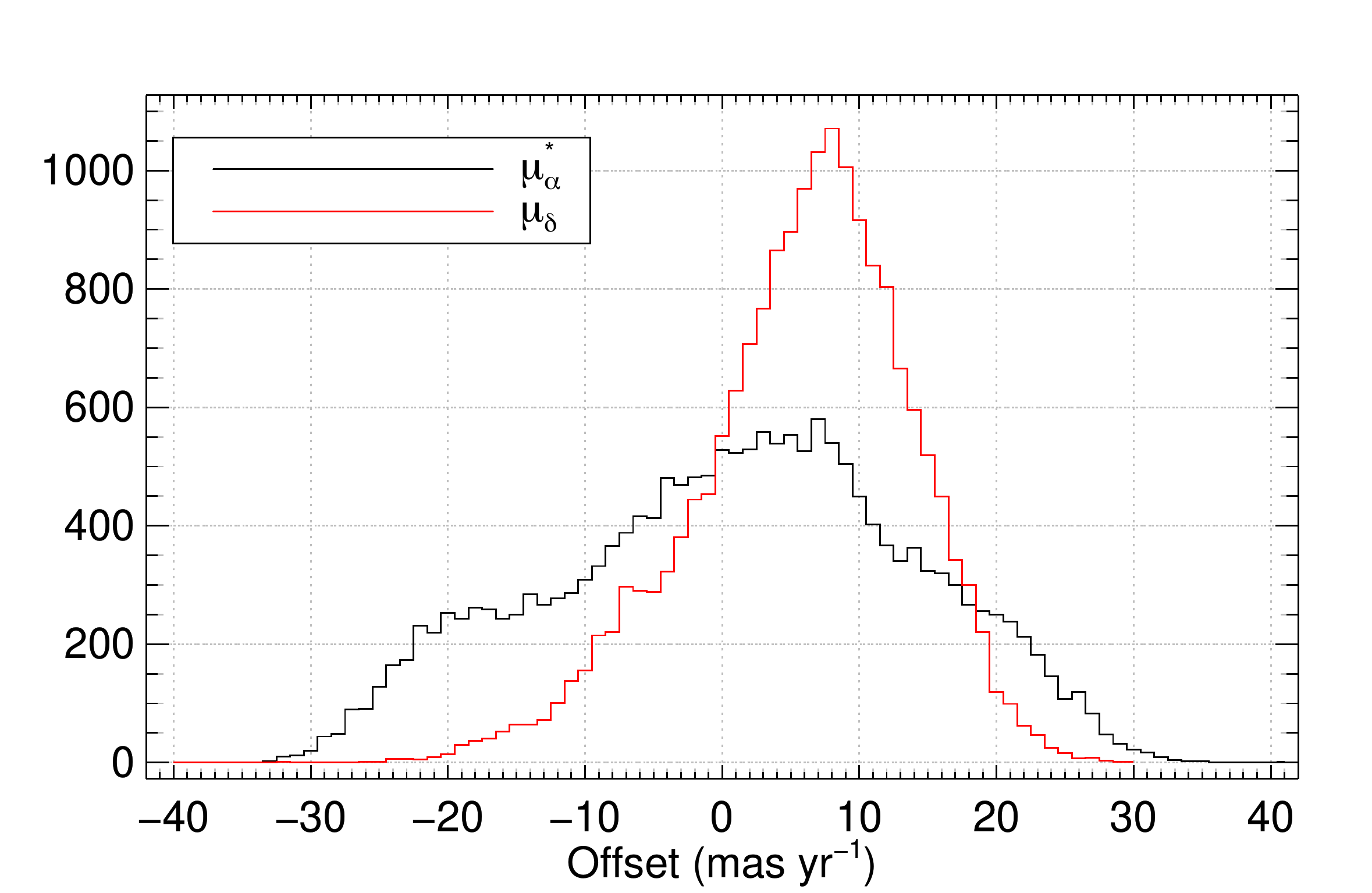}
    \caption{The distribution of median offsets between CatWISE2020 positions (top) and proper motions (bottom) and \textit{Gaia} DR2 in the 18,240 tiles.} 
    \label{fig:offset_distribution}
\end{figure}

As mentioned in \S\ref{sec:unwise}, CatWISE2020 processing did not apply the WCS adjustments to the unWISE coadds for the AllWISE epochs. This results in small systematic offsets between the CatWISE2020 Catalog and \textit{Gaia} DR2 in both position and motions. The R.A. and Dec offset with respect to the \textit{Gaia} DR2 counterpart (i.e. R.A.$_{\textit{Gaia}}$--R.A.$_{\rm CatWISE2020} \times \cos{\rm Dec_{CatWISE2020}}$ and Dec$_{\textit{Gaia}}$--Dec$_{\rm CatWISE2020}$), as well as the proper motion difference, was computed for all 2,179,410 sources in the comparison sample. Figure~\ref{fig:offset_distribution} shows the offset distribution, revealing a roughly Gaussian distributions in R.A. and Dec (top panel), with a 1-sigma dispersion of 46\,mas and 32\,mas respectively. The proper motion offsets (bottom panel) show a more complex distribution, since they are the result of the projected Solar motion through the Galaxy.

\begin{figure*}
\includegraphics[width=0.49\textwidth]{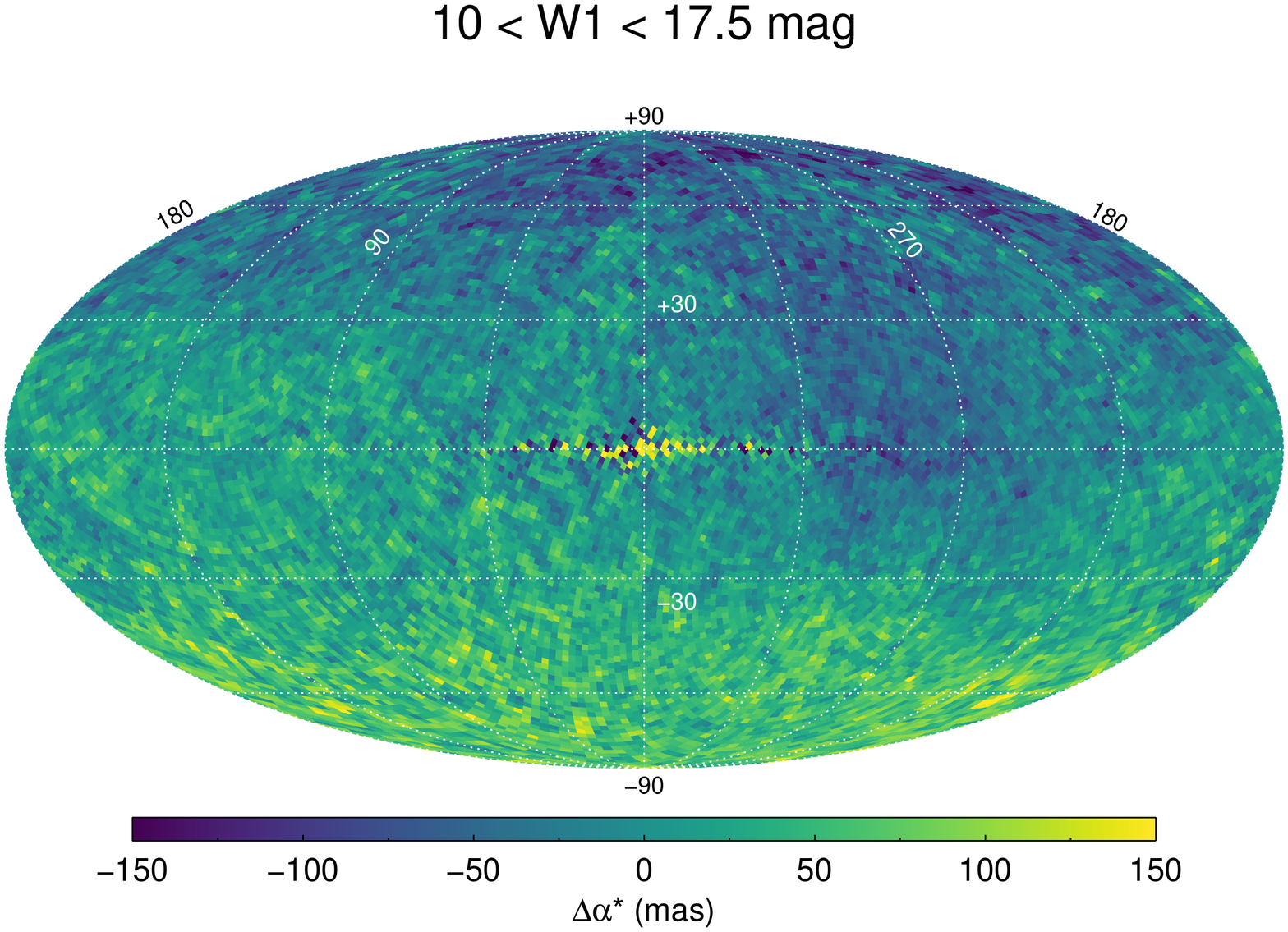}
\includegraphics[width=0.49\textwidth]{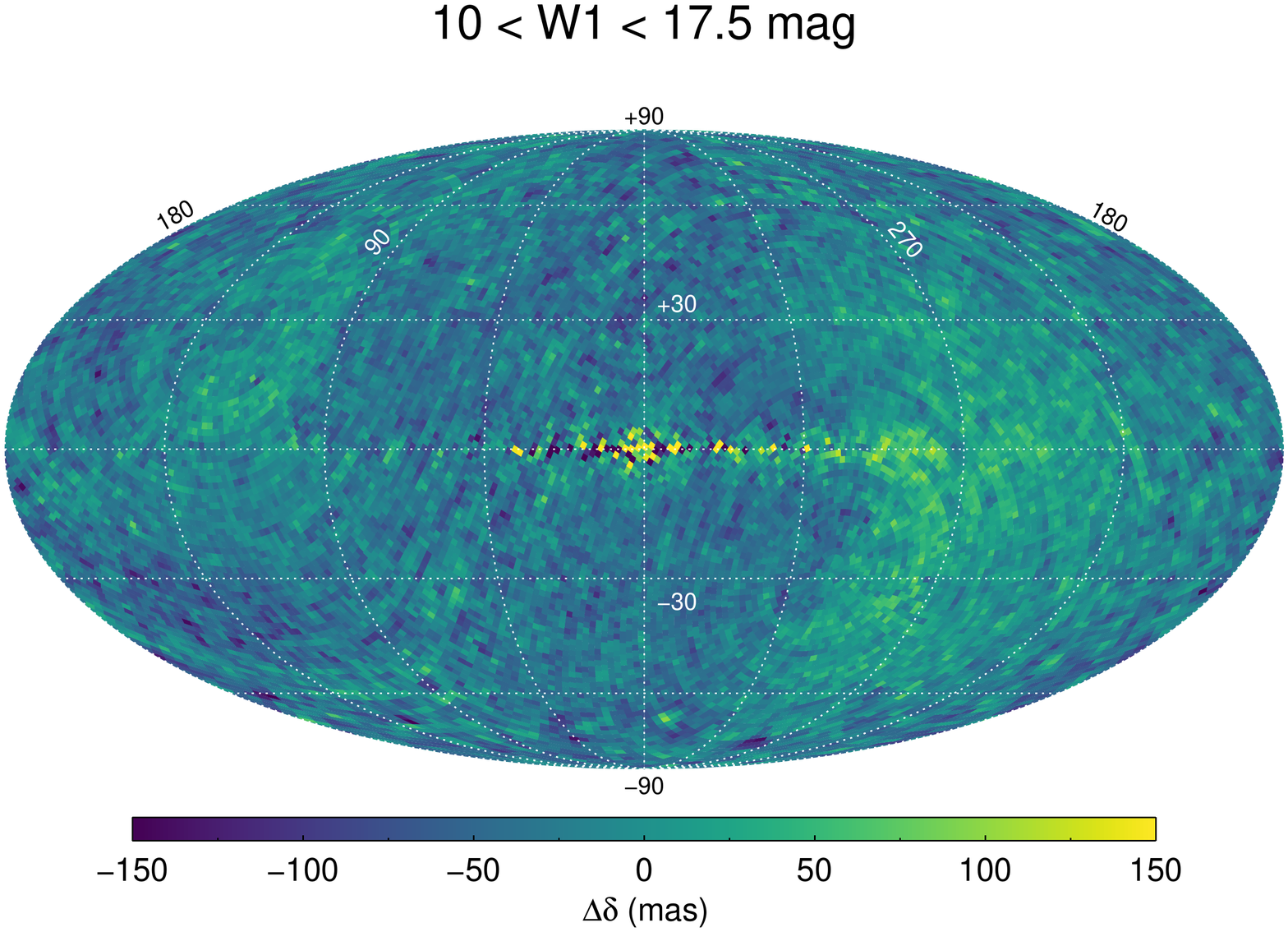}
\includegraphics[width=0.49\textwidth]{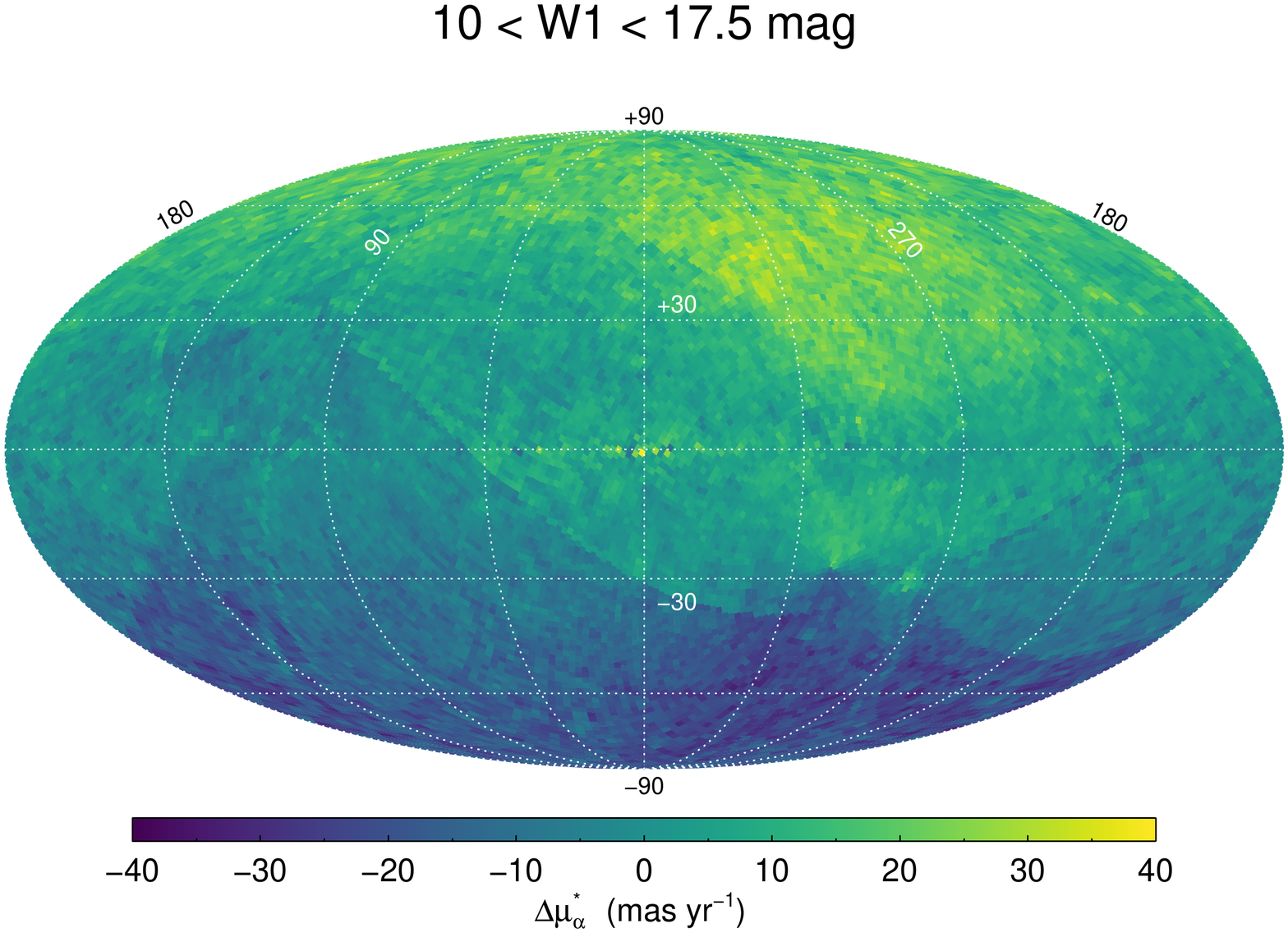}
\includegraphics[width=0.49\textwidth]{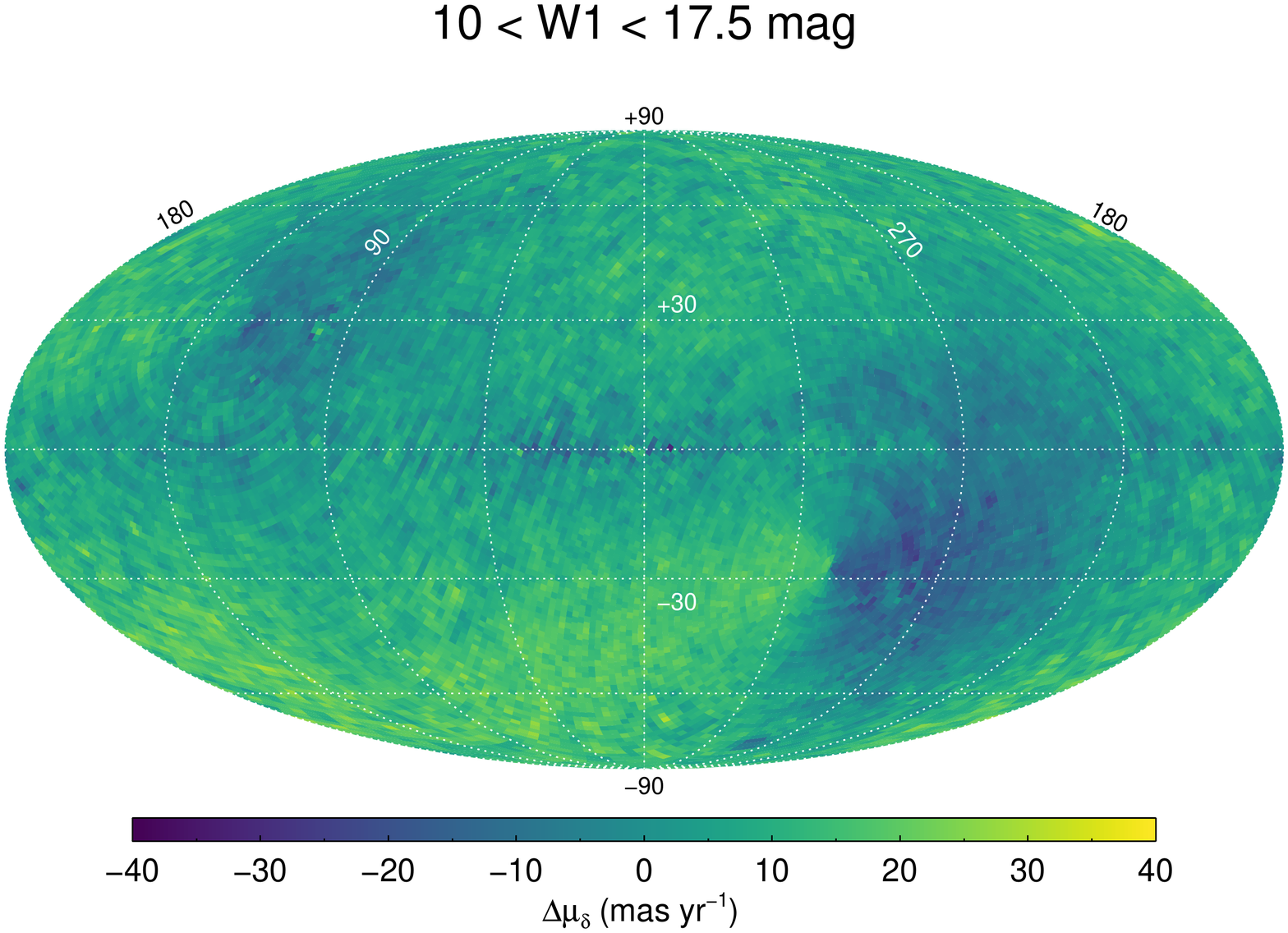}
\caption{Systematic offset between the CatWISE2020 Catalog and \textit{Gaia} DR2 positions (top row) and proper motion (bottom row). \label{fig:cw2020_astrometry_offset}}
\end{figure*}

Figure~\ref{fig:cw2020_astrometry_offset} maps the systematic offsets between the CatWISE2020 Catalog and \textit{Gaia} DR2, in Galactic coordinates. The systematic offset is not dependent on source density, i.e. there is no obvious degradation in high density regions. In the very center of the Milky Way the maps appear more noisy, but that is a spurious result of the low number of matches between the CatWISE2020 Catalog and \textit{Gaia} DR2, rather than a real astrometric effect. All maps show patterns that are consistent with the projection of the Solar motion through the Galaxy. The offsets are typically in the $\pm$150\,mas range in position, and in the $\pm$40\,mas\,yr$^{-1}$ range in proper motion. The median position and proper motion offsets in each tile are provided in a machine-readable table, and a sample is shown in Table~\ref{tab:astro_offsets}. For any application that requires the study of the kinematics of a large sample of objects, we recommend using these median values to correct motion and positions on a tile-by-tile basis. The tile in which a source is measured is encoded in the first 8 characters of the \textit{source\_id}, given in the first column of the CatWISE2020 Catalog and Reject Table. For example, source 0000p000\_b0-013835 was measured in tile 0000p000.

\begin{deluxetable*}{crrrrr}
\tablecaption{Systematic offsets between the CatWISE2020 Catalog and \textit{Gaia} DR2 (full table available on-line in machine readable format) \label{tab:astro_offsets}}
\tablehead{
\colhead{} & \multicolumn{5}{c}{\textit{Gaia} DR2 -- CatWISE2020 Catalog} \\
\colhead{Tile} & \colhead{$\Delta \alpha^*$} & \colhead{$\Delta \alpha$} & \colhead{$\Delta \delta$} & \colhead{$\Delta \mu_\alpha^*$} & \colhead{$\Delta \mu_\delta$} \\
\colhead{} & \colhead{deg} & \colhead{deg} & \colhead{deg} & \colhead{arcsec\,yr$^{-1}$} & \colhead{arcsec\,yr$^{-1}$} 
}
\startdata
0000m016 & 7.038531$\times 10^{-7}$ & 7.041276$\times 10^{-7}$ & 2.211593$\times 10^{-6}$ & $-$0.01956579 & 0.018327646 \\      
0000m031 & 9.905521$\times 10^{-6}$ & 9.920038$\times 10^{-6}$ & 6.129016$\times 10^{-6}$ & $-$0.017017405 & 0.0168396 \\
0000m046 & 1.5195505$\times 10^{-5}$ & 1.5244609$\times 10^{-5}$ & $-$1.5518368$\times 10^{-5}$ & $-$0.013825462 & 0.016300509 \\ 
0000m061 & 2.3382128$\times 10^{-5}$ & 2.3515273$\times 10^{-5}$ & $-$8.676556$\times 10^{-6}$ & $-$0.014948687 & 0.012481616 \\
0000m076 & 4.49237$\times 10^{-5}$ & 4.532183$\times 10^{-5}$ & $-$1.4328044$\times 10^{-5}$ & $-$0.015264964 & 0.016823826 \\
\ldots & \ldots & \ldots & \ldots & \ldots \\
\enddata
\end{deluxetable*}

\begin{figure*}
\centering
\includegraphics[width=\textwidth]{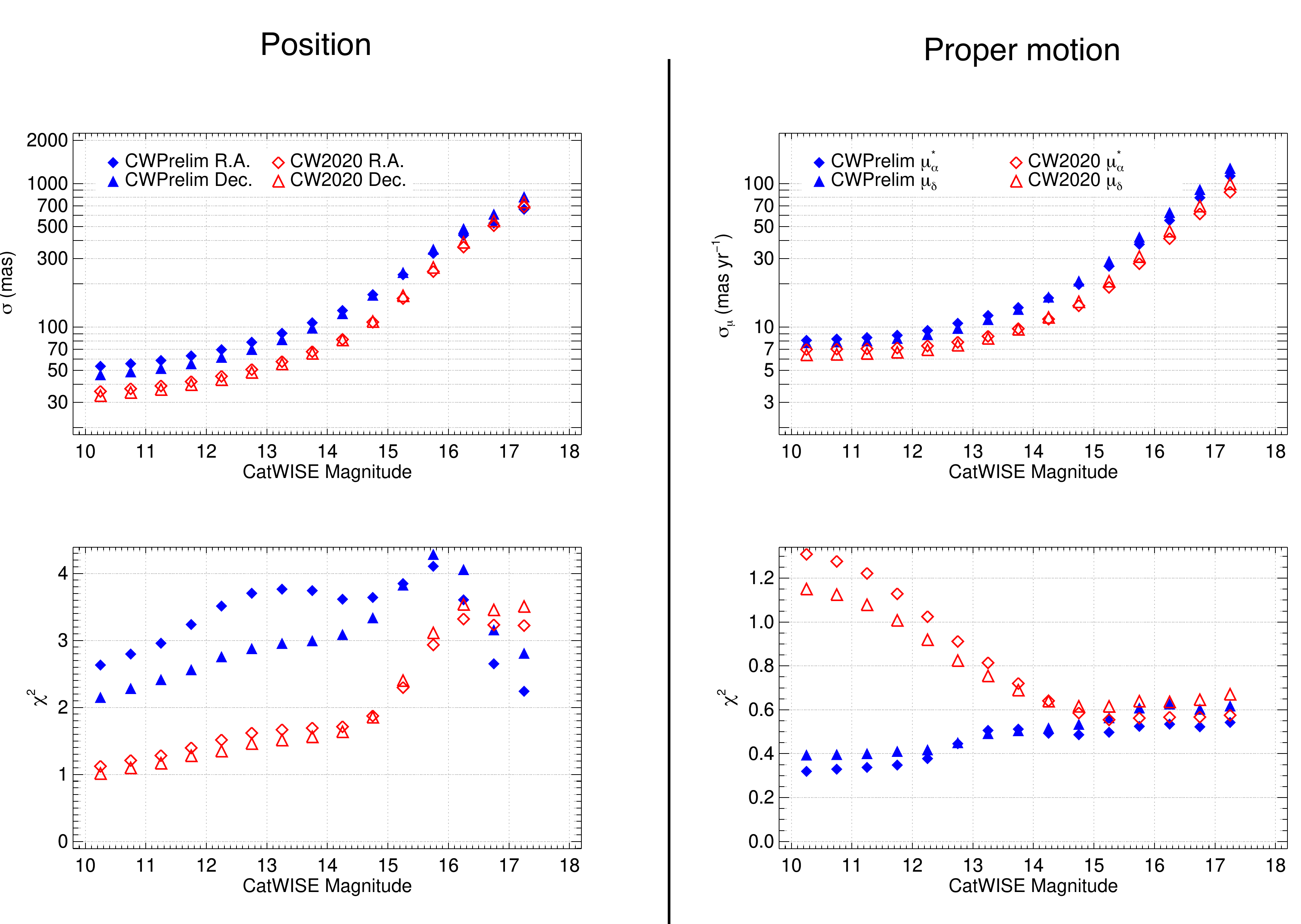}
\caption{CatWISE Preliminary and CatWISE2020 astrometric performance with respect to \textit{Gaia} DR2. CatWISE2020 positions and proper motions have been corrected for the systematic offsets, as discussed in \S\ref{sec:astrom_full_sky}. \textit{Top:} the 1-$\sigma$ dispersion between CatWISE and \textit{Gaia} R.A. (specifically, $\Delta\alpha^*$) and Dec. (left), and proper motion (right), for a subsample of $\sim$2.1 million sources in the $10<$W1$<17.5$\,mag range, uniformly distributed over the entire sky. \textit{Bottom:} the median $\chi^2$ computed taking into account CatWISE catalog uncertainties, \textit{Gaia} catalog uncertainties, and the uncertainty introduced by the translation of \textit{Gaia}'s positions to the CatWISE epoch. \label{fig:catwise_vs_gaia_fullsky}}
\end{figure*}

We applied the systematic offsets presented in Table~\ref{tab:astro_offsets} to the 2,179,410 sources in the comparison sample, and computed standard deviation between the CatWISE2020 motion-fit and \textit{Gaia} position and motion values. The results are summarized in Figure~\ref{fig:catwise_vs_gaia_fullsky}.

The positional accuracy floor for bright sources approaches $\sim35$\,mas, a factor $\sim1.4$ better than the performance achieved by the CatWISE Preliminary Catalog (Figure~\ref{fig:catwise_vs_gaia_fullsky}, top left panel). At fainter magnitudes, the CatWISE2020 Catalog also shows improved astrometric precision. The dispersion in the 15.0--15.5 mag bin is $\sim$165\,mas in the CatWISE2020 Catalog, while it is $\sim$220\,mas in the CatWISE Preliminary Catalog. In the faintest magnitude bin the dispersion in the CatWISE2020 Catalog is $\sim700$\,mas.

The motion accuracy floor for bright stars is just over 6 to 7\,mas\,yr$^{-1}$, as illustrated in the top right panel of Figure~\ref{fig:catwise_vs_gaia_fullsky}. At the faint end, the CatWISE2020 Catalog also shows improved performance with respect to the CatWISE Preliminary Catalog. In the 15.0--15.5 mag bin the one-sigma dispersion is $\sim$20\,mas\,yr$^{-1}$, compared to $\sim$26\,mas\,yr$^{-1}$ in the CatWISE Preliminary Catalog. In the faintest magnitude bin the dispersion in the CatWISE2020 Catalog is $\sim100$\,mas\,yr$^{-1}$.

The $\chi^2$ panels show different performance for positions and motions. The bottom left panel of Figure~\ref{fig:catwise_vs_gaia_fullsky} shows improved $\chi^2$ values for bright stars in the CatWISE2020 catalog with respect to the CatWISE Preliminary Catalog, with the $\chi^2$ values rising in fainter sources to the same level observed in the CatWISE Preliminary catalog. The motion $\chi^2$ (bottom right panel) for bright sources on the other hand are worse in the CatWISE2020 Catalog than in the CatWISE Preliminary Catalog, gradually improving towards fainter magnitudes and reaching the same level as the CatWISE Preliminry Catalog at W1,W2$\sim$15\,mag. The position $\chi^2$ are significantly improved most likely as a result of the better deblending, while motion $\chi^2$ are worse because of the smaller floor on the motion uncertainty imposed by the CatWISE2020 pipeline (\S\ref{sec:wphot}).

\begin{figure*}
\centering
\includegraphics[width=\textwidth]{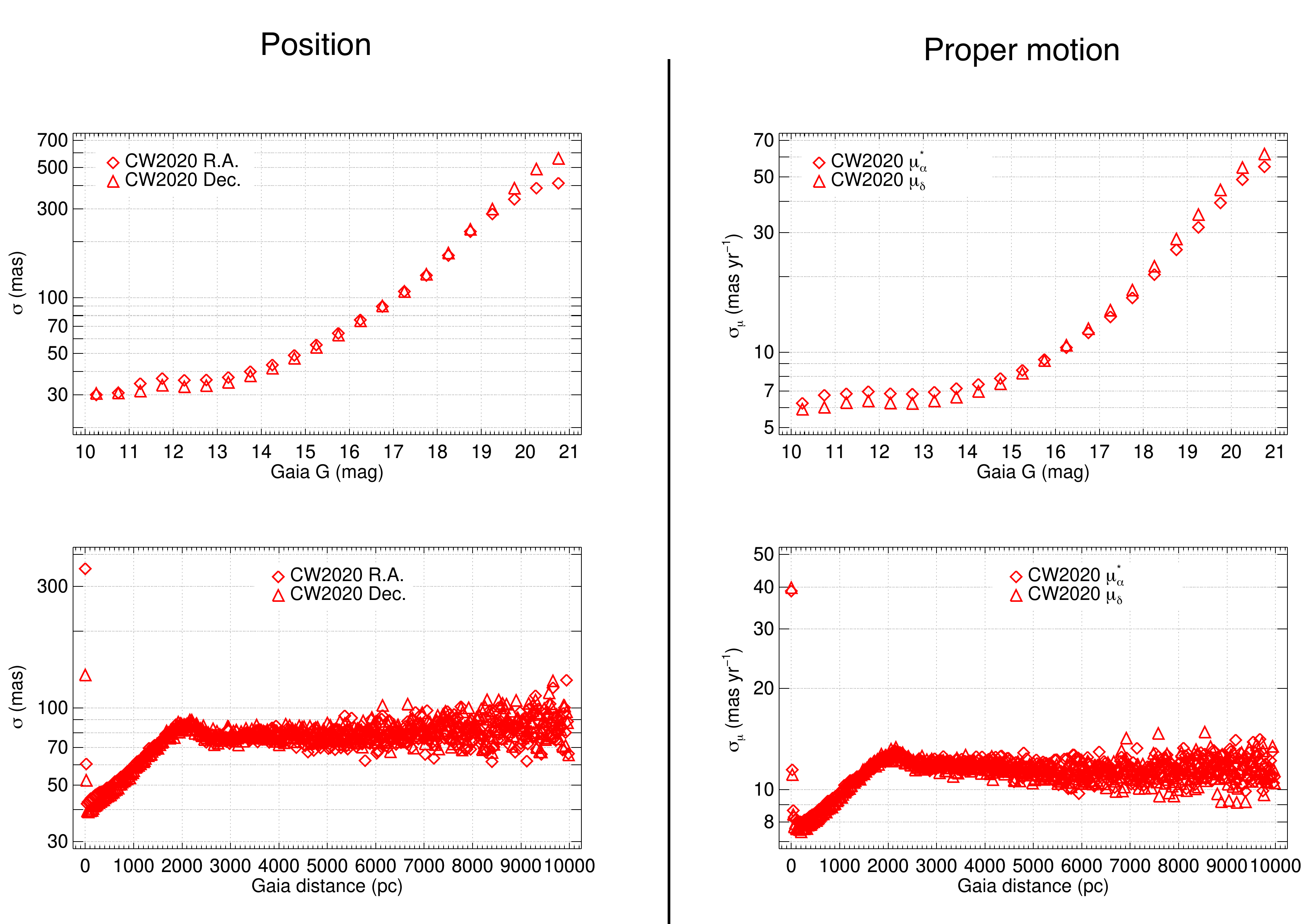}
\caption{CatWISE2020 astrometric performance with respect to \textit{Gaia} DR2. \textit{Top}: the 1-$\sigma$ dispersion between CatWISE2020 and \textit{Gaia} R.A. and Dec.  (left) and proper motion (right), as a function of \textit{Gaia} G magnitude. \textit{Bottom}: same as the top row, but as a function of \textit{Gaia} measured distance, in bins of 20\,pc. \label{fig:dist_gmag}}
\end{figure*}

\begin{figure}
    \centering
    \includegraphics[width=0.49\textwidth]{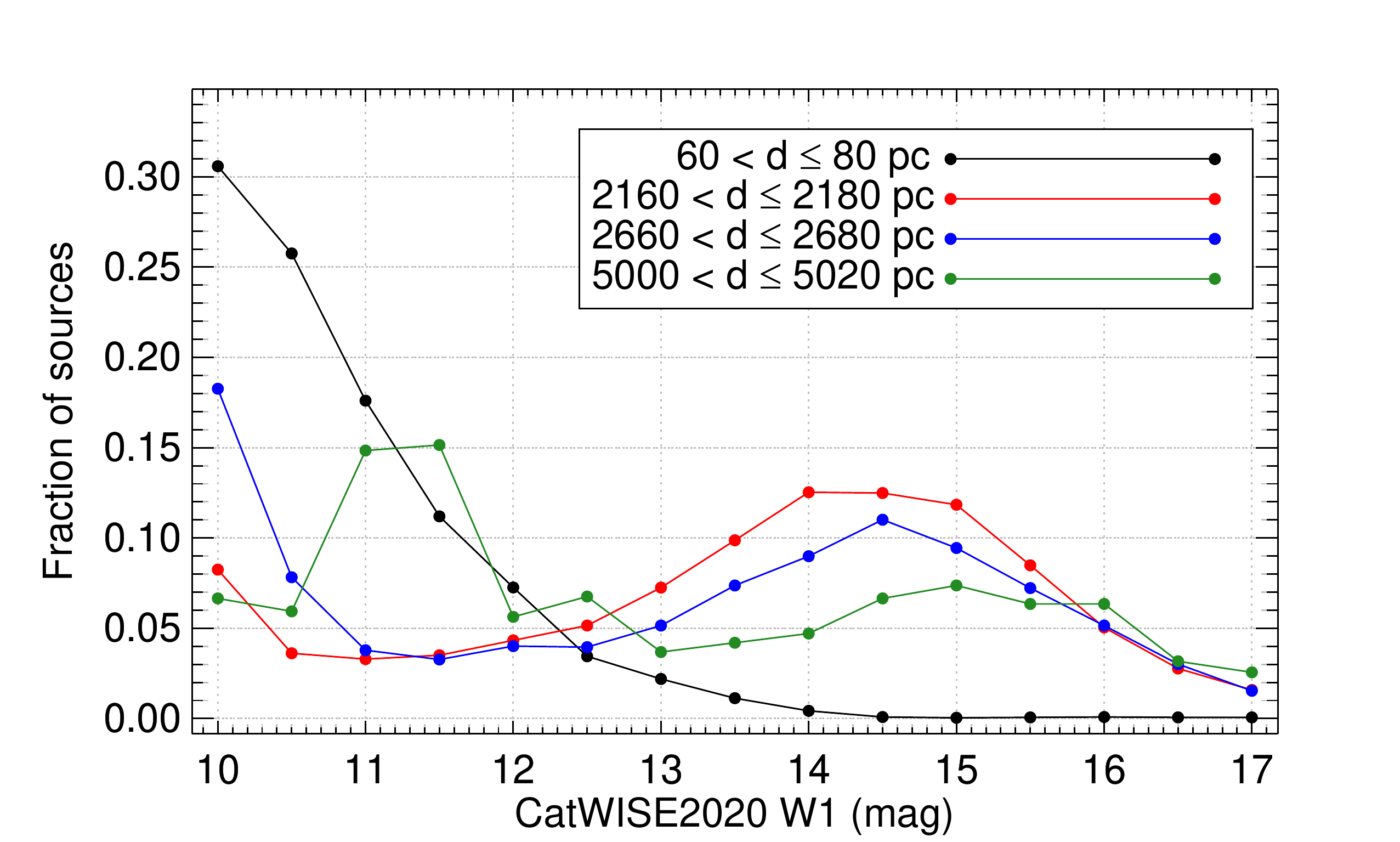}
    \includegraphics[width=0.49\textwidth]{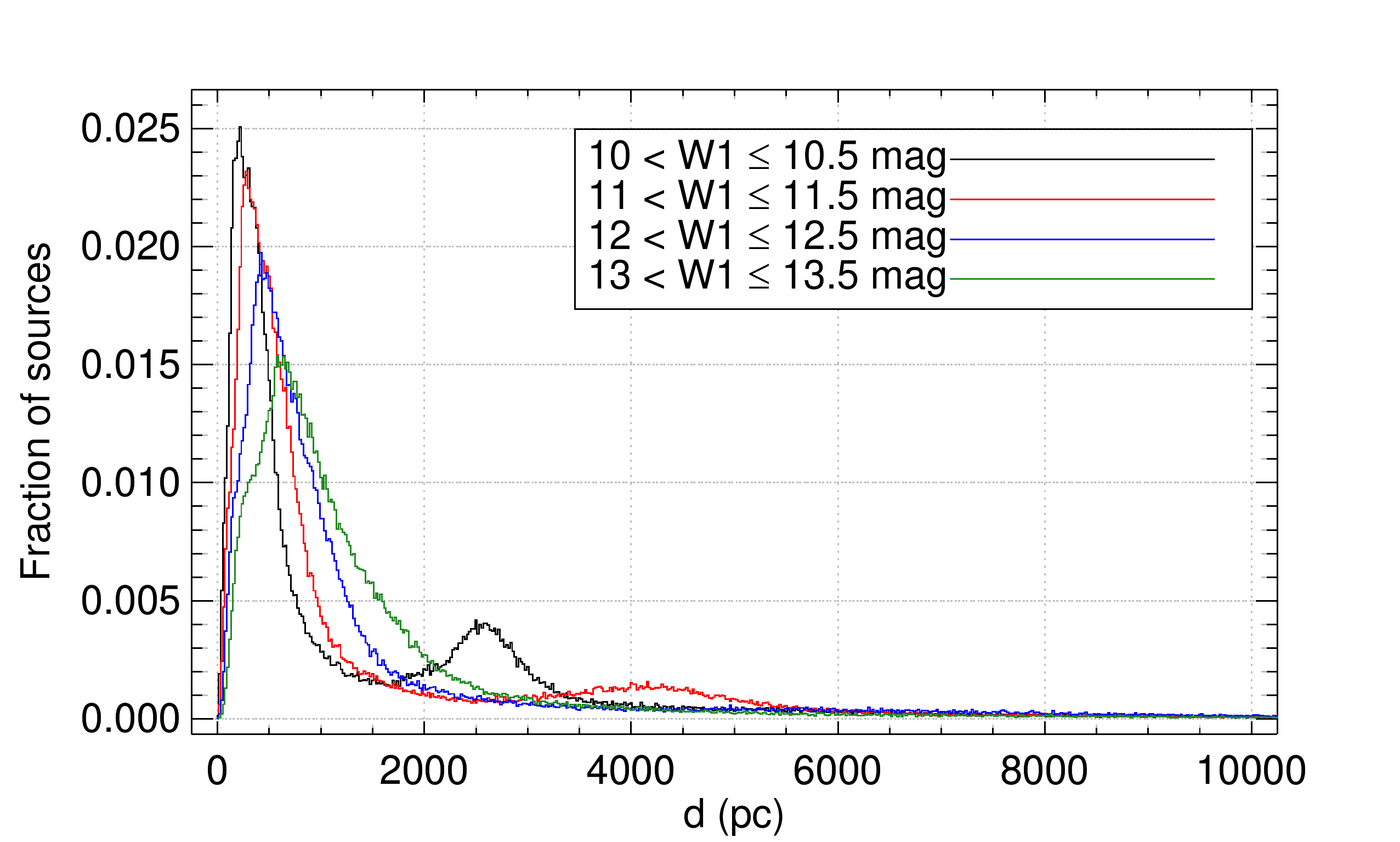}
    \caption{\textit{Top:} the W1 magnitude distribution for CatWISE2020 Catalog sources with \textit{Gaia} counterparts in four distance bins. The three nearest distance bins shown are chosen to be the three inflection points in the bottom row plots of Figure~\ref{fig:dist_gmag}. The fourth distance bin is chosen to represent the brightness distribution for sources with d\,$\gtrsim2.6$\,kpc. \textit{Bottom:} the distance distribution of sources in four magnitude bins. The distribution in the two brightest magnitude intervals show clear secondary peaks at d$>2$\,kpc, which are due to bright, distant RGBs.}
    \label{fig:brightness_distrib}
\end{figure}

To assess the possible role of increasing \textit{Gaia} errors, Figure~\ref{fig:dist_gmag} shows the astrometric performance of the CatWISE2020 Catalog as a function of \textit{Gaia} G magnitude (top row) and distance (bottom row). The top row plots show a behaviour similar to the one illustrated in Figure~\ref{fig:catwise_vs_gaia_fullsky}, with accuracy floors of 30\,mas and 6\,mas\,yr$^{-1}$ in positions and proper motions respectively, and accuracies of 600\,mas and 60\,mas\,yr$^{-1}$ at the faint end. The bottom row plots of Figure~\ref{fig:dist_gmag} on the other hand show a less intuitive trend. The very nearest distance bins are dominated by very bright, highly saturated sources. As a result, the measured performances with respect to \textit{Gaia} are $\sim$200\,mas in position and 40\,mas\,yr$^{-1}$ in proper motion. The astrometric performance improves very quickly as a function of distance, with minimum dispersion in the 60--80\,pc bin. Beyond 80\,pc the astrometry deteriorates again, as one might expect since in more distant bins the stars appear, on average, fainter. At $\sim$2.1\,kpc, however, the trend briefly reverses, reaching a local minimum at $\sim$2.6\,kpc. Beyond $\sim$2.6\,kpc the astrometric performance resumes very gradually deteriorating as a function of distance for positions, while for motions it slightly improves with distance. This pattern is a result of the change of the brightness distribution of stars in the CatWISE2020 Catalog as a function of distance. The top panel of Figure~\ref{fig:brightness_distrib} illustrates this point. The nearest bin is dominated by bright sources, and as such the astrometric performance is comparable to the astrometric performance in the brightest bins in Figure~\ref{fig:catwise_vs_gaia_fullsky}. The brightness distribution of more distant bins peaks at fainter magnitudes, and as such the astrometric performance degrades, once again following the well-demonstrated trend with brightness. Somewhat counterintuitively, however, beyond $\sim$2.1\,kpc the brightness distribution becomes bimodal, with a growing peak at bright magnitudes. This bright peak is most likely due to red giant stars. This can be seen as well in the bottom panel of Figure~\ref{fig:brightness_distrib}, where the distance distribution of $10 < {\rm W1} \leq 10.5$\,mag stars shows a secondary peak at d$\sim$2.6\,kpc. RGB stars have $-5 \lesssim M_{W1} \lesssim -2$\,mag \citep{Conroy2018}, roughly corresponding to the observed W1\,=\,10\,mag at this distance. Moving beyond $\sim$2.6\,kpc, the top panel of Figure~\ref{fig:brightness_distrib} shows that the bright peak moves towards fainter magnitudes and, therefore,  the accuracy of position measurements resumes degrading. While sources in a given distance bin become on average fainter with distance, their proper motion becomes on average smaller, counteracting the brightness-dependent astrometric degradation and resulting in a slow improvement of the proper motion performance as a function of distance.

\begin{figure*}
\includegraphics[width=0.5\textwidth]{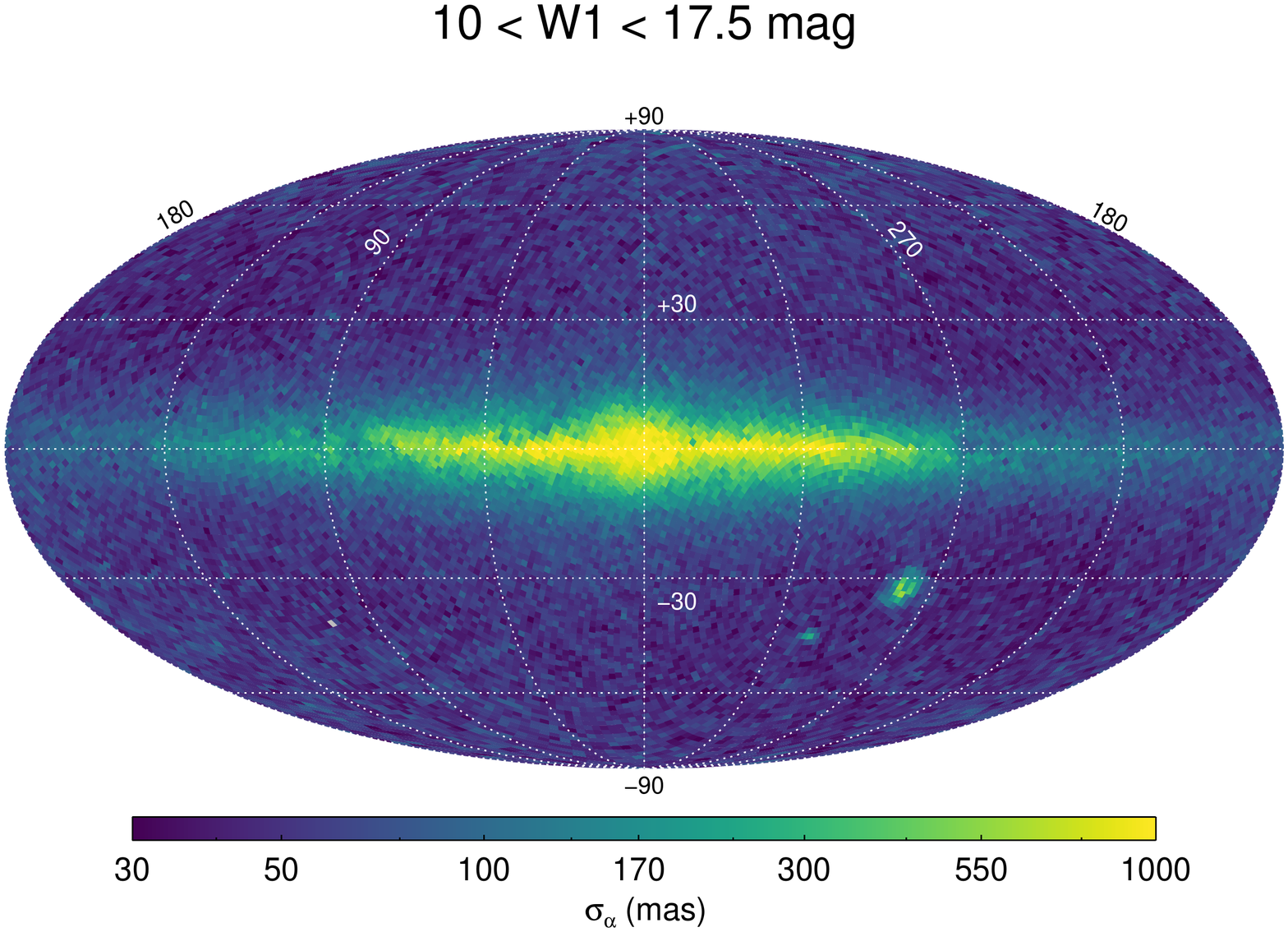}
\includegraphics[width=0.5\textwidth]{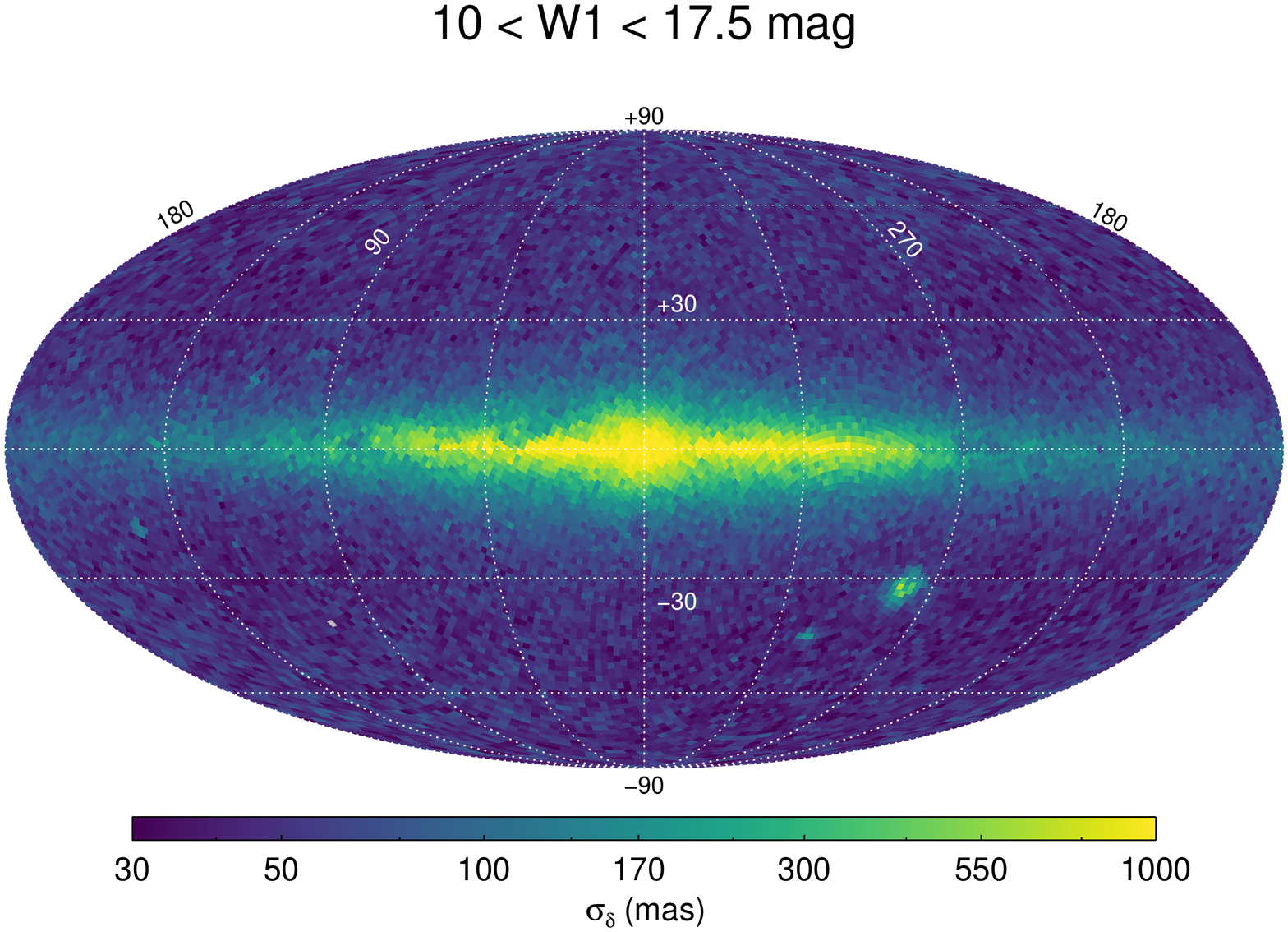}
\caption{1-$\sigma$ dispersion of the CatWISE2020 Catalog $\alpha$ (left) and $\delta$ (right) with respect to \textit{Gaia} DR2, for sources in the $10<$W1$<17.5$\,mag range. \label{fig:map_positions}}
\end{figure*}

\begin{figure*}
\includegraphics[width=0.5\textwidth]{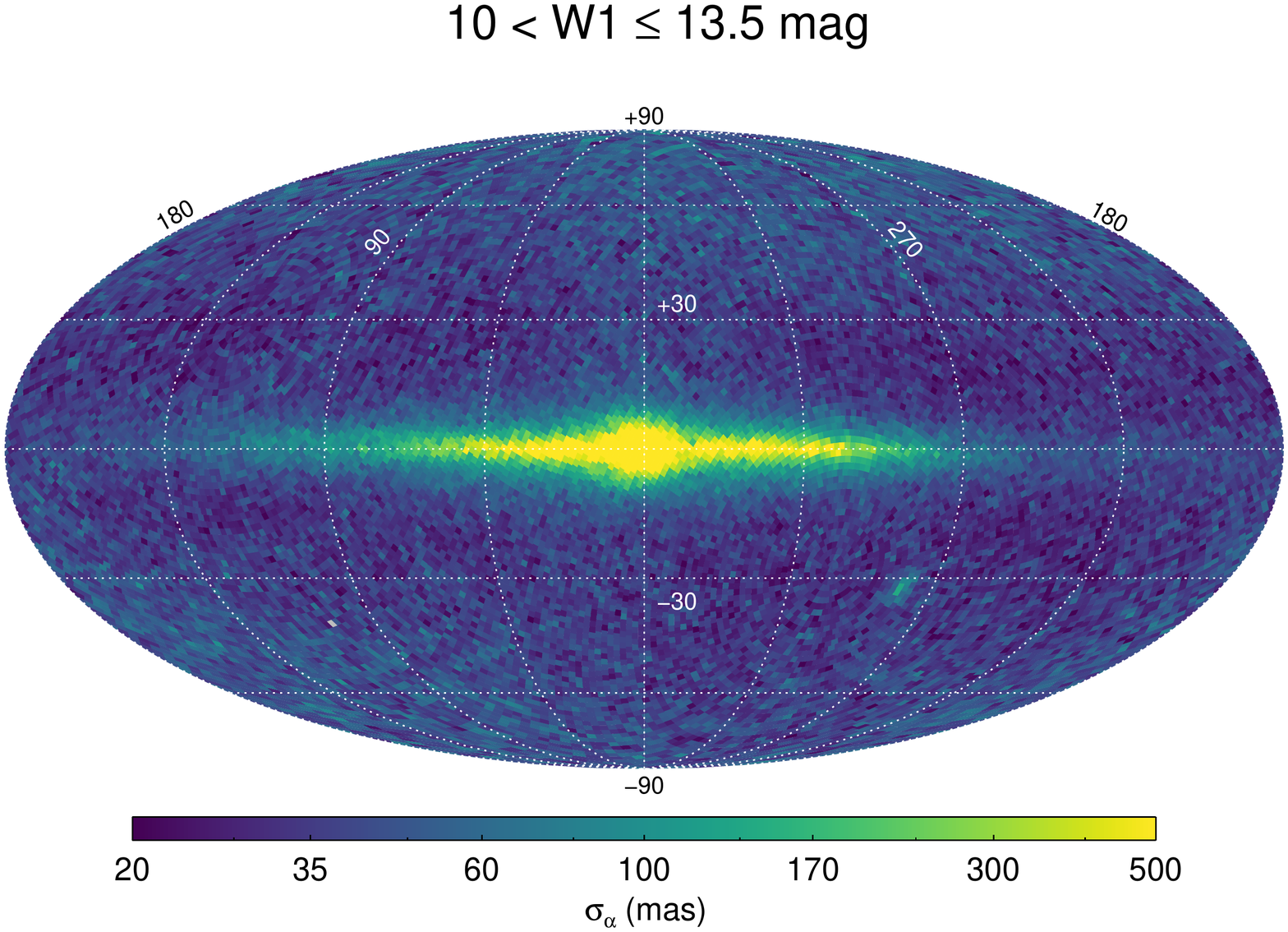}
\includegraphics[width=0.5\textwidth]{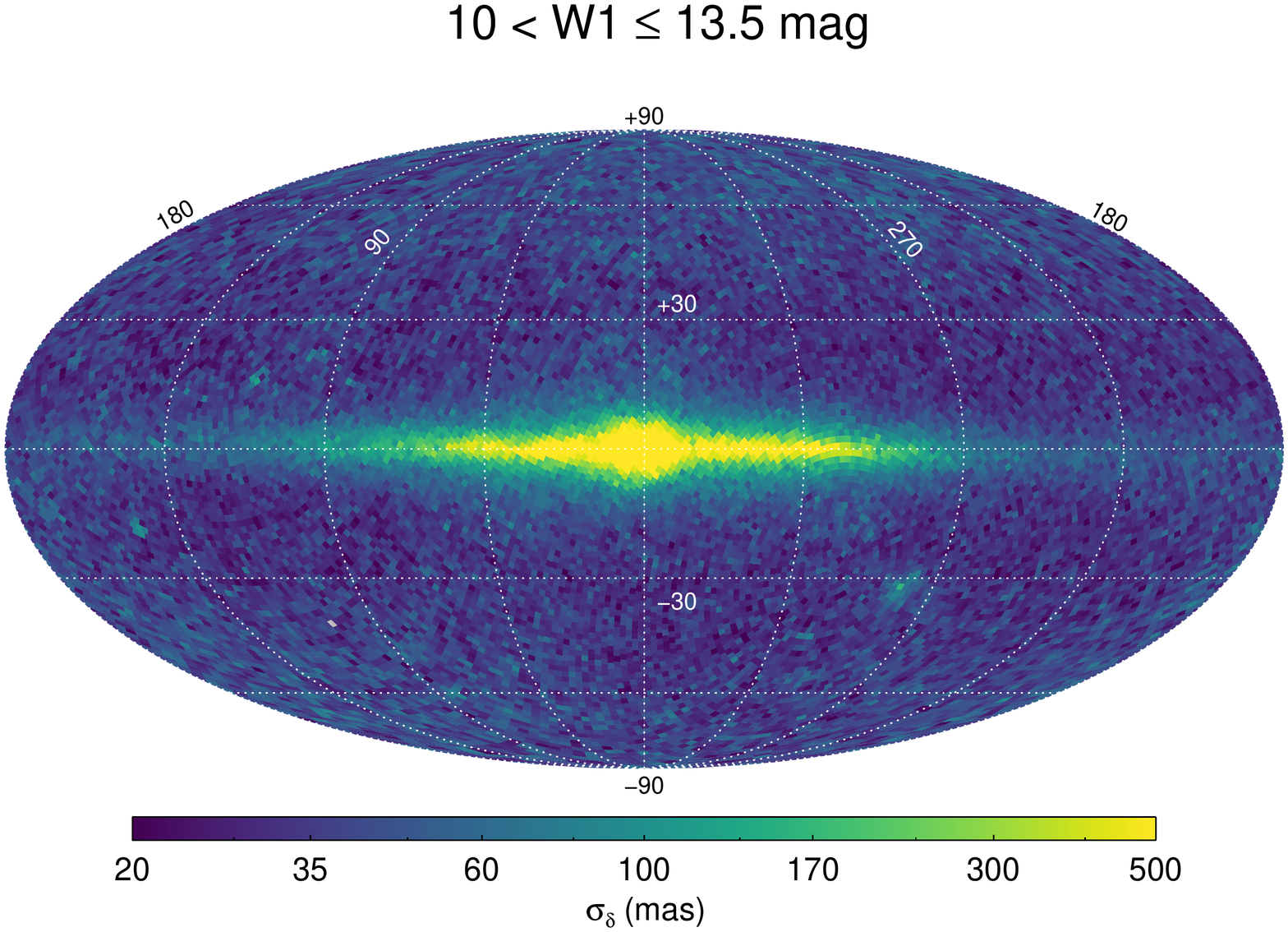}
\includegraphics[width=0.5\textwidth]{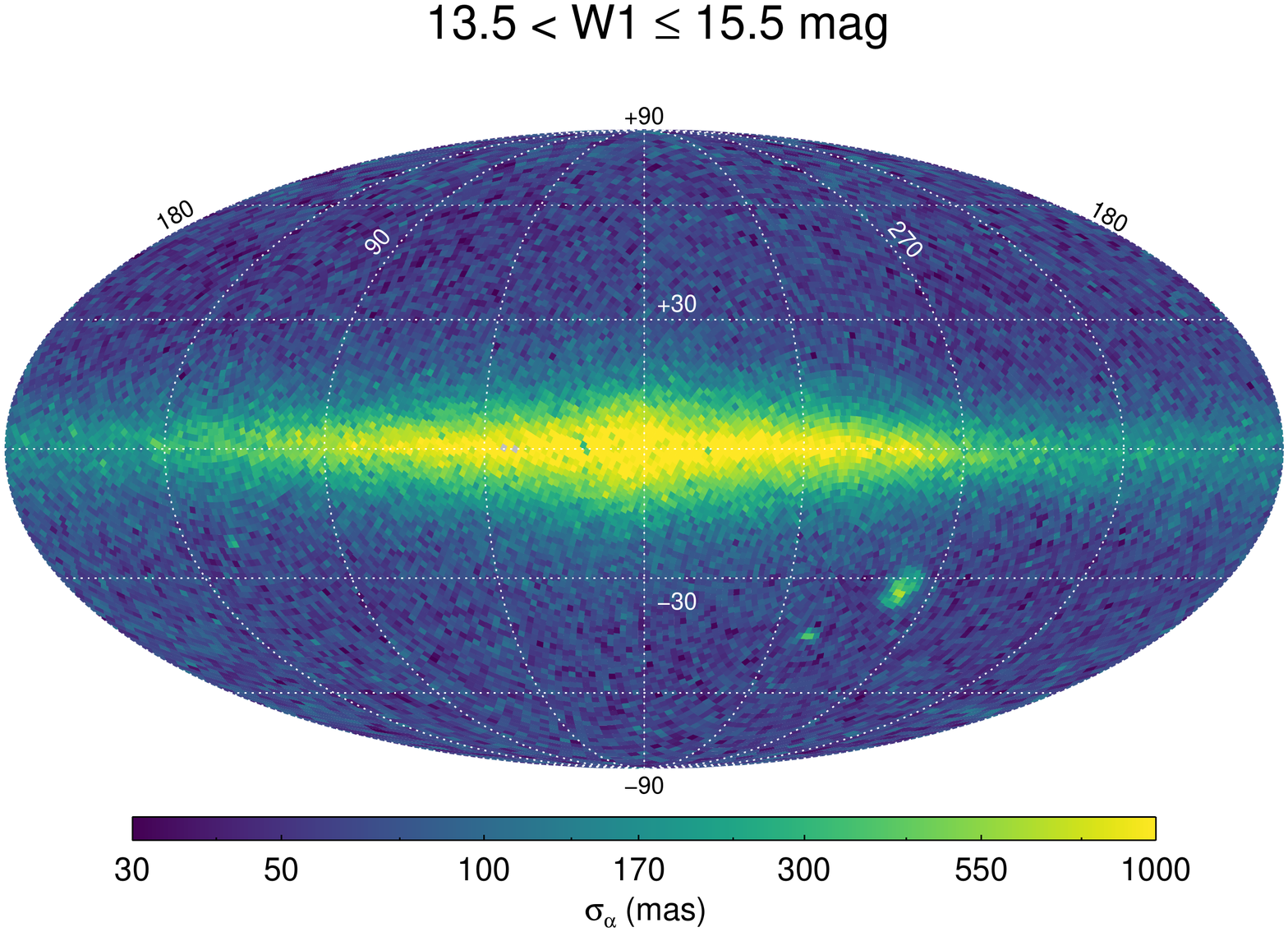}
\includegraphics[width=0.5\textwidth]{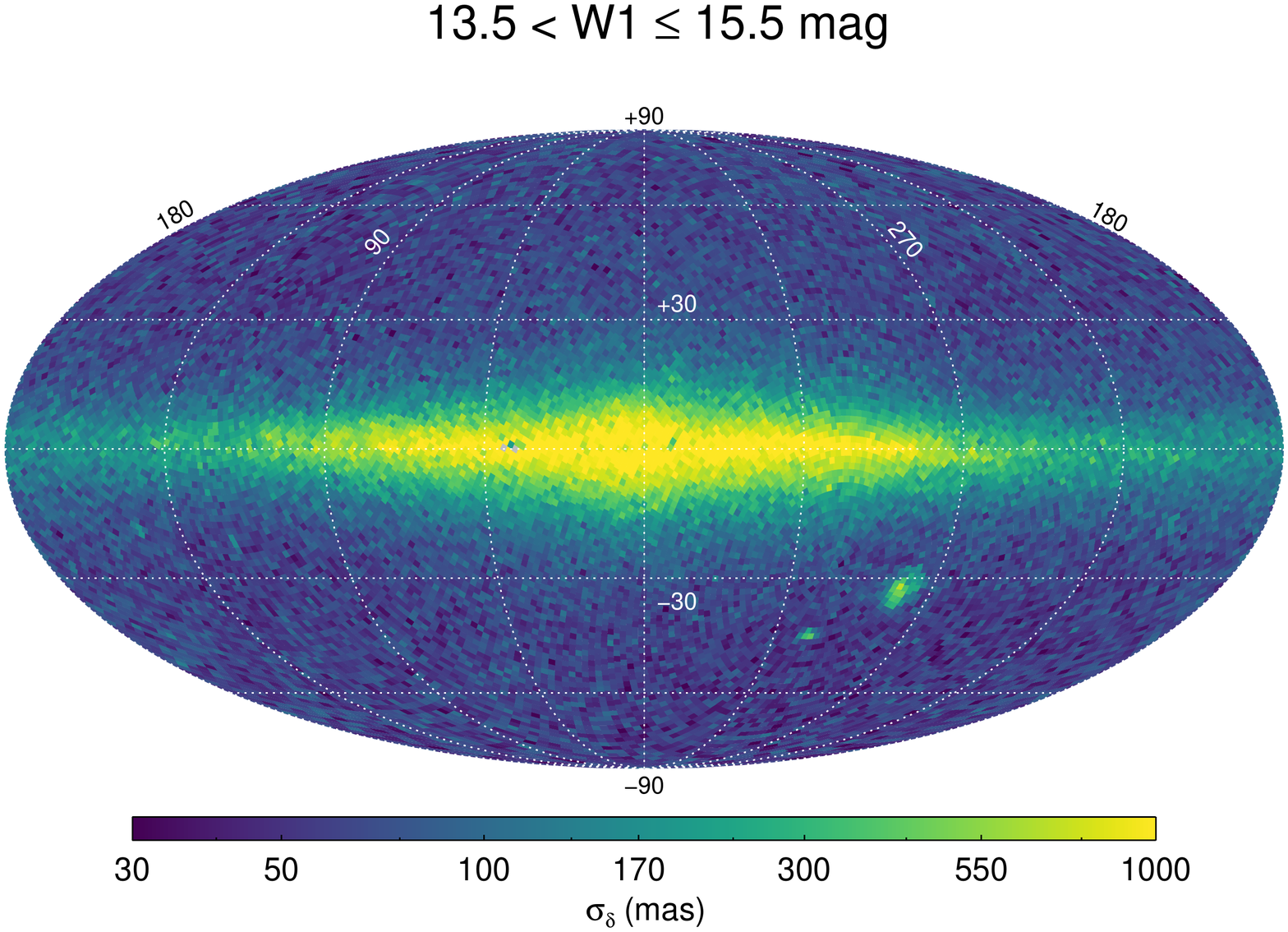}
\includegraphics[width=0.5\textwidth]{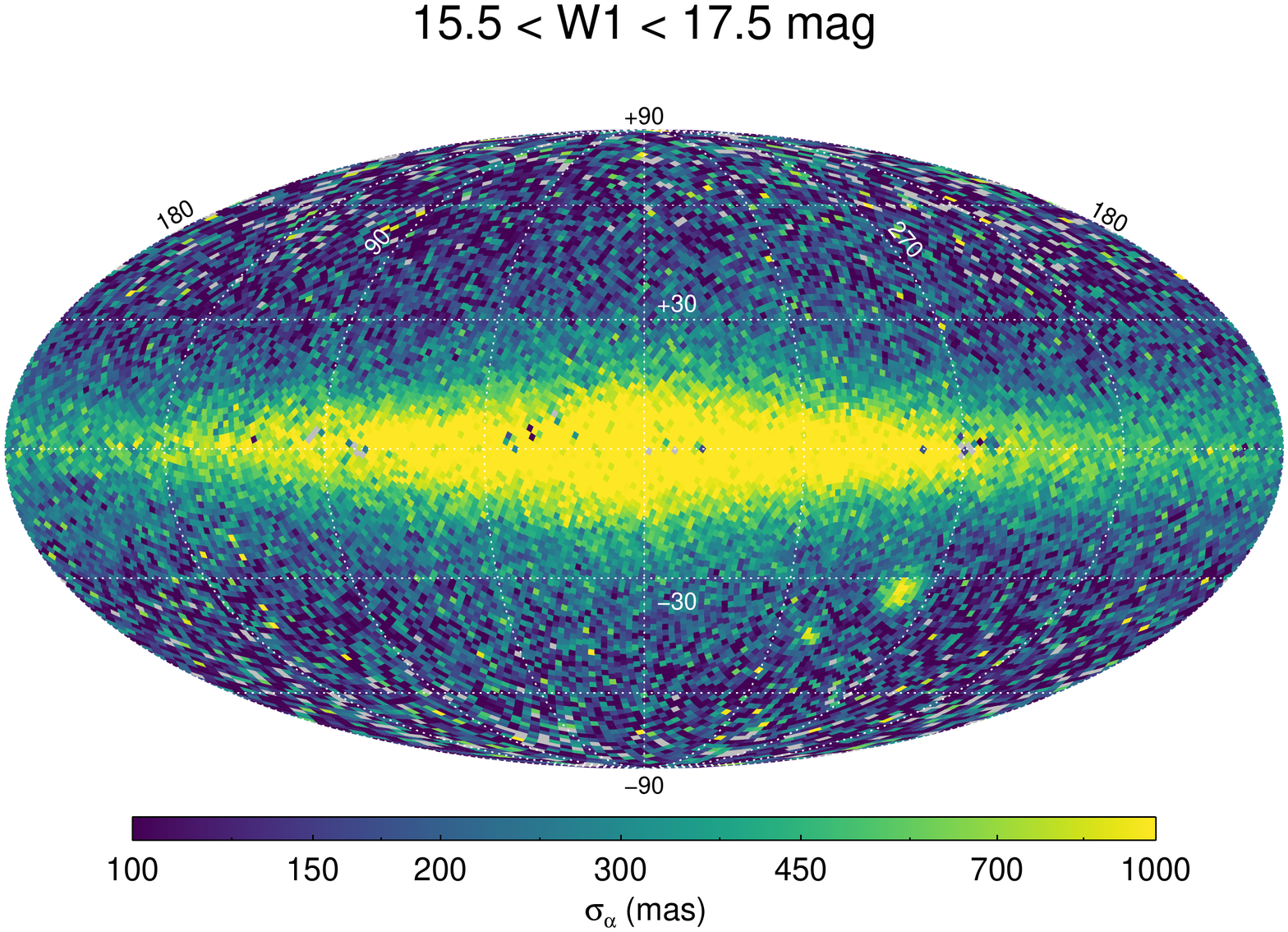}
\includegraphics[width=0.5\textwidth]{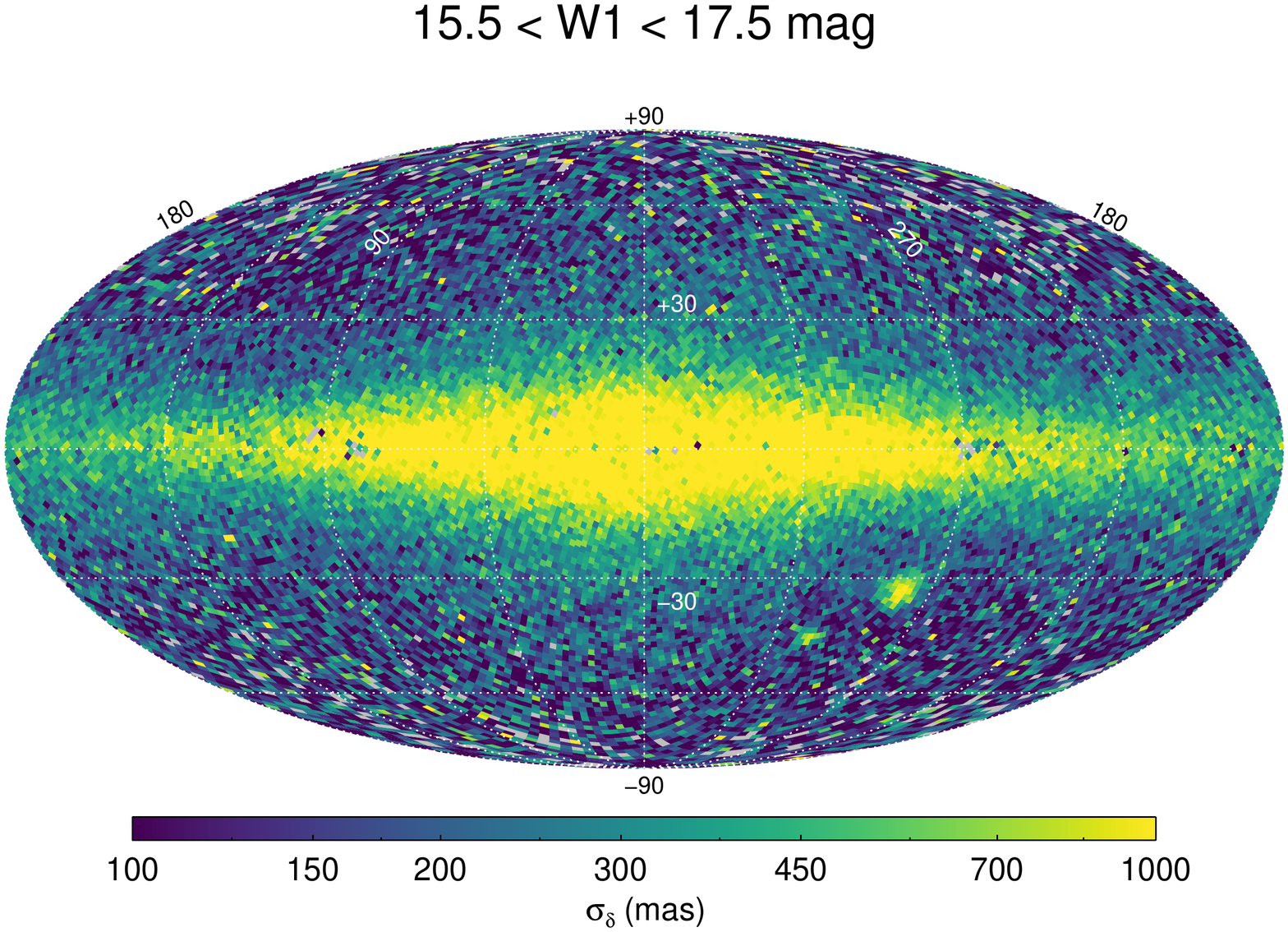}
\caption{Same as Figure~\ref{fig:map_positions}, but for three W1 magnitude ranges. Gray tiles are those where there were no sources in the CatWISE2020 Catalog in the given magnitude bin. \label{fig:map_positions_bins}}
\end{figure*}

\begin{figure*}
\includegraphics[width=0.5\textwidth]{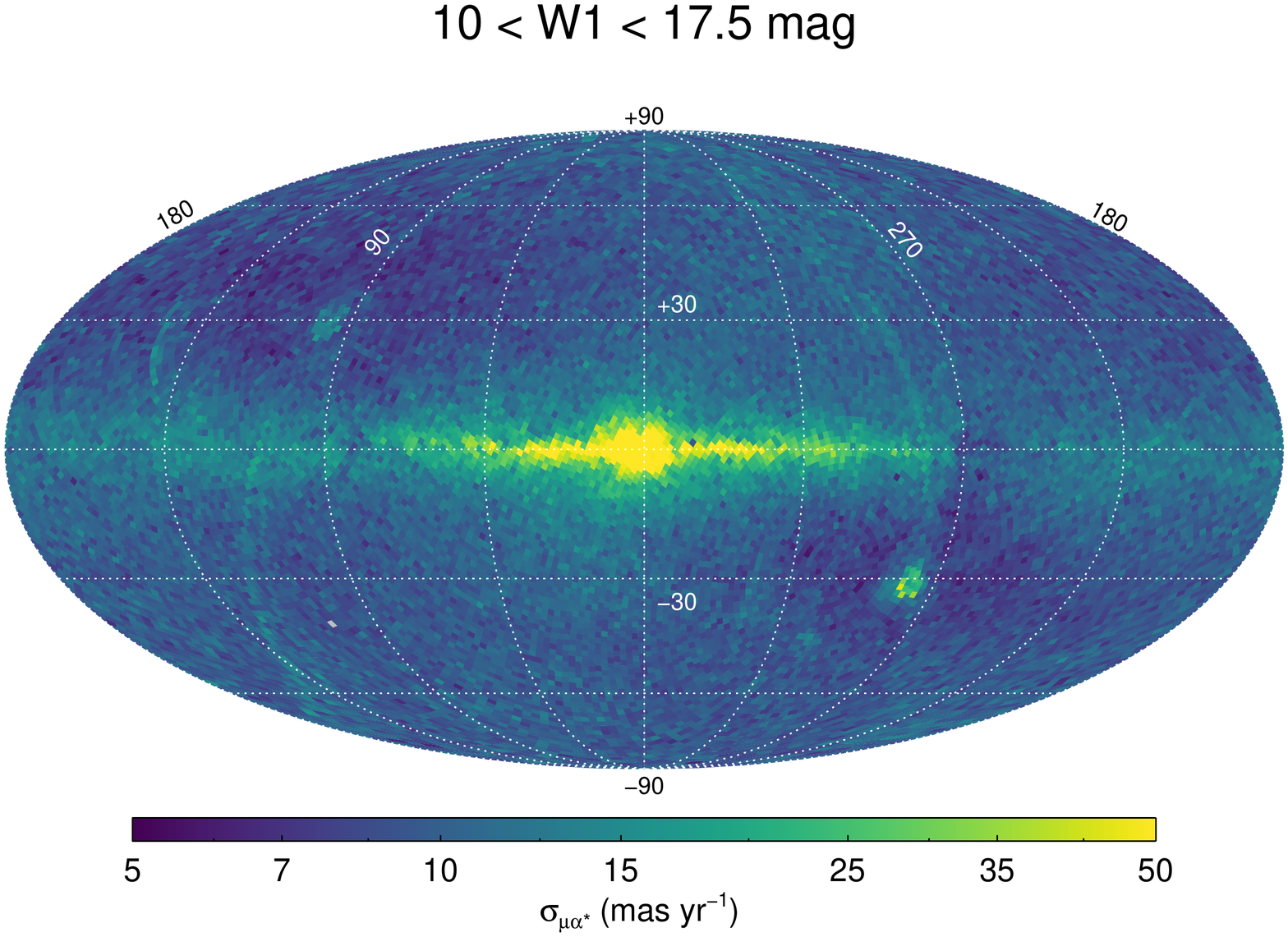}
\includegraphics[width=0.5\textwidth]{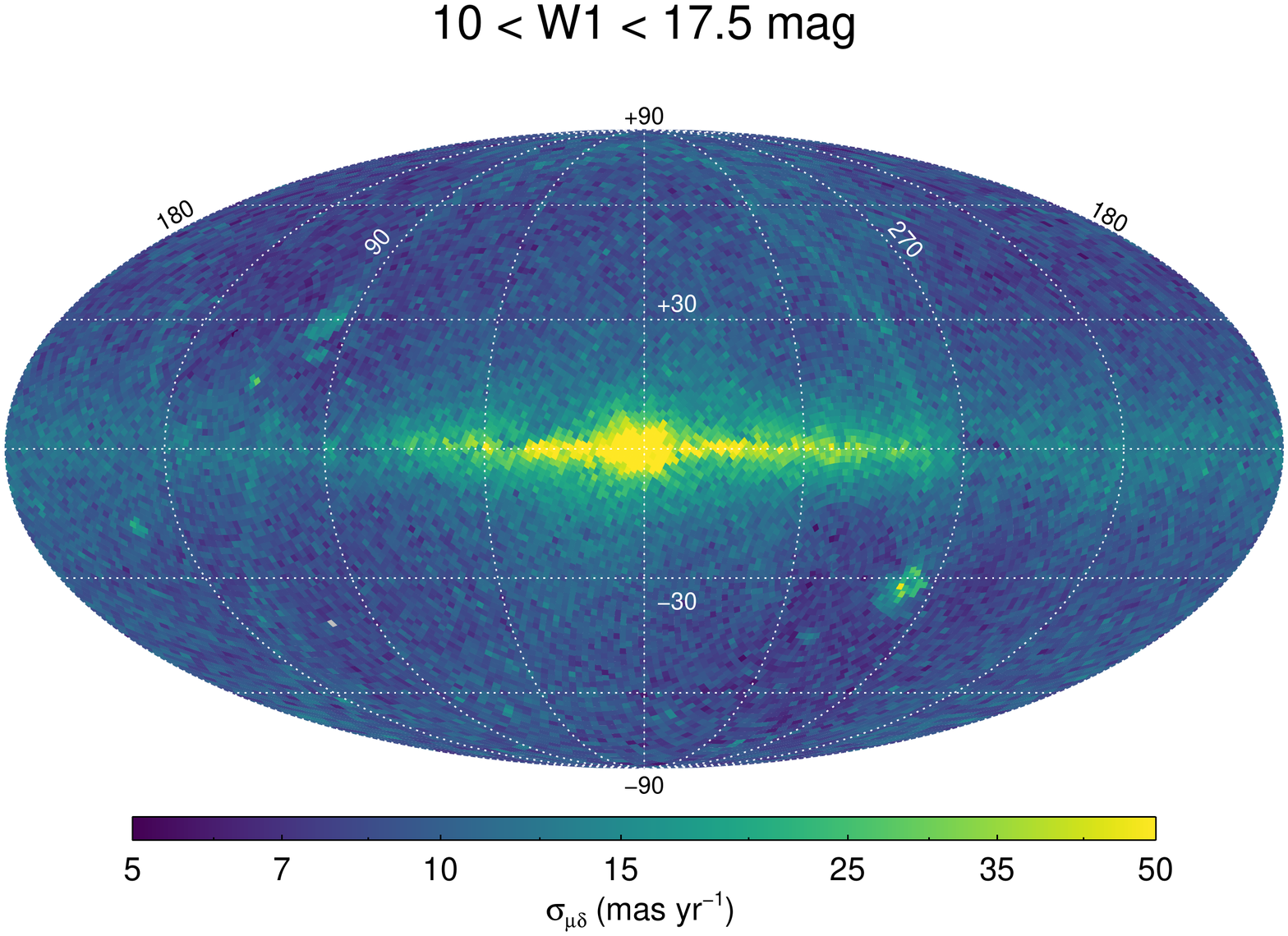}
\caption{Same as Figure~\ref{fig:map_positions}, but for the proper motion components.\label{fig:map_pms}}
\end{figure*}

\begin{figure*}
\includegraphics[width=0.5\textwidth]{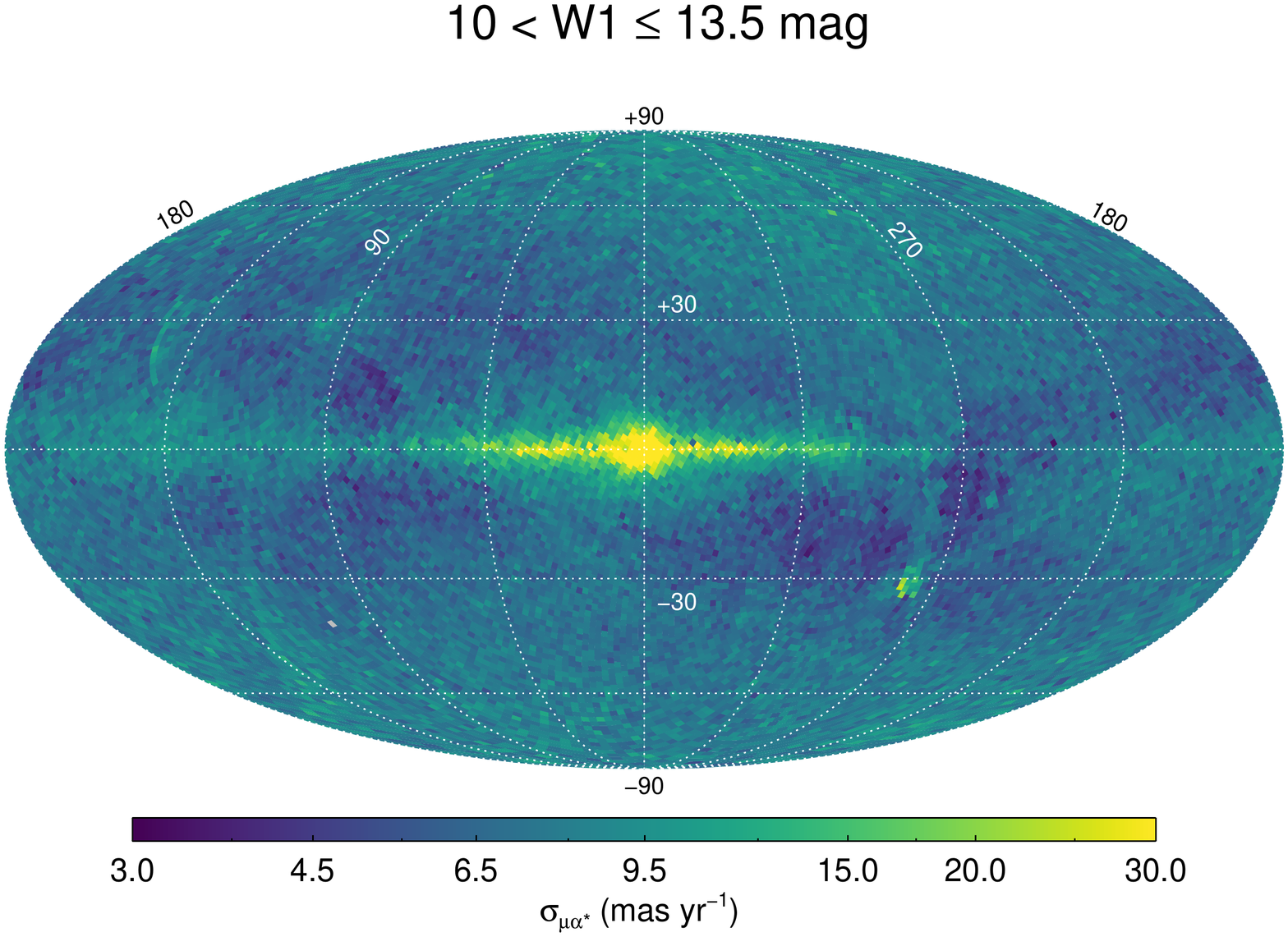}
\includegraphics[width=0.5\textwidth]{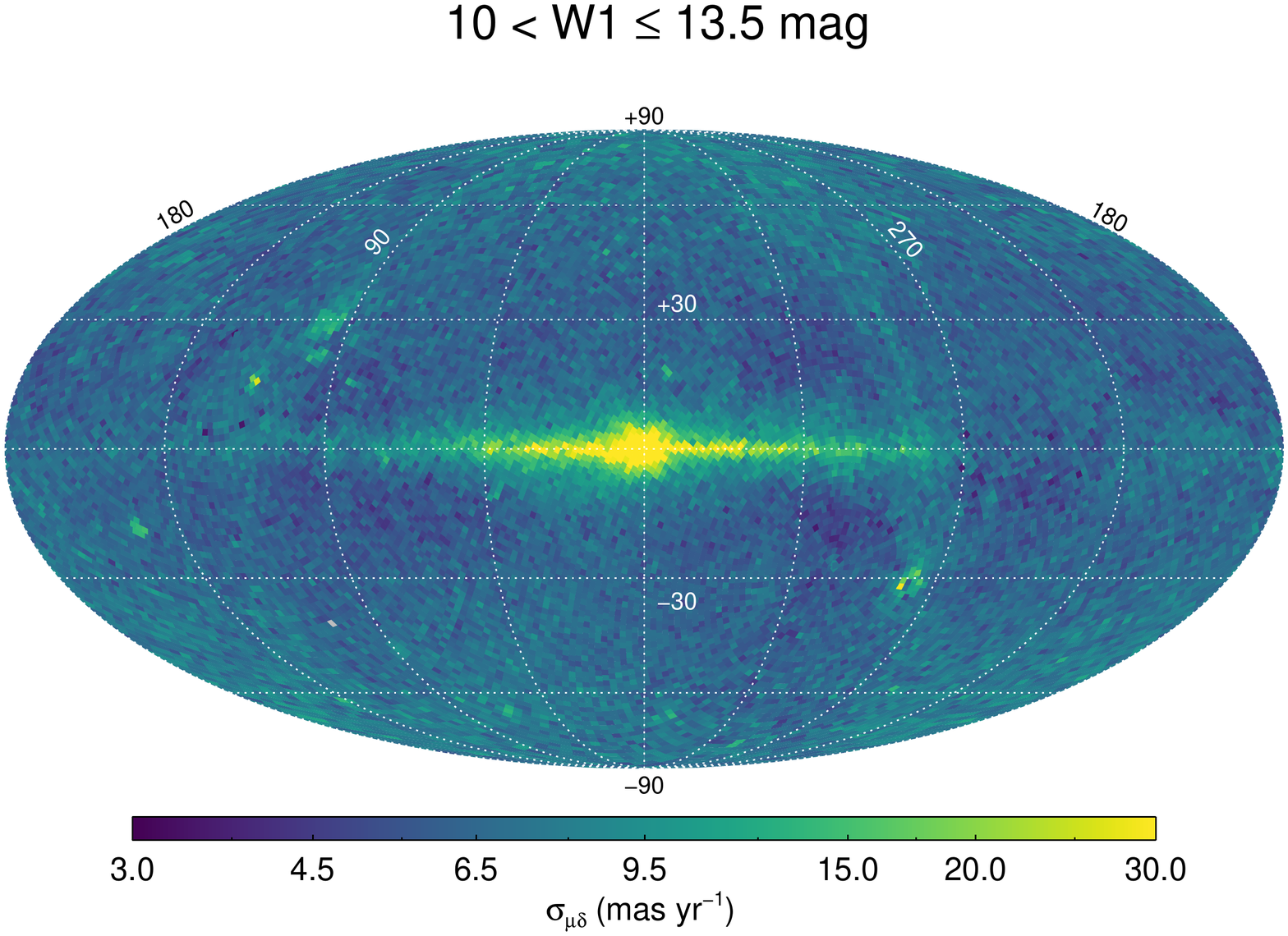}
\includegraphics[width=0.5\textwidth]{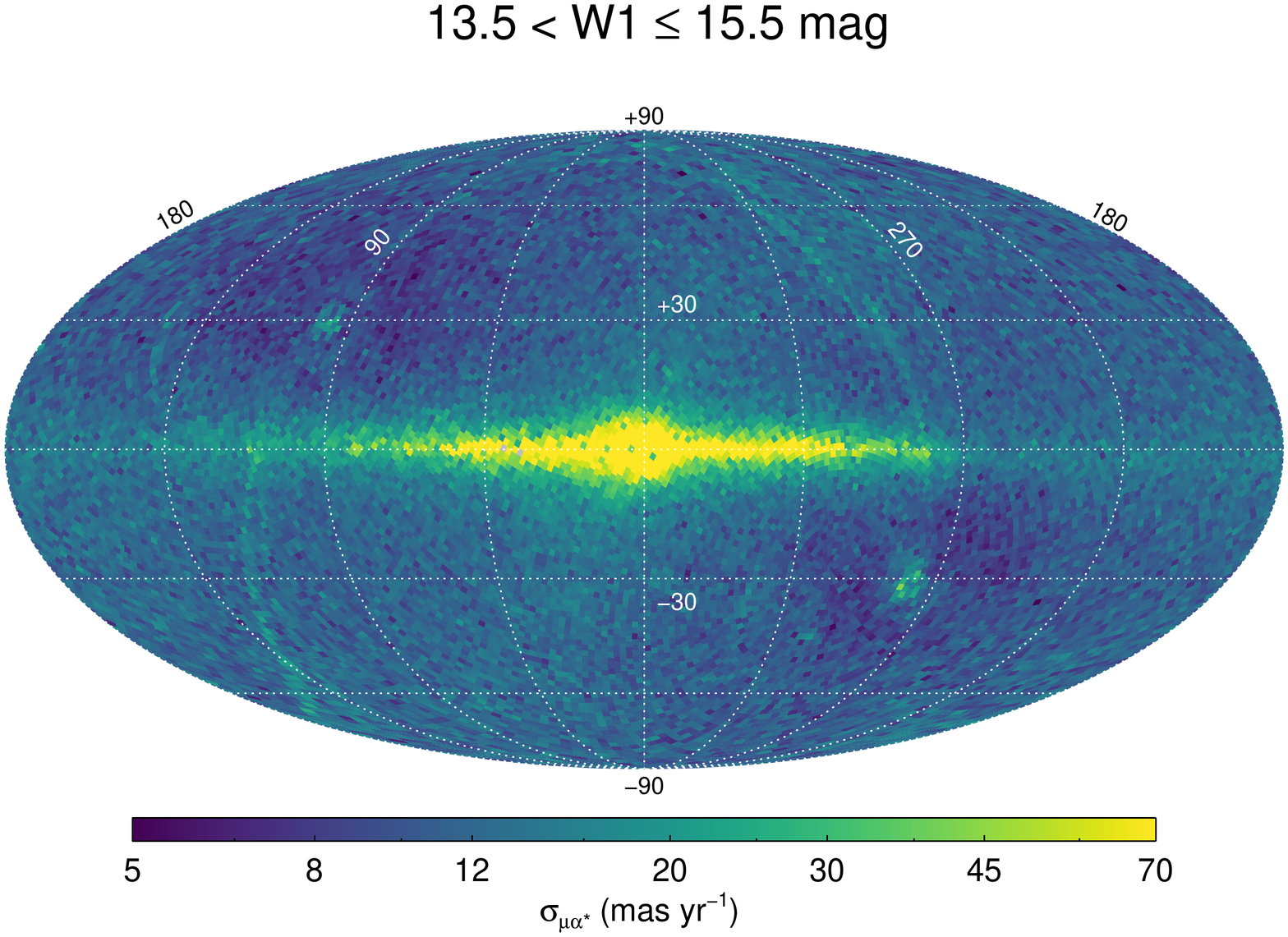}
\includegraphics[width=0.5\textwidth]{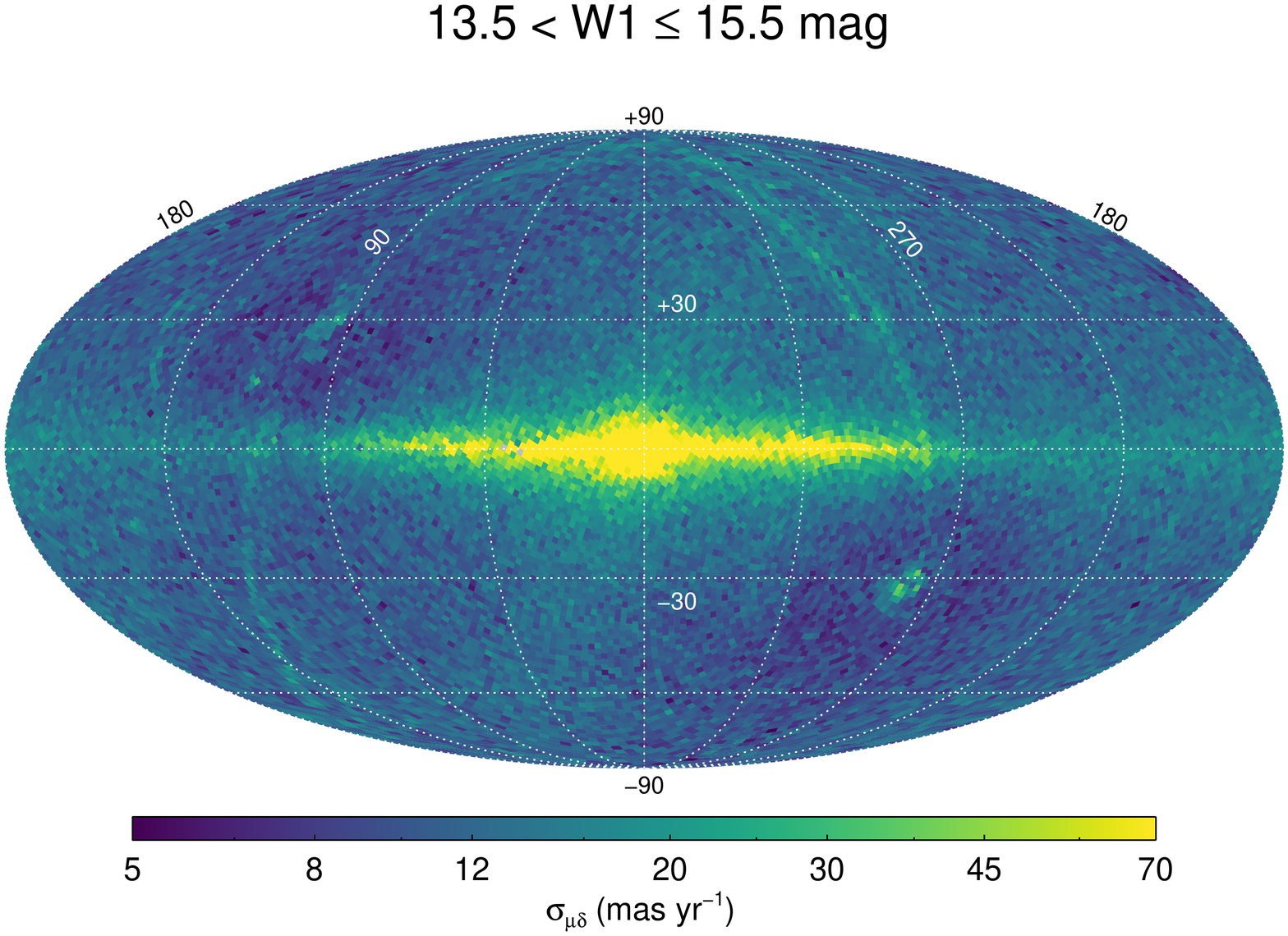}
\includegraphics[width=0.5\textwidth]{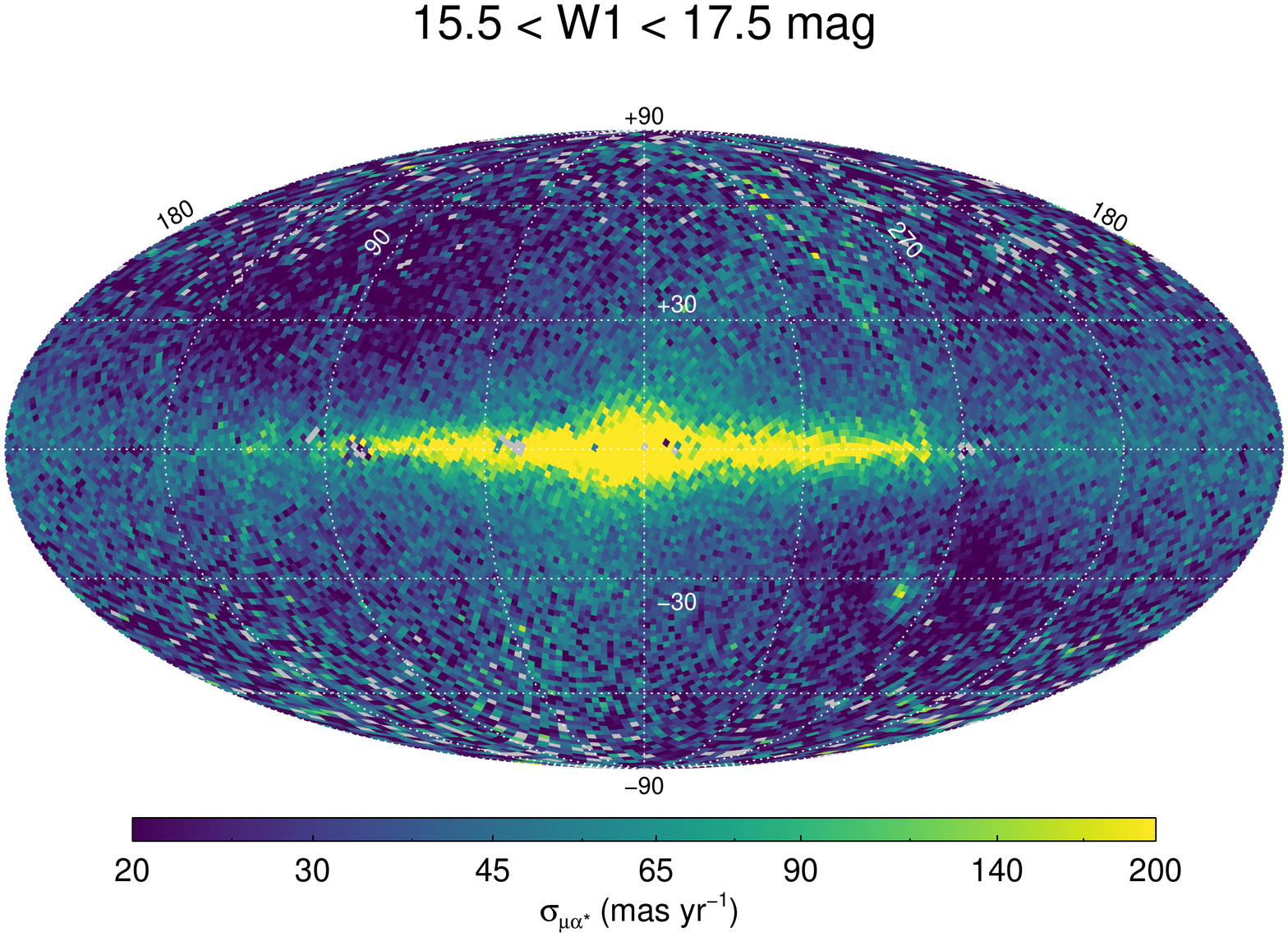}
\includegraphics[width=0.5\textwidth]{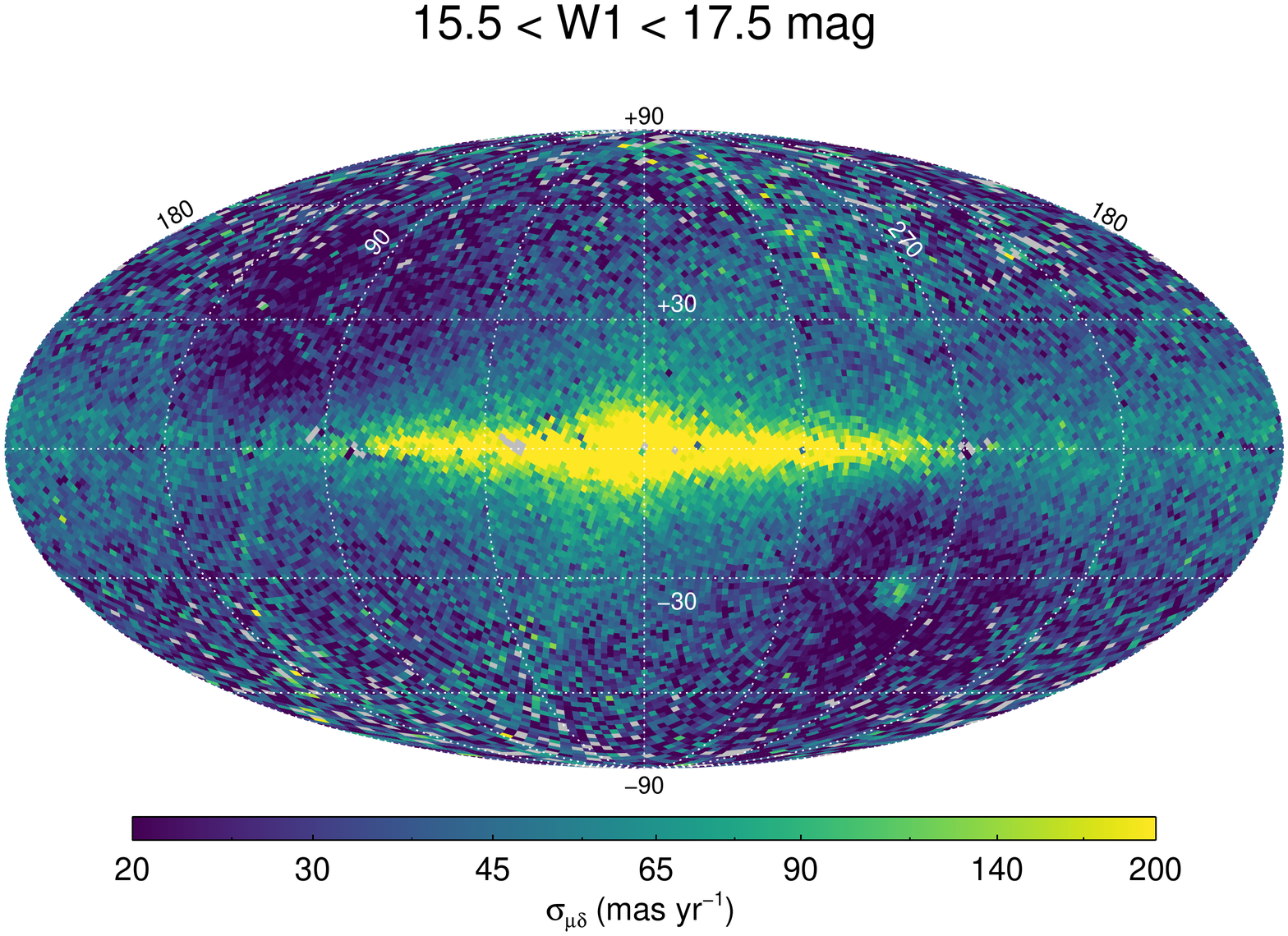}
\caption{Same as Figure~\ref{fig:map_pms}, but for three W1 magnitude ranges. Gray tiles are those where there were no sources in the CatWISE2020 Catalog in the given magnitude bin.\label{fig:map_pms_bins}}
\end{figure*}

The overall astrometric performance is, as one might expect, not uniform over the sky. Figures~\ref{fig:map_positions}--\ref{fig:map_pms_bins} show the 1-$\sigma$ dispersion in each tile with respect to \textit{Gaia} positions and motion components for the full magnitude range considered (Figures~\ref{fig:map_positions} and \ref{fig:map_pms}), and in three smaller magnitude intervals (Figures~\ref{fig:map_positions_bins} and \ref{fig:map_pms_bins}). The maps for the full magnitude range are smooth overall, indicating a fairly constant astrometric performance for the CatWISE2020 Catalog over the majority of the sky. The main features can be easily identified -- the Galactic plane (and in particular the bulge), and the Small and Large Magellanic Clouds (SMC and LMC). In those denser regions, the astrometric accuracy for the bright stars deteriorates to $\sim500$\,mas for positions and $\sim30$\,mas\,yr$^{-1}$ for motions, and to $\sim1,000$\,mas and $\sim200$\,mas\,yr$^{-1}$ (or worse) for the faint stars. 

The motion accuracy maps of Figure~\ref{fig:map_pms} and \ref{fig:map_pms_bins} show additional features, already noted in the CatWISE Preliminary Catalog, that appear to be related to the WISE survey strategy, and to the transition between the cryogenic and post-cryogenic phases of the mission. 

The maps for the faintest magnitude interval (bottom row of Figure~\ref{fig:map_positions_bins} and \ref{fig:map_pms_bins}) are more noisy, partly because of the low number of sources with a \textit{Gaia} counterpart in each tile, and partly because the CatWISE2020 pipeline detects only brighter sources in high density regions because of confusion noise. The astrometric performance of the CatWISE2020 Catalog remains fairly homogeneous over the sky even for the faintest sources, but the performance starts degrading at higher $b$ as one considers fainter and fainter sources. This is most likely due to the fact that the astrometry for faint sources is more susceptible to degradation due to blending.  

To further assess the astrometric performance of the CatWISE2020 pipeline on faint, red sources that are not seen by \textit{Gaia}, we compared the CatWISE2020 positions and motions to those measured in the Extended \textit{Gaia}-PS1-SDSS Proper Motion Catalog \citep[GPS1+;][]{Tian2020}. The GPS1+ catalog provides positions and proper motions for $\sim400$ million sources with $19.0\lesssim r \lesssim 22.5$\,mag over 3/4 of the sky, measured using the combination of \textit{Gaia}, PS1, SDSS, and 2MASS data.

We cross-matched our astrometric comparison sample of 2,735,892 stars with GPS1+ with a matching radius of 5\farcs 5, the same radius used in the comparison against \textit{Gaia} DR2. We then computed the one-sigma dispersion between the CatWISE2020 Catalog and GPS1+ positions and proper motions following the same procedure adopted for our comparison against \textit{Gaia} DR2. Since GPS1+ omits sources with r\,$\lesssim$\,19\,mag, the correct matching source is often missing for CatWISE2020 sources of W1\,$<$\,14.5\,mag. Therefore, for the rest of the analysis we only consider sources with W1\,$\geq$\,14.5\,mag.

\begin{figure*}
    \centering
    \includegraphics[width=\textwidth]{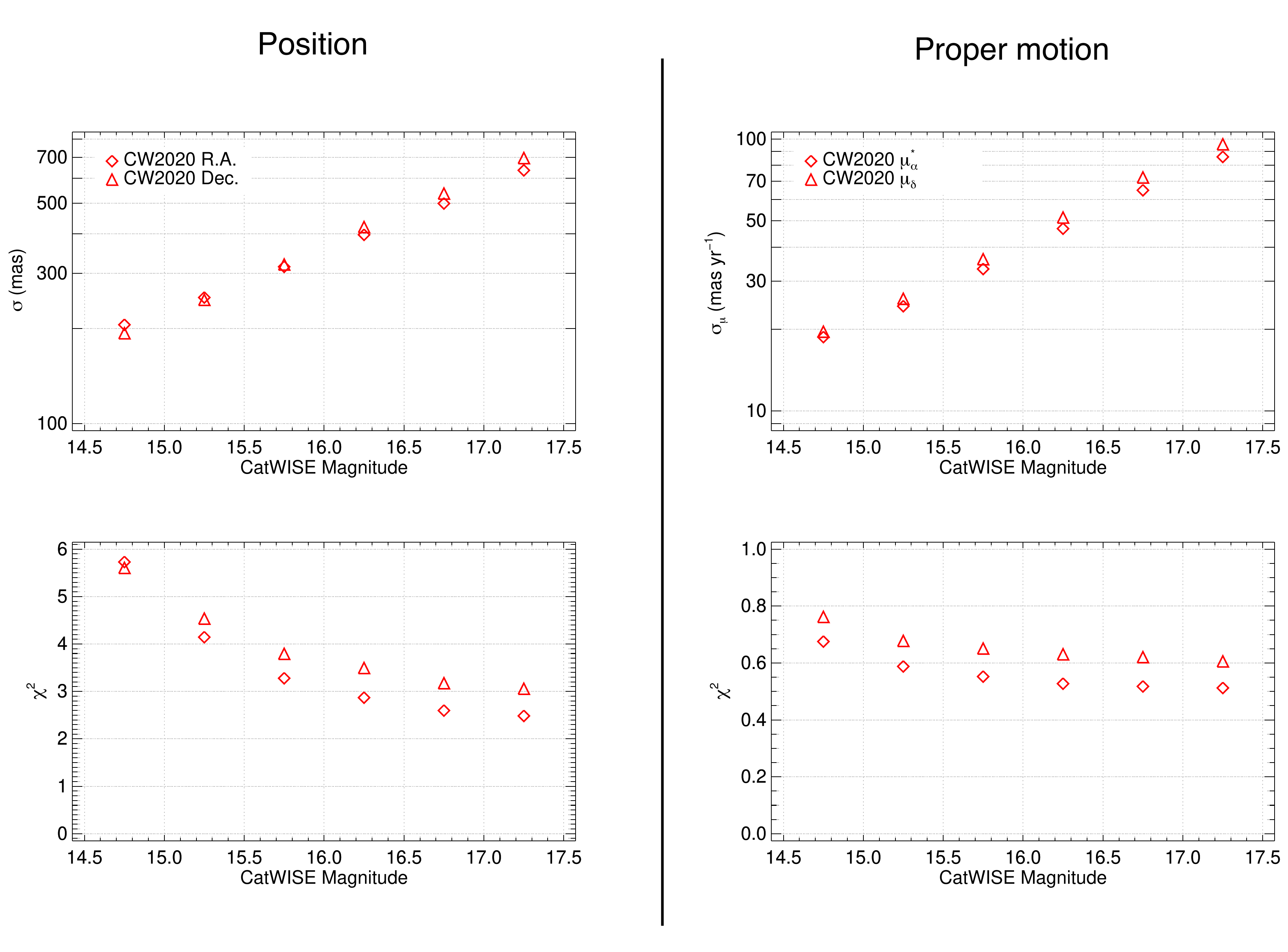}
    \caption{CatWISE2020 astrometric performance with respect to GPS1+. \textit{Top:} the 1-$\sigma$ dispersion between CatWISE2020 and GPS1+ R.A. and Dec. (left), and proper motion (right). \textit{Bottom:} the median $\chi^2$ computed taking into account CatWISE2020 Catalog uncertainties, GPS1+ catalog uncertainties, and the uncertainty introduced by the translation of GPS1+ positions to the CatWISE2020 epoch. \label{fig:gps1p}}
\end{figure*}

Figure~\ref{fig:gps1p} shows the results of the comparison. The one-sigma dispersion, presented in the top row, increase as a function of brightness, from $\sim200$\,mas for positions and $\sim20$\,mas\,yr$^{-1}$ for proper motions for the brightest sources, to $\sim650$\,mas for positions and $\sim90$\,mas\,yr$^{-1}$ for proper motions for the faintest sources. The performance at the faint end agrees well with the full-sky performance derived in the comparison against \textit{Gaia} DR2. At the bright end the omitted GPS1+ sources and resulting mismatches degrade the performance relative to the \textit{Gaia} DR2 comparison shown in Figures~\ref{fig:catwise_vs_gaia_fullsky}. The bottom row of Figure~\ref{fig:gps1p} shows that the median $\chi^2$ increase as a function of brightness, with a clearer trend for positions. Once again, this can be partly explained by the omission of bright GPS1+ sources and resulting mismatches in the brightest bins. 

\subsubsection{Astrometric Assessments in Selected Tiles \label{sec:astrom_4tiles}}

We assessed the astrometric performance of the CatWISE2020 Catalog in four representative tiles. These are the same four tiles we chose for the astrometric assessment of the CatWISE Preliminary Catalog \citep{Eisenhardt2020}, and are the following:

\begin{itemize}
    \item the COSMOS tile (tile 1497p015), representative of most of the sky, i.e. a field with average {\it WISE} coverage and average source density;
    \item the North Ecliptic Pole tile (NEP, tile 2709p666), a field with maximal {\it WISE} coverage and average source density;
    \item the South Ecliptic Pole tile (SEP, tile 0890m667) a field with maximal {\it WISE} coverage and high source density (the SEP tile contains part of the LMC); and
    \item the Galactic Center tile (GC, tile 2657m288) a field with average {\it WISE} coverage and maximal source density.
\end{itemize}

\begin{figure*}
\centering
\includegraphics[width=\textwidth]{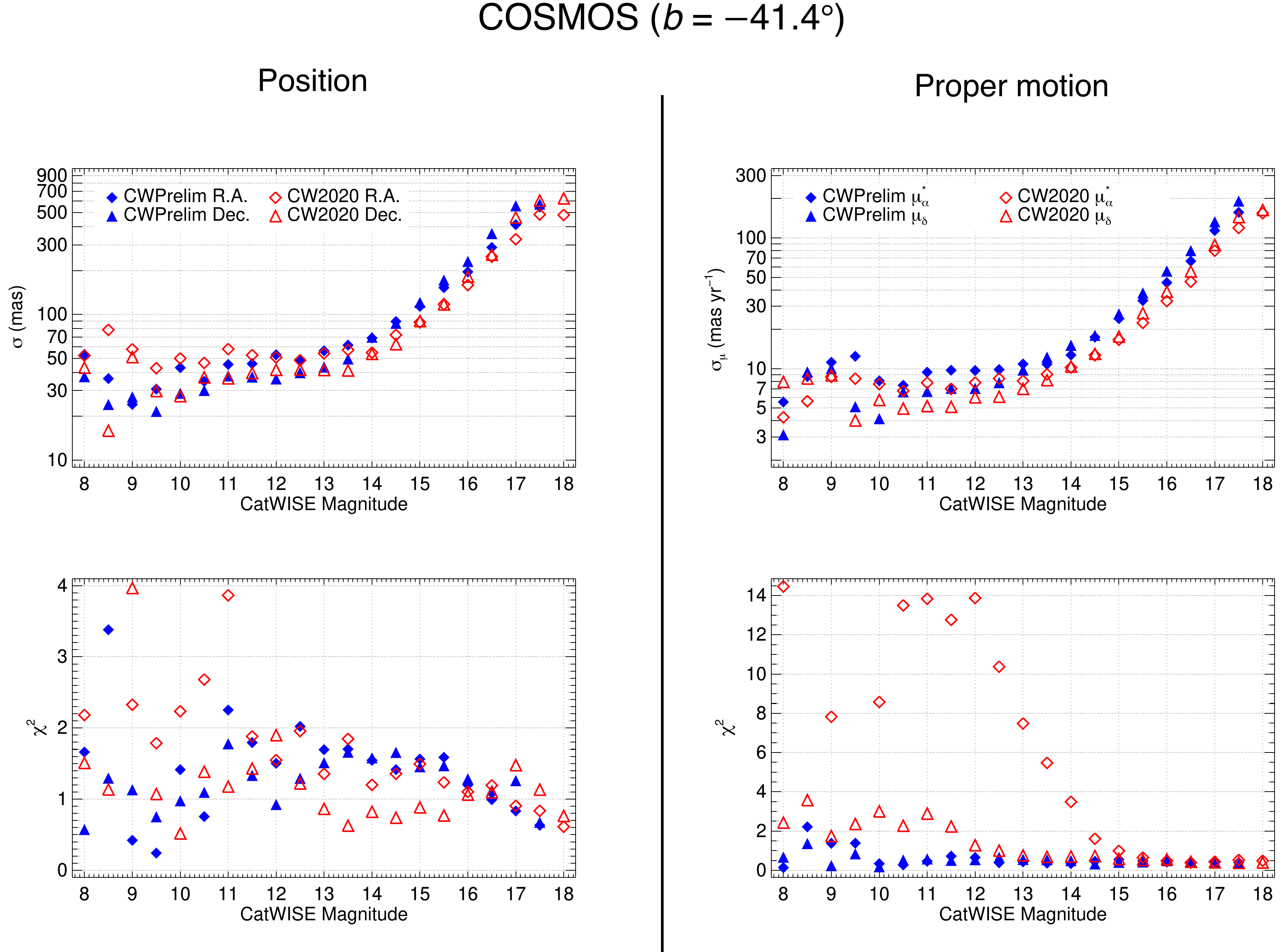}
\caption{CatWISE Preliminary Catalog (blue) and CatWISE2020 Catalog (red) astrometric performance with respect to \textit{Gaia} DR2 in the COSMOS tile (1497p015). The top row shows the 1-$\sigma$ dispersion between CatWISE and \textit{Gaia} R.A. (specifically, $\Delta\alpha\ \textrm{cos}(\delta)$) and Dec. (left), and proper motion (right). The bottom row shows the median $\chi^2$ computed taking into account CatWISE catalog uncertainties, \textit{Gaia} catalog uncertainties, and the uncertainty introduced by the translation of \textit{Gaia}'s positions to the CatWISE epoch.) \label{fig:catwise_vs_gaia_1497p015}}
\end{figure*}

\begin{figure*}
\centering
\includegraphics[width=\textwidth]{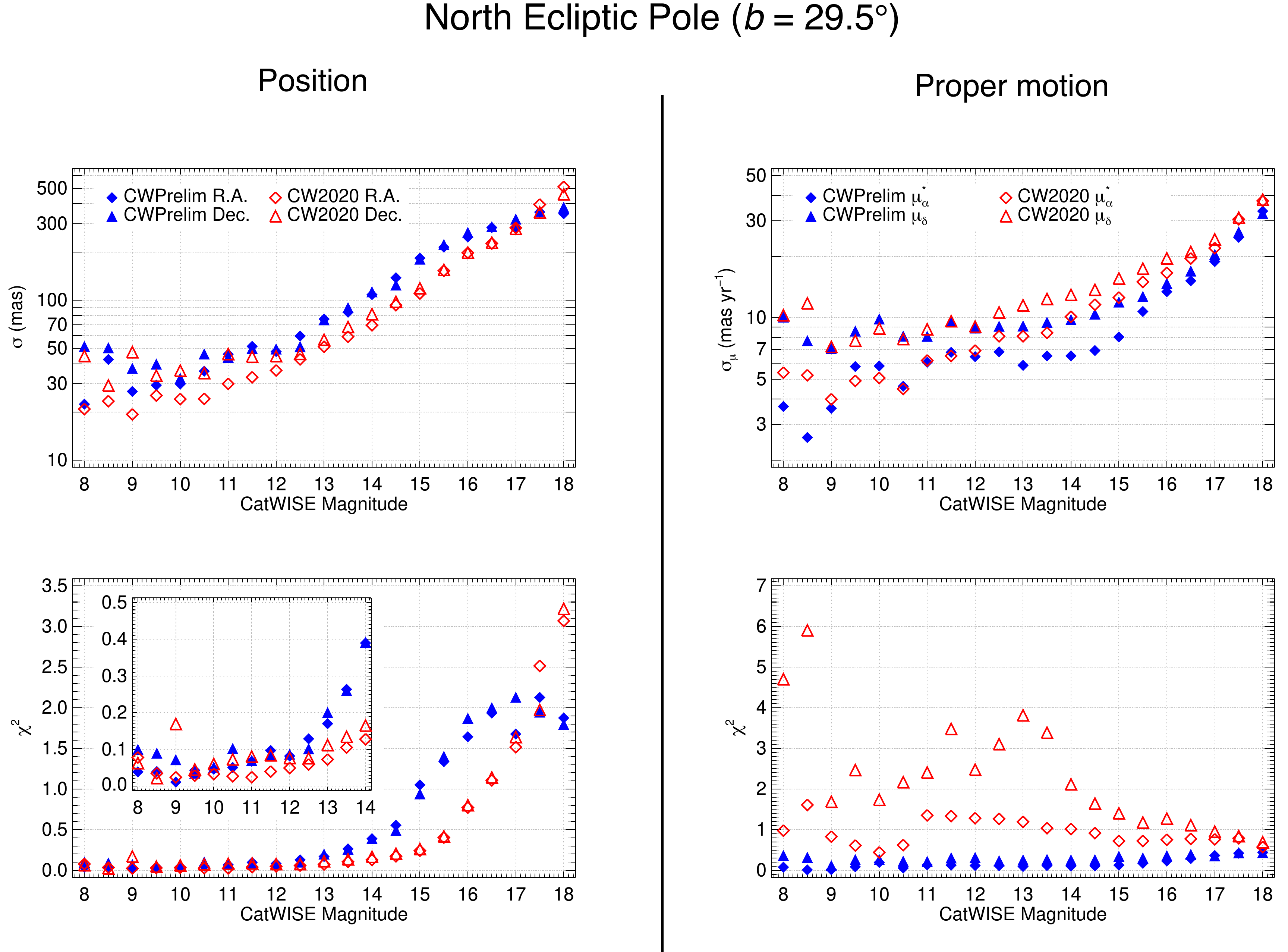}
\caption{Same as Figure~\ref{fig:catwise_vs_gaia_1497p015} but for the North Ecliptic Pole tile (2709p666) \label{fig:catwise_vs_gaia_2709p666}}
\end{figure*}

\begin{figure*}
\centering
\includegraphics[width=\textwidth]{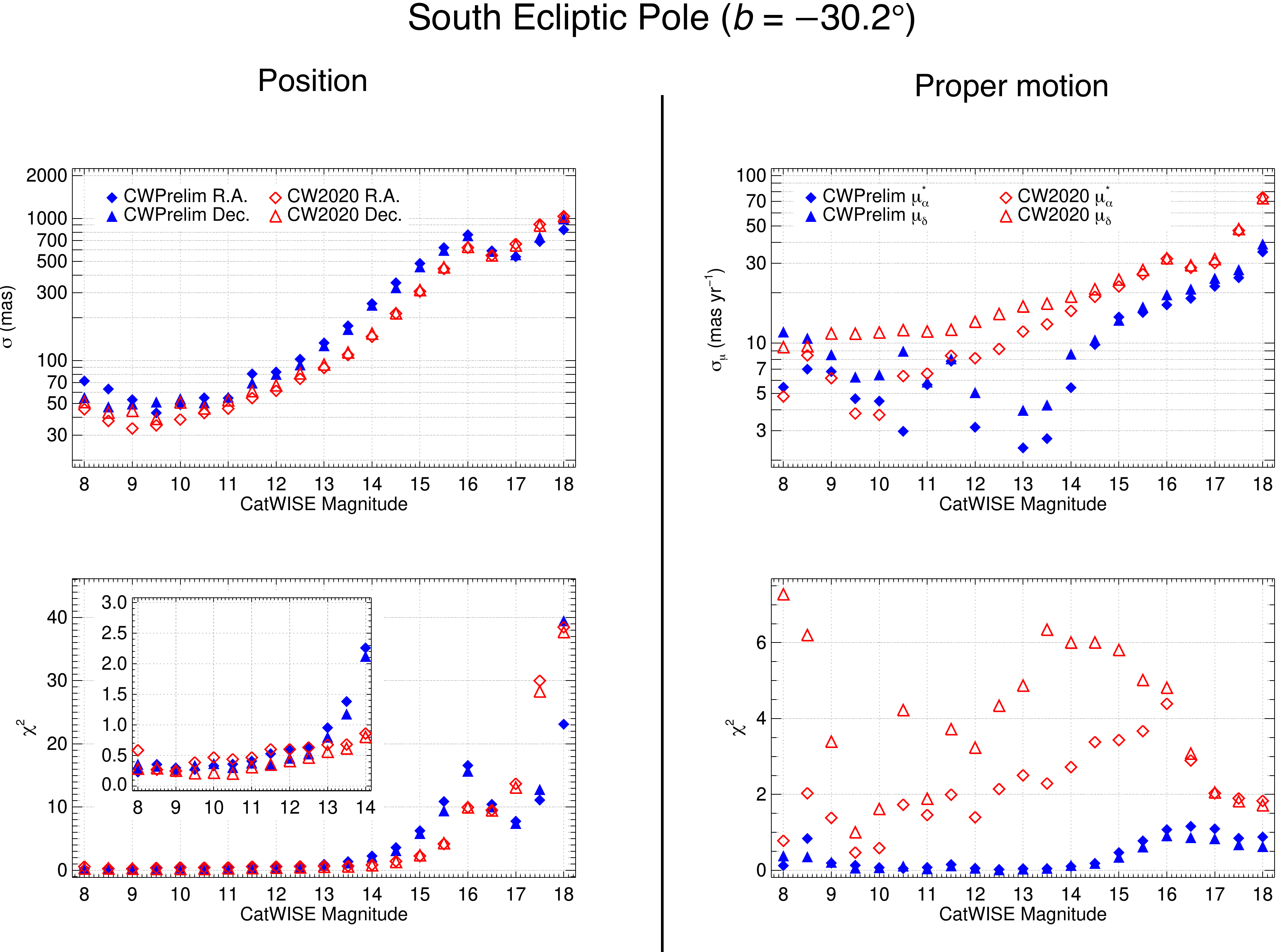}
\caption{Same as Figure~\ref{fig:catwise_vs_gaia_1497p015} but for the South Ecliptic Pole tile (0890m667). \label{fig:catiwse_vs_gaia_0890m667}}
\end{figure*}

\begin{figure*}
\centering
\includegraphics[width=\textwidth]{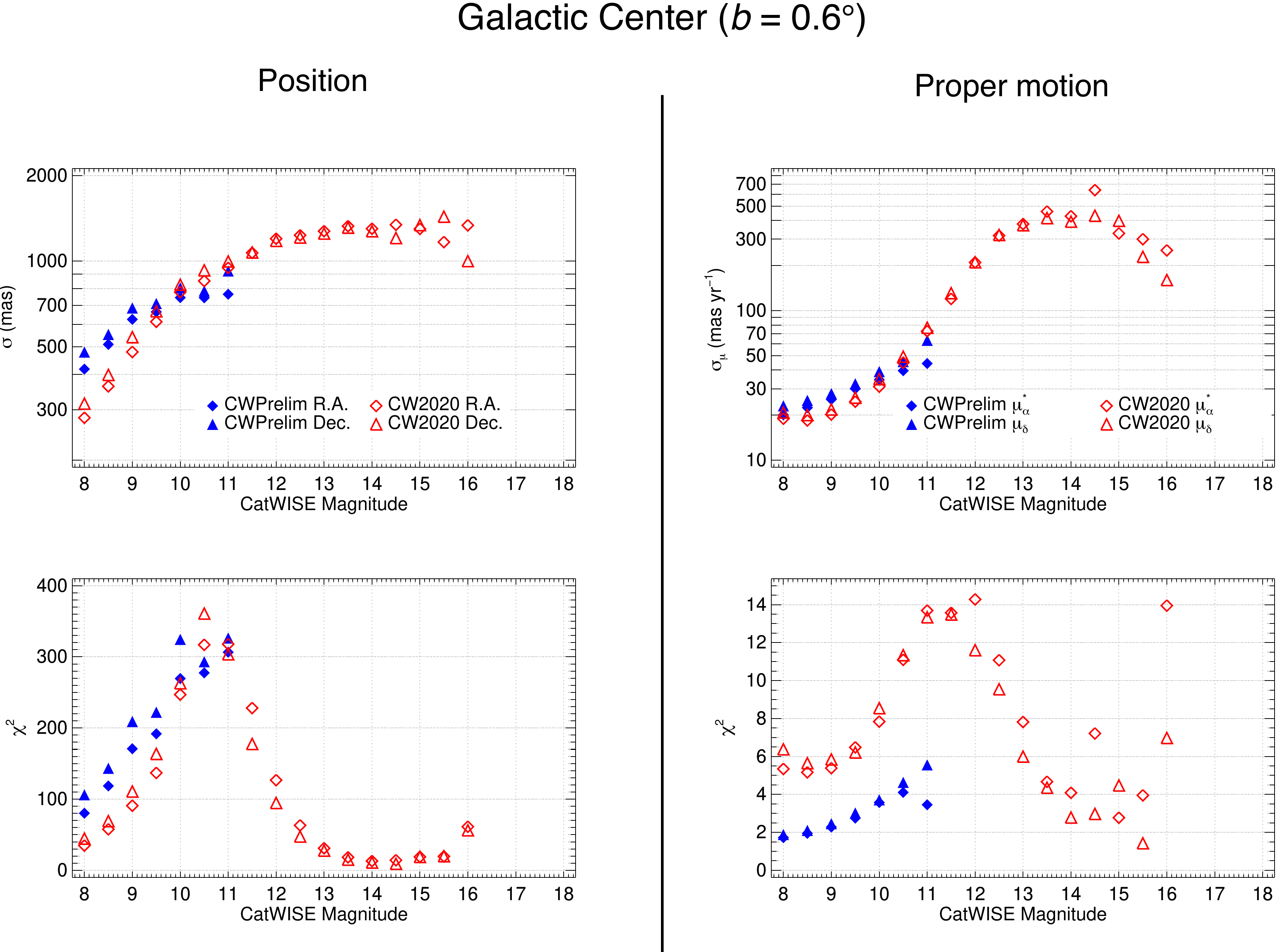}
\caption{Same as Figure~\ref{fig:catwise_vs_gaia_1497p015} but for the Galactic Center tile (2657m288) \label{fig:catwise_vs_gaia_2657m288}}
\end{figure*}

The astrometric performance of the CatWISE2020 Catalog was assessed following the same method described in \S\ref{sec:astrom_full_sky}. The results are shown in Figures~\ref{fig:catwise_vs_gaia_1497p015} to \ref{fig:catwise_vs_gaia_2657m288}. We quantitatively characterized the performance of the CatWISE2020 Catalog using the same ten metrics used in \citet{Eisenhardt2020}: 
\begin{itemize}

\item $\sigma_{\rm min}$ and $\sigma_{\mu,\,\rm min}$ are the accuracy floor for positions and motions, respectively, determined as the median dispersion with respect to \textit{Gaia} DR2 in the $8<$W1,W2$<10$\,mag interval. In the GC we restrict to $8<$W1,W2$<9$\,mag since the astrometric accuracy starts deteriorating significantly beyond W1,W2$\sim$9\,mag (see Figure~\ref{fig:catwise_vs_gaia_2657m288}). 

\item W1$_{\rm min}$, W2$_{\rm min}$, W1$_{\mu,\,\rm min}$, and W2$_{\mu,\,\rm min}$ are the W1 and W2 mag at which $\sigma_{\rm min}$ and $\sigma_{\mu,\,\rm min}$ are exceeded by no more than 20\,mas and 5\,mas yr$^{-1}$, respectively.

\item W1$_{500}$ and W2$_{500}$ are the W1 and W2 mag at which the accuracy on positions reaches 500\,mas.

\item W1$_{\mu,\,100}$ and W2$_{\mu,\,100}$ are the W1 and W2 mag at which the accuracy on motion reaches 100\,mas\,yr$^{-1}$.

\end{itemize}

The results for the four representative tiles are summarized and compared to the corresponding CatWISE Preliminary Catalog values in Table~\ref{table_ast_fields}.

\begin{deluxetable*}{lrlrlrlrl}
\tablecaption{CatWISE Astrometric Performance Evaluation
  Fields\label{table_ast_fields}}
\tablehead{
\colhead{} &            
\multicolumn{2}{c}{0890m667} &
\multicolumn{2}{c}{1497p015} &
\multicolumn{2}{c}{2657m288} &         
\multicolumn{2}{c}{2709p666} \\
\colhead{} & 
\multicolumn{2}{c}{SEP, LMC} & 
\multicolumn{2}{c}{COSMOS} & 
\multicolumn{2}{c}{GC} & 
\multicolumn{2}{c}{NEP}
}
\startdata
$l$ (deg) & \multicolumn{2}{c}{276.5} & \multicolumn{2}{c}{237.3} & \multicolumn{2}{c}{359.8} & \multicolumn{2}{c}{96.4} \\
$b$ (deg) & \multicolumn{2}{c}{$-30.2$} & \multicolumn{2}{c}{41.4} & \multicolumn{2}{c}{0.6} & \multicolumn{2}{c}{29.5} \\
$\beta$ (deg) & \multicolumn{2}{c}{$-89.6$} & \multicolumn{2}{c}{$-10.2$} & \multicolumn{2}{c}{$-5.4$} & \multicolumn{2}{c}{89.6} \\
\hline
 & Prelim. & 2020 & Prelim. & 2020 & Prelim. & 2020 & Prelim. & 2020 \\
\hline
Exp. & 7154 & 9230 & 90 & 140 & 86 & 133 & 7839 & 10000 \\
$\#$ & 71462 & 209518 & 58961 & 75639 & 63368 & 231239 & 61702 & 142975 \\
\hline
$\sigma_{\rm min}$ (mas) & 52.9 & 40.5 & 27.3 & 42.8 & 526.4 & 391.2 & 37.7 & 25.8 \\
W1$_{\rm min}$ (mag) & 11.0 & 11.5 & 12.5 & 14.5 & 8.0 & 8.5 & 12.0 & 12.0 \\
W1$_{500}$ (mag) & 15.1 & 16.0 & 17.0 & 17.5 & 8.4 & 9.0 & 18.5 & 18.0 \\
W2$_{\rm min}$ (mag) & 11.0 & 12.0 & 12.5 & 14.5 & 8.0 & 8.5 & 12.0 & 12.5 \\
W2$_{500}$ (mag) & 15.0 & 16.0 & 16.8 & 17.5 & 8.3 & 9.0 & 19.0 & 20.5 \\
\hline
$\sigma_{\mu, \rm min}$ (mas yr$^{-1}$) & 7.4 & 8.6 & 8.5 & 8.0 & 22.2 & 20.0 & 7.3 & 7.0 \\
W1$_{\mu, \rm min}$ (mag) & 14.5 & 13.5 & 13.5 & 14.5 & 9.0 & 9.0 & 15.5 & 14.5 \\
W1$_{\mu, 100}$ (mag) & 18.2 & 18.5 & 16.8 & 17.5 & $>11.0$ & 11.5 & $>19.0$ & 21.0 \\
W2$_{\mu, \rm min}$ (mag) & 14.5 & 13.5 & 13.5 & 14.5 & 9.0 & 9.0 & 15.5 & 14.0 \\
W2$_{\mu, 100}$ (mag) & $>$20.5 & 19.0 & 16.7 & 17.5 & $>$11.5 & 11.5 & $>$20.0 & 20.5 \\
\enddata
\tablecomments{ $l$, $b$, and $\beta$ are the Galactic longitude, Galactic latitude, and ecliptic latitude for the center of the tile, in degrees. Exp. indicates the number of exposures for the tile, $\#$ the number of sources (combining catalog and reject entries). The subsequent metrics are described in detail in \S\ref{sec:astrom_perf}.}
\end{deluxetable*}

In the COSMOS tile (Figure~\ref{fig:catwise_vs_gaia_1497p015}), the CatWISE2020 Catalog shows improved astrometric performance with respect to the CatWISE Preliminary Catalog in all metrics except the position precision floor. Table~\ref{table_ast_fields} shows that $\sigma_{\rm min}$ deteriorated from 27.3 to 42.8\,mas. Inspection of the top left panel of Figure \ref{fig:catwise_vs_gaia_1497p015} suggests that the asymptotic performance is actually comparable for the CatWISE2020 and Preliminary Catalog, with a precision floor of $\sim$40\,mas\,yr$^{-1}$. The CatWISE Preliminary Catalog however shows a dip in the measured motion sigma at W1,W2$\sim$9\,mag, which drives the $\sigma_{\rm min}$ to low values. Because the number of sources in this brightness regime is much lower compared to the fainter brightness regime (32 objects with W1,W2 $\leq$9.5\,mag), we suspect that that our metric may be affected by small number statistic fluctuations. The motion precision floor is not affected by the already mentioned systematic offsets, since in an individual tile the systematic offset is roughly constant over the entire tile area, and therefore does not impact the measured dispersion with respect to \textit{Gaia}. 

In the Galactic Center (Figure~\ref{fig:catwise_vs_gaia_2657m288}), the major gain of the CatWISE2020 Catalog with respect to the CatWISE Preliminary Catalog is the increased depth to which sources are recovered, with the CatWISE2020 Catalog now reaching down to 16th magnitude in W1 and W2. Most metrics also show improved or consistent performance, with the position precision floor ($\sigma_{\rm min}$) having improved significantly as a result of the more effective deblending. W1$_{\rm min}$ and W2$_{\rm min}$, as well as W1$_{500}$ and W2$_{500}$, are 0.5--0.7\,mag deeper in the CatWISE2020 Catalog. The motion metrics show comparable or slightly improved performance. The position $\chi^2$ are significantly improved as a result of the better deblending, while motion $\chi^2$ are worse because of the smaller floor on the motion uncertainty (\S\ref{sec:wphot}).

The NEP and SEP (Figure~\ref{fig:catwise_vs_gaia_2709p666} and \ref{fig:catiwse_vs_gaia_0890m667}) show improved performance for positions, but somewhat worse performance for motions. All position metrics either improved or remained constant in the CatWISE2020 Catalog, with the exception of W1$_{500}$ in the NEP, which is 0.5\,mag shallower. This metric, however, is close to the S/N 5 depth, and small number statistics noise is to be expected, because of the small number of CatWISE2020 sources that have a counterpart in \textit{Gaia} DR2 at such brightness. The motion metrics, on the other hand, are somewhat degraded. While the precision floor is at the same level achieved by the CatWISE Preliminary Catalog, the top right panel of Figures~\ref{fig:catwise_vs_gaia_2709p666} and \ref{fig:catiwse_vs_gaia_0890m667} shows that the CatWISE2020 motion dispersion is higher than in the CatWISE Preliminary Catalog across the entire magnitude range considered. Position $\chi^2$ are better at both poles, while motion $\chi^2$ are worse, once again because of the smaller uncertainty floor.

Overall, for faint sources all metrics in the CatWISE2020 Catalog retain the clear dependence on source density and coverage that was already noted by \citet{Eisenhardt2020} in the CatWISE Preliminary Catalog. The NEP (high coverage, average density) shows better W1$_{500}$, W2$_{500}$, W1$_{\mu,\,100}$ and W2$_{\mu,\,100}$ than the SEP (high coverage, high density) and COSMOS (average coverage, average density), which are in turn better than the GC (average coverage, high density).

\subsubsection{Fast Movers}
\label{sec:fast_movers}
In \citet{Eisenhardt2020} we discussed the performance of the CatWISE Preliminary Catalog on a sample of cold, fast moving brown dwarfs in the Solar neighborhood. Re-assessment of the CatWISE2020 Catalog performance on the same sample of brown dwarfs reveals that the motion accuracy is comparable in the two versions of the catalog. Moreover, thanks to a more effective deblending of partly blended sources, the CatWISE2020 pipeline successfully measures WISE J163940.83--684738.6 and WISEPC J205628.90+145953.3, the only two objects in our sample of 19 test objects that were missing from the CatWISE Preliminary Catalog.

One feature common to all of the fast moving brown dwarfs considered in this analysis is that their fast motion leads to multiple (spurious) detections, since they are essentially ``smeared'' in the full-depth unWISE coadds used for source detection. These multiple apparitions (up to seven for the fastest objects) are then passed through our photometry- and motion-measuring software. The first detection processed by \textsc{WPHOT} gets accurate motion and magnitude measurements, while the subsequent detections get progressively worse measurements. Those with the lowest S/N are typically discarded in the Reject Table, but the higher S/N ones remain in the catalog. For example the fastest moving brown dwarf, WISE~J085510.83--071442.5, appears five times in the CatWISE2020 Catalog, with three of those five apparitions having motion measurements consistent with the literature values.

As briefly discussed in \S\ref{sec:intro}, the performance of the CatWISE2020 Catalog on fast moving sources was further assessed by comparing the CatWISE2020 offset-corrected total measured motion for a sample of ultra-cool dwarfs within 20\,pc from the Sun to values reported in the literature \citep[and references therein]{Kirkpatrick2020}. The full sample consists of 224 objects, all of them with a counterpart in the CatWISE2020 Catalog, and with good quality proper motion measurements (i.e. $\mu_{\rm tot} / \sigma \mu_{\rm tot} \geq 3$). Figure~\ref{fig:fast_movers} shows a comparison of the CatWISE2020 measurements for these 224 objects to values reported in the literature. The agreement is excellent, and does not show strong dependence on total proper motion nor brightness, since the sample includes objects as bright as W2$\sim$7.3\,mag and as faint as W2$\sim$16.7\,mag. The one-sigma dispersion is $\sim$29\,mas\,yr$^{-1}$, while the median offset is $\sim$12\,mas\,yr$^{-1}$. Many of the 224 objects in this sample are too faint at optical wavelengths to be seen by \textit{Gaia}. Therefore, the CatWISE2020 Catalog crucially complements the ESA mission for late type stars and brown dwarfs. 

\section{Data Access}
\label{sec:access}
The merged files for the 18,240 tiles for the CatWISE2020 Catalog and Reject Table are available from IRSA\footnote{\url{https://irsa.ipac.caltech.edu/cgi-bin/Gator/nph-scan?mission=irsa&submit=Select&projshort=WISE}} in the WISE/NEOWISE Enhanced and Contributed Products area. IRSA's catalog search tools allow for complex search queries. IRSA also hosts the AllWISE Explanatory Supplement \citep{Cutri2013}, which provides full details on the AllWISE processing algorithms, and includes descriptions of the AllWISE Catalog columns, many of which are applicable to the CatWISE data products. \S\ref{sec:columns} provides additional information about CatWISE2020 columns.

The individual tile files have also been transferred to a data repository at the National Energy Research Scientific Computing Center\footnote{\url{https://portal.nersc.gov/project/cosmo/data/CatWISE/2020}} (NERSC), and are available in 18,240 pairs of gzipped ASCII files (one catalog and one reject file per tile) in IPAC table format, organized into 359 directories, one for each decimal degree of right ascension from 0\degree\ to 358\degree\ (there are no tiles whose ID begins with 359). Text files providing the format and a brief description of the columns in the catalog and reject files are also provided there. As for the CatWISE Preliminary Catalog, the catalog and reject files for tiles near the ecliptic poles \citep[listed in Table 1 in][]{Eisenhardt2020}, where a single PSF per band was used for processing, include the string ``opt0" in their names. Files for tiles where different PSFs were used for ascending and descending scans include the string ``opt1" in their names. 

Current information about CatWISE data products and links to the data on IRSA and NERSC are provided at the CatWISE website \url{https://catwise.github.io}.

\acknowledgments

We thank the anonymous referee for their careful review and thoughtful suggestions. CatWISE uses data products from WISE, which is a joint project of the University of California, Los Angeles, and the Jet Propulsion Laboratory (JPL)/California Institute of Technology (Caltech), funded by the National Aeronautics and Space Administration (NASA), and from NEOWISE, which is a JPL/Caltech project funded by NASA.  Characterization of CatWISE performance uses data from {\it Gaia} and from {\it Spitzer}. CatWISE is led by JPL/Caltech, with funding from NASA's Astrophysics Data Analysis Program (ADAP), and is also supported in part by ADAP grant NNH17AE75I at Lawrence Berkeley Laboratory. FM acknowledges support from the NASA Postdoctoral Program at the Jet Propulsion Laboratory, administered by Universities Space Research Association under a contract with NASA. FM also acknowledges support from grant 80NSSC20K0452 under the NASA Astrophysics Data Analysis Program. FM would like to thank Dongwei Fan for help with accessing the GPS1+ data via che China-VO.

\appendix

\section{Caveats}
\label{sec:caveats}

The CatWISE2020 Catalog contains a number of features that users should be aware of. Among these are:
\begin{itemize}
    \item The CatWISE2020 Catalog represents a dramatic improvement over the CatWISE Preliminary Catalog in terms of depth and completeness in the Galactic plane. However, the CatWISE2020 Catalog astrometric performance degrades as source density increases. Figures~\ref{fig:map_positions}--\ref{fig:map_pms_bins} illustrate this.
    \item The completeness for bright sources is low (Figure \ref{fig:brightCompleteness}). Users interested in complete samples brighter than 4.5\,mag in WISE bands should use AllWISE. However, the CatWISE2020 Catalog has much better reliability for bright sources than does the CatWISE Preliminary Catalog (Figure \ref{fig:brightReliability}). 
    \item Small systematic offsets are present in CatWISE2020 astrometry with respect to {\it Gaia} DR2 (see \S\ref{sec:unwise}, \S\ref{sec:astrom_perf}, and Figure~\ref{fig:cw2020_astrometry_offset}). Because adjustments were made to the WCS for the AllWISE epochs for the CatWISE Preliminary Catalog but not for the CatWISE2020 Catalog, these systematic offsets are different between the two catalogs. Table~\ref{tab:astro_offsets} provides the corresponding systematic corrections to the CatWISE2020 Catalog position and motion for each tile.
    \item Because of coordinate singularities, CatWISE tile Point Spread Functions \citep[see \S 3.2 of][]{Eisenhardt2020} within a few degrees of the equatorial poles used an unnecessarily large range of rotation angles, resulting in smearing of these PSFs.
    \item Magnitude and position uncertainties occasionally round to 0.
    \item The aperture magnitudes are the result of averaging fluxes in the CatWISE2020 Catalog, while in the CatWISE Preliminary Catalog, the aperture magnitudes from the ascending and descending epochs were averaged.
    \item In the NERSC version of the CatWISE2020 data products, the standard aperture magnitude uncertainties, \textit{w1sigm} and \textit{w2sigm}, are often 0.0 or 1.0 because of an error in the CatWISE2020 pipeline when averaging fluxes. The corresponding magnitudes, \textit{w1mag} and \textit{w2mag}, are correct, as are the individual aperture magnitude uncertainties. Users who are interested in standard aperture magnitudes should use the \textit{w1sigm\_2} and \textit{w2sigm\_2} for the uncertainties, as those are identical to what the \textit{w1sigm} and \textit{w2sigm} should have been. In the IRSA version, this error was corrected on 2020 Aug. 25, and \textit{w1sigm} and \textit{w2sigm} values retrieved after this date are correct. 
    \item A floor of 1 mas yr$^{-1}$ was imposed on the tabulated motion uncertainties for the CatWISE2020 data products, while for the CatWISE Preliminary data products the floor was 10 mas yr$^{-1}$. The lower CatWISE2020 uncertainty floor is now significantly smaller than the minimum measured scatter with respect to {\it Gaia} motion, as illustrated in the right column of Figures~\ref{fig:catwise_vs_gaia_1497p015} to \ref{fig:catwise_vs_gaia_2657m288}.
    \item CatWISE2020 source designations ({\it source\_name}) include a lower case letter suffix to distinguish sources that would otherwise have the same designation. In the CatWISE Preliminary Catalog, the first such source (usually the brightest) did not receive a suffix while subsequent such sources did, beginning with ``b". In the CatWISE2020 data products, the first such source receives an ``a" suffix.
    \item Sources in the updated unWISE catalog detection list may be omitted from the CatWISE2020 Reject Table if they are too near tile edges, probably due to the fitting region used by WPHOT being truncated. This does not affect the CatWISE Preliminary Reject Table because the MDET software does not find sources this close to tile edges.
    \item As discussed in \S\ref{sec:mdet}, resolved galaxies result in multiple detections by \textit{crowdsource} and hence multiple entries in the CatWISE2020 Catalog. Users interested in local overdensities in the galaxy distribution (such as for clustering analyses or detection of galaxy clusters) should be aware of this feature and take precautions to avoid misinterpretation. 
\end{itemize}

\end{document}